\documentclass[12pt,preprint]{aastex}
\begin{document}

\title{The Change of the Orbital Periods Across Eruptions and the Ejected Mass For Recurrent Novae CI Aquilae and U Scorpii}
\author{Bradley E. Schaefer\affil{Physics and Astronomy, Louisiana State University, Baton Rouge, LA 70803}}

\begin{abstract}

I report on the cumulative results from a program started 24 years ago designed to measure the orbital period change of recurrent novae (RNe) across an eruption.  The goal is to use the orbital period change to measure the mass ejected during each eruption as the key part of trying to measure whether the RNe white dwarfs are gaining or losing mass over an entire eruption cycle, and hence whether they can be progenitors for Type Ia supernovae.  This program has now been completed for two eclipsing RNe; CI Aquilae (CI Aql) across its eruption in 2000 and U Scorpii (U Sco) across its eruption in 1999.  For CI Aql, I present 78 eclipse times from 1991-2009 (including four during the tail of the 2000 eruption) plus two eclipses from 1926 and 1935.  For U Sco, I present 67 eclipse times, including 46 times during quiescence from 1989-2009, plus 21 eclipse times in the tails of the 1945, 1999, and 2010 eruptions.  The eclipse times during the tails of eruptions are systematically and substantially shifted with respect to the ephemerides from the eclipses in quiescence, with this being caused by shifts of the center of light during the eruption.  These eclipse times are plotted on an O-C diagram and fitted to models with a steady period change ($\dot{P}$) between eruptions (caused by, for example, conservative mass transfer) plus an abrupt period change ($\Delta P$) at the time of eruption.  The primary uncertainty arises from the correlation between $\Delta P$ with $\dot{P}$, such that a more negative $\dot{P}$ makes for a more positive $\Delta P$.  For CI Aql, the best fit is  $\Delta P=-3.7^{+9.2}_{-7.3} \times 10^{-7}$.  For U Sco, the best fit is $\Delta P=(+43 \pm 69) \times 10^{-7}$ days.  These period changes can directly give a dynamical measure of the mass ejected ($M_{ejecta}$) during each eruption with negligible sensitivity to the stellar masses and no uncertainty from distances.  For CI Aql, the one-sigma upper limit is $M_{ejecta}<10 \times 10^{-7}$ M$_{\odot}$.  For U Sco, I derive $M_{ejecta}=(43 \pm 67) \times 10^{-7}$ M$_{\odot}$.

\end{abstract}

\section{Introduction}

Recurrent novae (RNe) are interacting binaries where a non-degenerate companion star pours matter onto a white dwarf (WD) through Roche lobe overflow, which accumulates on the surface until the pressure increases to the point where a runaway thermonuclear explosion sends out an expanding shell (Payne-Gaposchkin 1964; Bode and Evans 2008).  Only ten RNe are currently known in our galaxy (Schaefer 2010).  The nova event on RNe are identical to those of ordinary classical novae, but RNe are different from classical novae in that their recurrence time scale is short enough that multiple eruptions have been detected. 

A short recurrence time scale (say, shorter than a century or so) requires two conditions.  First, the WD must be near the Chandrasekhar mass, with something like a mass $M_{WD}\gtrsim1.3$ M$_{\odot}$, so that the critical pressure can be achieved with a relatively small accreted mass.  Second, the accretion rate must be unusually high, with a rate of $\dot{M}\sim10^{-7}$ M$_{\odot}$ yr$^{-1}$, so that even this small critical mass can be accumulated quickly.

On the face of it, with a near Chandrasekhar mass WD collecting mass at a high rate, the WD will soon reach the Chandrasekhar mass and explode as a Type Ia supernova.  With this logic, the RNe systems are a strong candidate for being the progenitors of Type Ia supernovae.  The progenitor problem has been one of the more important challenges in stellar astrophysics for many decades (Parthasarathy et al. 2007; Livio 2000; Branch et al. 1995; Trimble 1984; Whelan \& Iben 1973).  In the last decade, the identity of the progenitor systems has taken on an over-riding importance for many aspects of cosmology.  The reason is that the Type Ia supernovae are now the premium tool for precision cosmology, but there is the outstanding problem that the evolution of the standard candle relation (as we look to a more distant metal-poor Universe) depends on the progenitor class.  The evolution effects might be comparable to the cosmological effects (Dom\'inguez et al. 2001), so the progenitor problem must be solved before supernova can be used for precision cosmology.  With this, the testing of RNe as progenitors becomes a critical part of solving the progenitor problem.

The primary problem with the above logic is that the RN WDs not only accrete mass continuously, but they also eject substantial amounts of mass each eruption.  So the real question is whether the WD is gaining mass as averaged over each eruption cycle.  That is, is the ejected mass ($M_{ejecta}$) greater or less than the mass accreted between eruptions?  For an average accretion rate of $\dot{M}$ and an inter-eruption time interval of $\Delta T$, the question is whether $M_{ejecta} \le \dot{M} \Delta T$?  If `yes', then the WD is gaining mass over each eruption cycle, the WD mass will soon increase to the Chandrasekhar limit, and a Type Ia supernova is inevitable.  Out of the three quantities ($M_{ejecta}$, $\dot{M}$, and $\Delta T$), the poorest known is the ejecta mass.  So the evaluation of RNe as progenitors is largely come down to measures of $M_{ejecta}$.

The traditional method for determining $M_{ejecta}$ is to measure the nebular emission-line fluxes, derive the emission measure, estimate the nebular volume, and then derive the ejecta mass.  For a review, see Gallagher \& Starrfield (1978).  Unfortunately, this method is highly model dependent, requiring assumptions on the poorly known distance, the filling factor, the composition, and that a steady state is present.  In addition, the volume of the shell depends on the cube of the expansion velocity, but it is unclear as to which velocity to adopt from an emission line profile (e.g., HWZI or HWHM) and this leads to additional uncertainties of roughly two cubed.  Typical uncertainties in these assumptions leads to order-of-magnitude errors, so it is not surprising that published values for individual novae events range up to two orders-of-magnitude discrepant (Schwarz 2002; Shaviv et al. 2002; Vanlandingham et al. 2002: Gallagher \& Starrfield 1978).  Theoretical models have similar scatter.  So in all, we can only agree with the Gallagher \& Starrfield review that ``The masses of material expelled by active novae are poorly known."  If we are to ever understand whether RNe are progenitors, we must have a much more accurate method to measure $M_{ejecta}$.  An analysis of the methods and a summary of the specific reported values for both CI Aql and U Sco are presented in Appendix A.

	A solution for getting an accurate value of $M_{ejecta}$ is to measure the change in the orbital period across the eruption event ($\Delta P$), as this will yield a confident dynamical measure of $M_{ejecta}$.  The idea is that by Kepler's Law (and the conservation of angular momentum) the orbital period ($P$) will change when the system's mass changes (due to the loss of the ejected mass) so that a measure of $\Delta P$ will directly give the mass loss.  Schaefer \& Patterson (1983) derive that $M_{ejecta}=(M_{wd}/A)(\Delta P/P)$, where $M_{wd}$ is the white dwarf mass and $A$ is a parameter depending on the fraction of matter captured by the companion and the specific angular momentum of the ejecta.  The value of $A$ is known to be unity to within perhaps 10\%.  The orbital period will change from its pre-eruption value ($P_{pre}$) to its post-eruption value ($P_{post}$), with $\Delta P = P_{post} - P_{pre}$.  The value of $P_{post}$ is generally easy to measure accurately over long time spans with either spectroscopic or photometric time series.  The white dwarf mass will always be known to perhaps 30\% uncertainty or $\sim 10\%$ for RNe.  In all, if we can simply measure the value of $P_{pre}$, then we can get an accurate measure of $M_{ejecta}$.

Schaefer \& Patterson (1983) realized that $P_{pre}$ could be measured for novae with relatively long orbital periods and with deep eclipses by going back to archival photographic plates.  That is, with approximate eclipse times (from the times of plates showing the nova at the minimum of the eclipses) spread over many decades, the $P_{pre}$ can be measured with high accuracy.  At the time, only one known nova (BT Mon, Nova Mon 1939) could possibly have this program completed.  Schaefer \& Patterson found that $P_{pre}=0.3338010$ day, $P_{post}=0.3338141$ days, $\Delta P$ was 40 parts per million with the period {\it increasing} across the eruption, and $M_{ejecta}=3\times10^{-5}$ M$_{\odot}$.  This single measure of $M_{ejecta}$ is the only reliable value known.

In 1987, I realized that the way to measure more reliable values of $M_{ejecta}$ is to look at RNe, as only in these cases can we know which stars to examine {\it before} their next eruption.  (Livio 1991 later independently had the same idea.)  But at the time, the majority of RNe had no known orbital periods, and those that did were all uselessly long for this task.  So I started a program for seeking photometric orbital modulation.  This succeeded with the discovery of $P$ for five RNe (Schaefer 1990; Schaefer et al. 1992;  Schaefer 2009).  One of the early discoveries was that the RN U Scorpii (U Sco) was a deeply eclipsing binary with a 1.23 day orbital period.  The deep eclipses and manageable period enables the program of getting an extremely accurate measure of $P_{pre}$ (after the 1987 eruption).  An early report with an accurate period was in Schaefer \& Ringwald (1995), with all subsequent eclipse times coming out only in 2011 with this paper.  All it takes is to measure many eclipse times, await the subsequent eruption (which came in 1999), and then measure the post-eruption period.  In 1988, I knew that this would play into the even-then venerable progenitor problem, as well as that this would be a decades-long program.

In 2000, the inconspicuous old nova CI Aquilae (CI Aql) had a {\it second} observed nova eruption and joined the ranks of the RNe.  The first eruption was poorly observed in 1917 (Schaefer 2010), while Schaefer (2004) discovered a previously unrecognized eruption in 1945.  By great good fortune (actually, more due to systematic long and hard work), CI Aql had already been identified as a deep eclipsing binary (Mennickent \& Honeycutt 1995).  They recorded pre-eruption light curves (from which eclipse times can be derived) from 1990-1995.  Thus, CI Aql had a $P_{pre}$ measured, its eruption had already occurred, so all that was needed was to measure an accurate $P_{post}$.

This paper records the results of my program, started 24 years ago, to measure the period change across RN eruptions so as to measure $M_{ejecta}$, with application to the Type Ia supernova progenitor problem.  Two on-line tables will give essentially all the magnitudes in quiescence for both U Sco and CI Aql.  Eclipse light curves and templates are given in Section 3, with eclipse times in Section 4.  The physics of the changes in the orbital periods, as well as the primary tool of the O-C diagram, are presented in Appendix B.    Sections 5 and 6 derive the $P_{pre}$, $P_{post}$, and $\Delta P$ values for CI Aql and U Sco.  In Section 7, I derive the $M_{ejecta}$ for both RNe.  The higher analysis of these results are reserved for a separate paper.

\section{The Observations}

Since 1987, I have accumulated a large number of observations of CI Aql and U Sco in quiescence.  Journals of the observations are given in Table 1 for CI Aql and Table 2 for U Sco.  Unfortunately, my U Sco observations from 1987 (Schaefer 1988) were not recoverable from magnetic tape, and it is ironic that century older astronomical plates still show the U Sco field.  Most of the observations were made as time series photometry, where repeated CCD images were taken in rapid succession throughout eclipses.  For CI Aql, I report a total of 4960 magnitudes on 387 nights.  For U Sco, I report a total of 2382 magnitudes on 116 nights.  Each CCD image underwent the standard processing (bias subtraction, flat fielding) and quality checks (e.g., bad columns, poor signal-to-noise ratio).  Photometry was all performed with the APPHOT package in IRAF, which is fine for the uncrowded stars.  The RN magnitudes were all performed differentially with respect to a suite of three comparison stars, with explicit coordinates and BVRIJHK magnitudes as given in Schaefer (2010).

The resulting magnitudes are presented in full in Table 3 for CI Aql and in Table 4 for U Sco.  The first column lists the heliocentric Julian Date (HJD) for the middle of each exposure.  The second column gives the photometric band for each reported magnitude.  (Standard bands are indicated with one letter, while unfiltered CCD images with the zeros set by the V-band and R-band comparison star magnitudes are identified as `CV' and `CR' respectively.)  The third column gives the derived magnitude for the RN.  The fourth column gives the one-sigma error for the magnitude, where the statistical error (as reported by APPHOT) is added in quadrature with the estimated systematic error of 0.015 mag.  The last column gives the orbital phase as calculated by the best linear ephemeris (see below) for the last decade.

I have also spent much time looking through archival plate collections for magnitudes in quiescence.  The goal was to find old plates that showed the RNe in eclipse so as to get very old eclipse times.  I have now examined all the plates for CI Aql and U Sco at Harvard College Observatory.  I have also obtained scans of various deep Schmidt plates from the Palomar Observatory, the European Southern Observatory, and the Anglo-Australian Observatory, with these providing a small number of early magnitudes.  In addition, all the archives at the Maria Mitchell Observatory (see Robinson, Clayton, \& Schaefer 2007), Sonneberg Observatory, Cerro Tololo Schmidt, and the Tautenberg Schmidt were exhaustively searched, without finding any useful images.  All these photographic measures are also recorded in Tables 3 and 4.

For U Sco, I have one additional eclipse time, during quiescence in June 1990, from R. Wade (Pennsylvania State University) by private communication, based on continuum fluxes in a series of spectra.  With a four night run on the Cerro Tololo 4-m telescope, Wade covered one minimum on the first night, an ingress on the second night, and the remaining two nights were clouded out.  The spectra were taken with the Folded Schmidt camera plus the TI CCD with grating 450 (II) and the Corning 9780 order filter.  Almost all exposure times were 1500 seconds.  The observed magnitudes were taken with IRAF software by measuring across the spectra from 0.47-0.48 micron wavelength.  The long slit was simultaneously on U Sco and the star labeled `Check' (Schaefer 2010).  The U Sco magnitude is with respect to that of Check, which has a B magnitude of 18.60.  The scatter from outside eclipse suggests a one-sigma uncertainty of around 0.1 mag.  On the first night, only five measures are near the bottom of the eclipse, one point giving the egress, and several points giving the ingress.  A parabola fitted to the last eight points of the first night give a chi-square of 10.4 for 4 degrees of freedom, and a time of minimum of HJD 2448043.7262$\pm$0.0042.  The ingress light curve on the second night is not useable because the non-standard magnitude is not calibrated for the minimum or shape of the eclipse light curve. 

\section{Eclipse Light Curves and Templates}

Eclipse light curves have already been presented in Schaefer (2010) and Hachisu, Kato, \& Schaefer (2002) for CI Aql, and in Schaefer (1991; 2010) and Schaefer \& Ringwald (1995) for U Sco.  In this section, I will present new light curves with various views as well as templates of average light curves.  I hope that modelers of these RNe can use these templates as explicit tests for their models.

Figure 1 shows the folded light curve of CI Aql from 1991 to 1996.  We see a prominent secondary eclipse, a prominent ellipsoidal effect, a symmetric light curve with the two maxima equal, and no large amplitude flickering.  Figure 2 shows the folded light curve of CI Aql in three intervals throughout 2001 to illustrate the changing brightness as CI Aql fades in the late tail of its eruption in 2000. Flickering is apparent in the tail, and the light curves in the tail are consistent in shape with each other, suggesting that the tail light is some uneclipsed isotropic fading source.  Figure 3 shows 3221 magnitudes in quiescence  (all the V-band magnitudes from 2002 to 2009).  Many points come out.  First, CI Aql has occasional flickering up to a third of a magnitude.  Second, this flickering only appears outside of eclipse, apparently between orbital phases of 0.13-0.84, with this result suggests the hot spot as being the source of the flickering.  Third, CI Aql is brighter by 0.08 mag at the phase 0.25 elongation than at the phase 0.75 elongation.  This unusual trait is shared by other RNe (Schaefer 2010).  Fourth, CI Aql does appear to have a secondary eclipse with a duration comparable to the primary eclipse and an amplitude of around 0.1 mag.  Figure 4 shows superposed close-ups of eight well-observed eclipse light curves.  There is no significant evidence for flickering during the eclipse.  However, the magnitudes at minimum vary up and down by about 0.2 mag. Therefore, a substantial part of the V-band light is coming from a region that is not eclipsed.  This is consistent with the calculation of Hachisu, Kato \& Schaefer (2003) that the back edge of the accretion disk is not eclipsed, so that modest changes in the accretion rate (and hence the disk brightness) will cause the light at minimum to vary somewhat.

Figure 5 shows the folded light curve of U Sco from 1988 to 1997 in the B-band.  We see a deep eclipse with no  flickering, yet for which the egresses and ingresses do not track identically from eclipse to eclipse.  Outside of eclipse, the brightness level apparently changes substantially with flickering light added on top.  No secondary eclipse is visible  This pre-eruption light curve appears indistinguishable from the B-band light curve for 2001 to 2009.  Figure 6 shows the folded light curve from 2001 to 2009 in the I-band.  Again, there is no flickering during eclipse.  As in the other bands, the brightness levels for the uneclipsed system at the start of ingress and the end of ingress vary by up to half a magnitude.  A prominent secondary eclipse is visible, as the companion star contributes a significant fraction of the system light in the I-band.  Figure 7 shows the folded light curve for the most homogenous set of data that I have, the eight eclipses from 2008 to 2009 in the I-band.  We see that the eclipses are flat bottomed from phases -0.010 to +0.010, pointing to total eclipses.  Nevertheless, the magnitude at minimum does apparently jitter around by about a tenth of a magnitude.

Table 5 presents various light curve templates for both U Sco and CI Aql.  That is, an average magnitude is given as a function of phase for various cases.  The cases are for CI Aql from 1991-1996 and 2002-2009 in the V-band, and for U Sco in the B-band and the I-band.  The template is intended to follow along the middle of the distribution of magnitudes at a given phase.  This can be problematic for the case where light curves on different dates follow significantly different light curves, so that if few dates are available in some phase range, then the template can be distorted by the randomness of whether `bright' or `dim' nights are dominant.

\section{Eclipse Times}

The purpose behind the eclipse timing is to measure some reproducible fiducial orbital phase, from which we can find the orbital period.  The time of minimum light is when the maximum luminosity near the WD is covered up.  That is, the eclipse minimum is the time of the conjunction between the companion star and the {\it brightest} spot of the same size.  During quiescence, the brightest spot might be the hot spot, where the accretion stream hits the accretion disk, or it might be the inner disk, where the temperatures are highest.  The point is that the eclipse minimum has some small offset from the geometric conjunction between the WD and the companion.  And there will be some small jitter in the orbital phase of the minima due to changes in the relative brightness of the accretion components.  But we can only accept this unwanted jitter, as there is no way measure the time of conjunction.

The best method for determining the time of minimum depends on the number of magnitudes and their coverage of the eclipse.  I will now describe the procedures for the case where the minimum is well covered (Section 4.1), the case where only the ingress or egress is covered (Section 4.2), the case where only one magnitude is available (Section 4.3), the case where the eclipse is on the tail of the eruption light curve (Section 4.4), and in the case of the few reported eclipse times in the literature (Section 4.5).  To end this section, I will collect all the minimum times (Section 4.6).

	\subsection{Good Coverage of the Minimum}

A simplistic way to determine the time of minima is to take the time of the faintest point in the light curve.  But this method is poor because ordinary Poisson fluctuations throughout a fairly flat-bottomed minimum will cause the faintest point to be displaced from the true minimum by up to the half-duration of the minimum.  So we should have a method to determine the minima times that somehow averages over many points near the minimum.  Another possible method is to bisect the times when the ingress and egress pass through some given magnitude.  The Poisson fluctuations and binning can be largely eliminated by fitting straight lines to the appropriate portions of the ingress and egress.  This method has the advantage of using a fast changing part of the light curve, so that the times can be accurately determined.  But this method requires that the light curve be symmetric around the time of minimum, and this is wrong at least far from the minimum.  If we picked some bright magnitude then the systematic errors will be large and if we pick a magnitude near the minimum then the measurement uncertainties along the relatively flat portions of the light curve will be large.  So we need a method that concentrates near the minimum (so as to have minimal effect of the asymmetries in the light curve), gets above the minimum (to where the rate of change of the light curve is substantial) and uses many points (to minimize Poisson variations).

I determine the time of minimum light by fitting a parabola to the near-minimum portions of the eclipse light curve.  The minima of any function will be well-approximated by a parabola over a sufficiently small range, so the idea is to perform the fit over a sufficiently small range such that the asymmetric terms are negligible.  (If I try to fit higher order terms, then the minimum time is sensitive to the order and to the chosen time interval for the fit.  With variable coverage in time and with the eclipse shapes changing, the use of higher order terms only adds large systematic uncertainty.)  This method has the advantages of being applicable to any light curve that covers the time of the minimum, of being insensitive to the high frequency noise from Poisson fluctuations, and of being insensitive to asymmetries in the light curve.

In practice, I determine the best fit parabola by the usual chi-square minimization.  That is, a parabolic model with three parameters (the time of minimum, the faintest magnitude, and the curvature) is varied until the chi-square comparison between the model and the observations is best.  The data points for inclusion in this fit are determined by the times over which the model is fainter than some cutoff magnitude.  This cutoff is usually taken to be 0.1-0.2 mag brighter than the minimum, so that enough time is included for the curvature around the minimum to dominate and yet not so long that any asymmetries will be apparent.  Once a good minimum is achieved, then the points included in the fit are frozen and the fits is optimized again, with this being required so as to avoid effects where the chi-square is changing due to differing points being included in the fit.  The one-sigma uncertainty on the time of minimum is found by varying the time until the chi-square has risen by unity from its minimum value.

In practice, I find that many of the eclipses have best fits for which the reduced chi-square is substantially smaller than unity.  This is seen only in light curves where the quoted error for each point is dominated by the arbitrary systematic error that I added in.  (For well exposed images, the statistical uncertainties, as reported by IRAF, are always unrealistically small, so I have added a typical systematic error of 0.015 mag in quadrature as an effort to make the quoted error bars more realistic.  This value is based on the typical scatter for standard stars in a plot of the reduced magnitude versus airmass, with the ultimate cause of this scatter being unknown.)  This problem of a low reduced chi-square is fully consistent with the real systematic uncertainty being more like 0.010 mag for many of the observations.  With the problem coming from a sometimes slightly conservative choice in the level of systematic errors, the quoted error bars on the minimum time will be slightly too large for some eclipses.

The parabola fitting method has a free parameter of the cutoff magnitude.  This cutoff should not be too close to the minimum or else there will be too few points with too little curvature to well determine the time of minimum.  Nor should this cutoff be too near the out-of-eclipse level or else the asymmetries will make large systematic uncertainties in the derived time of minimum.  Within some middle range, how sensitive is the derived minimum time to the choice of the cutoff magnitude?  To test this, I have taken a typical light curve (for the 2009 May 1 eclipse of CI Aql) with coverage throughout the eclipse.  For this case, my normal analysis had used a cutoff of V=16.55 mag and found a minimum magnitude of 16.735 and a minimum time of JD 2454953.7887$\pm$0.0003.  For cutoff magnitudes of 16.70, 16.65, 16.60, 16.55, 16.50, 16.45, and 16.40, the fractional part of the best fit Julian Date is 0.7890$\pm$0.0009, 0.7889$\pm$0.005, 0.7887$\pm$0.0003, 0.7887$\pm$0.0003, 0.7886$\pm$0.0002, 0.7887$\pm$0.0002, and 0.7885$\pm$0.0001 resepctively.  As expected, the statistical uncertainty gets smaller as the cutoff is raised.  (The systematic uncertainty from the light curve asymmetry will get larger as the cutoff is raised.)  Importantly, the variations in the derived minimum time due to changing the cutoff are always small compared to my final error bars, which is to say that the best fit eclipse time is insensitive to the choice of the cutoff for any reasonable range.

As a further test of the parabola fitting method as well as of the asymmetry in the light curve, I have performed the bisector method for various well observed eclipses.  In the same case as in the previous paragraph (for the 2009 May 1 eclipse of CI Aql), I have fitted line segments to portions of the ingress and egress.  With these fits, I get the time when the light curve passes through various fiducial magnitudes, and these times are then bisected to get a minimum time.  For fiducial magnitudes of 16.70, 16.65, 16.60, 16.55, 16.50, 16.45, and 16.40, the fractional part of the Julian Date for the minimum is 0.7889, 0.7885, 0.7886, 0.7884, 0.7882, and 0.7880.  We see a systematic but small shift as the fiducial magnitude is raised, and this is a demonstration of the level of change caused by the asymmetries in the light curve.  For the middle range of fiducial magnitudes, the changes are always small compared to the final quoted uncertainties.  From the above, I conclude that the parabola fitting method is robust and produces accurate (or slightly too large) error bars.

	\subsection{Coverage of Ingress or Egress}

A number of eclipses were covered with fast time series photometry, but only a portion of the eclipse was covered.  This was usually due to clouds or twilight intervening.  Without substantial coverage on both sides of the minimum, the parabolic fitting procedure does not work reliably.  Nevertheless, these data still provide strict constraints on the time of the minima.  The general procedure that I will use to derive the minima times will be to determine the average offsets between when the eclipse light curve passes through some fiducial magnitude as based on light curves with full coverage (see the previous subsection) and then apply these offsets to the partial-coverage light curves (this subsection).

From Figures 4 and 7, it looks like the shape of the eclipse light curve is nearly constant.  (This is not perfect, as both RNe show some variations from orbit-to-orbit.)  When all the light curves are shifted to a common minimum, the light curves fall on top of each other with only moderate scatter.  So for example for CI Aql, when the ingress fades through a magnitude 0.2 mag brighter than the minimum that time is 0.0216 day before the minimum on average, and when the egress brightens through the same magnitude that time is 0.0203 day after the maximum on average.  This can be placed into an equation as
\begin{equation}
T_{min} = T_m + \Delta T, 
\end{equation}
where $\Delta T$ is the offset in time from when the brightness passes through the fiducial magnitude ($T_m$), with this value a function of $m_{fid}$, the branch (ingress or egress), and the star.  I will take the fiducial magnitude to be 0.1, 0.2, 0.3, and 0.4 mag brighter than the observed minimum magnitude ($m_{min}$).  This simple offset in time can often be improved by taking account for the slope of the light curve as it passes through the fiducial magnitude.  That is, if the eclipse light curve is `flatter' (with a weaker parabolic term), then the slope will be shallower and the time offset will be greater.  This relation can be quantified as 
\begin{equation}
T_{min} = T_m + (A + S_m B), 
\end{equation}
where $T_{min}$ is the derived time of minimum, $T_m$ is the time when the eclipse light curve passes through some fiducial magnitude $m_{fid}$, $A$ and $B$ are two constants that depend on $m_{fid}$, and $S_m$ is the observed slope of the light curve (in magnitudes per day) when it passes through $m_{fid}$.  $T_m$ and $S_m$ are both calculated from the light curve by fitting a simple line to the magnitudes within a small range centered on $m_{fid}$.  In general, Eq. 2 is better than Eq. 1, other than the case where the fiducial magnitude is too close to the minimum so that the measurement uncertainties dominate.  Also, for the case of U Sco in the B and V bands, I have too few fully-covered light curves to measure A and B, so I can only use Eq. 1.

The constants in Equations 1 and 2 can be derived from the well-observed eclipses, for both ingress and egress, for fiducial magnitudes 0.1, 0.2, 0.3, and 0.4 mag brighter than the eclipse minimum.  For CI Aql, the constants have been derived from 19 eclipses with good coverage from 2002-2008.  For U Sco, the constants have been derived for 3, 1, and 17 eclipses with good coverage from 1994-2009.  The average values of $\Delta T$, A, and B are reported in Table 6 for both branches for both RN for all four fiducial magnitudes.  From these eclipses, I have also calculated the RMS scatter in the deviation between the derived minima times versus those derived by parabola fitting, $\sigma$, with these being the accuracy in the derived eclipse minimum time for any one measure.  

If the light curve includes a recognizable minimum, then the fiducial magnitude can be determined, the observed light curve will yield the time and slope for when the RN passes through this fiducial magnitude, and then Eq. 1 or Eq. 2 plus the values in Table 6 can be used to derive the minimum time and its one-sigma uncertainty.

Amongst the eclipse light curves with partial coverage (so that a parabola fit is not reasonable), many do cover the minimum, so that the minimum magnitude can be measured to better than 0.01 mag.  For these, the fiducial magnitudes are known to good accuracy and the one-sigma uncertainties in Table 6 represent the entire error bars.  For many partial light curves, the minimum is not covered, so the minimum magnitude is not known and the fiducial magnitudes have corresponding uncertainty.  So for example, if we only observed a descending branch, then we could assume the minimum equals the average minimum, take the fiducial magnitude to be some value, extract a time and a slope from the light curve, and then deduce the time of minimum.  Alternatively, if we take the minimum to be fainter than average, then our fiducial magnitude would be fainter, the time when the ingress crosses that fiducial magnitude would be later, and the derived minimum time would be later.  Therefore, when the minimum magnitude is not observed, an additional uncertainty will arise from not knowing the correct fiducial magnitudes.  For cases where the minimum is not directly observed, the best that I can do is simply to take the average for that RN and filter.  For CI Aql in the V-band, this is 16.74$\pm$0.05.  For U Sco, the average $m_{min}$ is 19.98$\pm$0.14, 18.96$\pm$0.05, and 18.18$\pm$0.07 in the B, V, and I bands respectively.  I have taken account for this by the propagation of errors, where the uncertainty in the minimum magnitude is taken to be the RMS scatter of the minima.

For each light curve, I measure the times and slopes for as many of the fiducial magnitudes as possible (for 0.1, 0.2, 0.3, and 0.4 mag above minimum for ingress or egress), and derive a time of minimum for each.  If one whole branch plus the minimum is observed, then we will have four derived times each accurate to 0.0019-0.0066 day (see Table 6), then these times can be combined as a weighted average.

As a test for the procedure described in this subsection, I have applied it to the ten well-observed CI Aql light curves from 2009.  These ten eclipses have accurate minima times from parabola fitting.  The deviations of the derived minima times from the procedure in this section versus the accurate times from parabola fitting will provide a good measure of the real accuracy of the procedure.  For the case where the minimum magnitude is known from the observed minimum (as when one of the branches plus the minimum is observed), I have 61 measures of the deviations, each from a particular eclipse and fiducial magnitude. The RMS of the deviations is 0.0024 day.  For the case where the average minimum magnitude (16.71$\pm$0.07) is adopted, the RMS of the deviations is 0.0036 day.  For the case where I take weighted means of minimum time estimates from multiple fiducial magnitudes, I find that the RMS deviations are 0.0022 day for when the minimum magnitude is known and 0.0027 day for when the average case is adopted for the minimum magnitude.  These improvements are smaller than expected if the individual measures are independent.  I take this to mean that the various measures along each branch are not independent, with the implication that one fiducial magnitude is almost as good as averaging the results from up and down the light curve.  Another implication is that the formal error bars obtained by the weighted average from multiple measures will always be too optimistic.  With this experience, I will adopt one-sigma uncertainties of 0.0022  for the case where the minimum magnitude is well-measured and 0.0027 days for the case where it is not measured.

	\subsection{Single Magnitudes During Eclipse}

Certainly the best way to get an eclipse time is to have full coverage across a minimum, or at least with substantial portions of the ingress or egress.  Unfortunately, many of the key old eclipses were only observed with {\it one} image.  For CI Aql, we have all of the pre-eruption data as these single images (either from the Harvard plates or from RoboScope).  There is still good information as to the time of the eclipse minimum, even though the accuracy will be poorer than the situations with time series.  So for example, if an old plate shows the RN at the magnitude usually taken as the eclipse minimum, then the middle time for that plate will be the best estimate of the minimum time with some to-be-determined error bar.  Or if an old CCD image shows the RN halfway between the usual minimum and maximum levels, then the time of minimum will be roughly the middle exposure time plus-or-minus a quarter of the eclipse duration.  In most cases, the ambiguity of knowing whether the image is during the ingress or egress will be resolved by comparison with other plates similar in time.  With this, the many pre-eruption RoboScope images of CI Aql can be used to determine the eclipse times with fairly good accuracy, and the few Harvard plates showing CI Aql in eclipse are so far back in time that the long lever arm will finely constrain the O-C diagram despite moderately large uncertainties in the eclipse times.   

For the RoboScope data on CI Aql (Mennickent \& Honeycutt 1995), the magnitudes are almost all once per night.  The folded light curve (see Fig. 1) easily resolves whether the image is on the descending or rising branch.  The estimate of the time of minimum is as easy as interpolating the observed magnitude in the CI Aql 1991-1996 template (Table 5), and adding the corresponding time offset to the observed time.  I have selected all images with the phase within 0.068 of eclipse by the original Minnickent \& Honeycutt ephemeris.  The uncertainty in how much brighter this measured magnitude is above the minimum will be the quoted photometric uncertainty (typically 0.02-0.04 mag) added in quadrature with the vertical scatter in the eclipse light curves (0.05 mag, compare with Figure 4).  This uncertainty is then propagated through the eclipse light curve template to get the uncertainty in the time of the minimum.  The resultant error bars range from 0.0040 to 0.0117 days (with one at 0.0263 days) depending on the observed magnitude.

For the Harvard plates, the typical exposure time is one hour, and this smears out the recorded light curve (Schaefer \& Patterson 1983).  CI Aql was positively detected on 12 plates in quiescence, and one plate had a very faint limit.  (This is in addition to the 20 plates at Harvard which recorded CI Aql in eruption, see Schaefer 2010.)  The typical B-band magnitude at maximum light is 17.0-17.2 mag.  So the two plates taken with $B>17.5$ (plate MF10536) and with $B=17.75$ (plate A17852) are certainly near the eclipse minimum.  As such, the middle time for the plate exposure is taken as the time of the eclipse.  The uncertainty in this estimate will be comparable to the half exposure time (0.02 days) plus the half duration of minimum ($\sim$0.01 days), which I will take to be 0.03 days

	\subsection{Eclipse Times During the Tails of Eruptions}

After the nova shell becomes optically thin (at the end of the initial fast fall in the light curve), the binary system and its eclipses are visible.  The RNe can be well observed during the tail of its eruption, partly because observers have the interest of the eruption to motivate sustained photometry and partly because the star is brighter than in quiescence so it is easier to observe.  This photometry might be either a fast time series by a single observer that happens to cover the time of an eclipse or as single magnitudes taken on many nights (possibly by many independent observers) folded on the period to provide a phased light curve.  A problem with combining magnitudes over many nights is that the uneclipsed nova light is fading, so any folded light curve must be detrended first.  A problem with combining magnitudes from many observers is that often each observer will have some small offset so as to convert their quoted magnitudes onto some standard magnitude system.  These problems are simple to overcome, yet they will nonetheless lead to some increased scatter in any combined light curve.

For the case of the early tail in the 2000 eruption of CI Aql, the folded and detrended light curve (Fig. 8) shows significant and substantial roughly-sinusoidal modulation on the orbital period.  For days 40-200 after the peak, we see a well defined sinusoidal envelope roughly half a magnitude in width that contains most of the points.  The best fit sinusoidal amplitude is 0.16$\pm$0.02 mag and the phase of minimum is -0.037$\pm$0.017.  That is, the minimum in the light curve comes 0.0229$\pm$0.0105 days early in comparison with the post-eruption ephemeris, and this gives a minimum epoch of HJD 2451792.7070$\pm$0.0105.  The scatter about the best fit sine wave is much larger than the observational uncertainty (typically 0.15 mag for visual observations), and I have to add in quadrature a value of 0.26 mag so as to get a reduced chi-square equal to unity.  The cause of this larger scatter is undoubtedly a combination of imperfect detrending and intrinsic variations of CI Aql that are on faster timescales than the trend line can take out.  The light curve does not show any primary eclipse.  I interpret this as being caused by an emission region that is substantially larger than the binary orbit (so any eclipse of its center would cause only a small drop in light) and that is transparent enough so that the inner regions can be seen (with the irradiation of the inner hemisphere on the companion providing the modulation with the orbital period).  That is, the outer ejected shell has already faded to insignificance in the continuum light, the luminous and extended envelope would provide most of the light, the binary would primarily be visible due to the hot illuminated hemisphere of the companion star, and the accretion disk is still disrupted by the nova wind that is creating the extended envelope.

I have observed two fast time series of CI Aql in its late decline phase (August 2001, see Fig. 2).  Both are well observed minima, for which I used the parabola fitting method to find the minimum time.  As the light curve had not yet returned to the quiescent level, these two times will have to be treated as eruption times, with potential systematic offsets when compared to the true queiscent conditions.

The 1945 eruption of U Sco was discovered as part of an exhaustive search through the Harvard archival plates, with the (perhaps surprising) result that a full eruption was well-covered on 37 plates (Schaefer 2010).  Until the 2010 eruption, these plates provided the only late coverage of the U Sco light curve, with 20 magnitudes covering the time from 33 to 47 days after the peak.  (The exact HJD of the plates has been confirmed from the original logbooks as well as the calculated values on the plate envelopes.)  This late coverage is during a time when eclipses are prominent (Schaefer et al. 2010), so this provides an opportunity to seek an eclipse time in the tail of the 1945 eruption.  With the B-band magnitudes reported in Schaefer (2010), I have determined a trend line, with this being subtracted out.  The folded de-trended light curve (Figure 9) shows an obvious eclipse, and we can even see the slow ingress known from the 2010 eruption at the same time after peak (Schaefer et al. 2011).  The points in the minimum and along the egress are all from one night (2 July 1945).  I have performed a chi-square fit to compare the measured magnitudes to the light curve template derived for the same time after peak from the 2010 eruption (Schaefer et al. 2011).  The measurement errors for the magnitudes is 0.15 mag, and to this I have added 0.10 mag in quadrature so as to represent the orbit-to-orbit scatter known from the 2010 eruption light curve.  The only free parameter in this fit is the phase of the minimum.  The best fit phase of minimum corresponds to HJD 2431639.300.  The one-sigma uncertainty is $\pm$0.009 days.

The 1987 eruption of U Sco (Schaefer 2010) has few magnitudes on the plateau phase (during which eclipses would be visible).  These points happen to include no values with phase between 0.78 and 1.08, so no eclipses could have been detected.

The 1999 eruption of U Sco has many magnitudes after the fast early decline (Schaefer 2010).  Several eclipses are visible in the de-trended light curve.  One eclipse at 19 days after the peak has already been reported by Matsumoto et al. (2003), as will be discussed in the next section.  Another eclipse is visible 25 days after the peak, but only ingress is recorded.  With the minimum magnitude uncertain by perhaps even half a magnitude (due to questions on subtracting the trend as well as the depth of the eclipse at that epoch), the uncertainty in the derived time of minimum (cf. Section 4.2) will be much too large to be of any use.  The only other confident eclipse is 30 days after the peak.  Here, the light curve is seen to be rising, with the faintest magnitude being 1.21 mag fainter than the accurately known trend line, so the first observations is certainly very close in time to the minimum.  I take the time of minimum to be the time of this first observation, and give it an uncertainty appropriate for the sampling and the observed duration of minima from 2010.  Thus, I get the minimum time to be HJD 2451265.3060$\pm$0.0100.

	\subsection{Eclipse Times From the Literature}

Eclipse times in quiescence have already been published for both CI Aql and U Sco in Schaefer (1990) and Schaefer \& Ringwald (1995), with all of these times now being superseded by the analysis in this paper.  Twenty well-observed eclipse times in the tail of the 2010 eruption of U Sco are reported in Schaefer et al. (2011), and these are adopted here.  The only other eclipse times in the literature are few in number, isolated in measure, and all in the tail of eruption light curves.  Due to the shifts of the eclipse times during the eruptions, these published times cannot be combined with eclipse times from quiescence so as to derive any orbital period change.  Without the realization of the time shifts, earlier papers (Matsumoto et al. 2003; Lederle \& Kimeswinger 2003) had claimed to measure period changes, and these are completely erroneous.

For CI Aql, Matsumoto et al. (2001) present a large number of magnitudes in many filters, and the orbital modulation can be clearly seen.  However, they do not present any minimum times, nor do they give their magnitudes so others can derive minima times, and their plotted light curves are not adequate to determine the minimum times with any accuracy.  Their folded light curve from some unstated time during the 300 day long plateau shows an amplitude of 0.6 mag, with an apparent shallow eclipse.  They present a new ephemeris for the eclipses, but it is never stated whether the original Mennickent \& Honeycutt (1995) data was also used for the equation.  The only apparently useful time is their specified epoch for their new ephemeris, HJD 2451701.2086$\pm$0.0089, which presumably represents an observed minimum time.  However, this time is just 29 days after peak at a time when no eclipses were visible.  As such, I cannot accept their epoch as a reliable observed eclipse time, even though they must have many good eclipse times that are unreported.

For CI Aql, Lederle \& Kimeswinger (2003) present several long time series taken in 2001 and 2002, as taken with the Innsbruck 60-cm telescope plus a CCD camera with a V filter.  They also present a nice and detailed physical model that reproduces their light curves and derives the parameters of the CI Aql system.  But they do not present any individual minima times, nor do they present any photometry data, while their plotted light curves are inadequate to get any accurate minima times.  The only useful time that I can pull out of their paper is their specified epoch for their new ephemeris, Julian date 2452081.5022$\pm$0.0046, which presumably has a heliocentric correction, and which presumably represents an observed minimum time.

For U Sco, Thoroughgood et al. (2001) identified an eclipse as based on seven flux density measures from spectra during the 1999 eruption. The eclipse is at a time 51 days after the peak, which corresponds to the middle of the second plateau phase (Schaefer et al. 2010).  They fitted a parabola to the full minimum, and reported the heliocentric Julian date of the minimum as 2451286.2143$\pm$0.0050.

For U Sco, Kato (2001) measured 22 magnitudes over 0.14 days (early in the 1987 eruption) to find random fluctuations, for which he picked out one to identify as an eclipse minimum.  But this claimed minimum is not different from other minima in the light curve, so its identification as an eclipse is poor.  Also, this time series was taken just seven days after the peak, at a time when we know that the nova shell is optically thick and eclipses are not visible (Schaefer et al. 2011).  So I reject this time as a valid eclipse time.  Kato (2001) makes a further claim to have recognized an eclipse in a folded light curve based on data from the VSOLJ.  However, the claimed eclipse has much too short a duration to be the real eclipse.  And these observations were collected from 5-11 days after the peak, so they cannot represent eclipses.  Again, I reject the times from this claim.

For U Sco, Matsumoto et al. (2003) present many series of fast photometry throughout the 1999 eruption.  They claim to identify a secondary eclipse at a time six days after the peak.  However, I only see a usual fluctuation for which they converted a marginally significant inflection in the decline into an apparent minimum by subtracting out an unrealistically steep trend.  In any case, we now know that the nova shell was optically thick at six days after the peak, so no eclipse could be visible (Schaefer et al. 2011).  Therefore,  I reject this claimed secondary minimum time.  Matsumoto et al. (2003) also report the time of one primary minimum, based on piecing together an egress and the next ingress.  Their reported time is HJD 2451254.211, which is 19 days after the peak, on the first plateau.  No error bar in the eclipse time is quoted, so I will adopt $\pm$0.01 days as appropriate for their sampling.

	\subsection{Eclipse Minima Times}

With these observations and analysis procedures, I have tabulated all known eclipse times for both CI Aql and U Sco.  These are presented with identical formats in Table 7 for CI Aql and Table 8 for U Sco.  The first column gives the UT date of the observed minimum.  The second column gives the Observatory and/or telescope used to make the observations, or alternatively the data collecting organization and the observer.  The third column lists the heliocentric Julian Date (HJD) of the observed time of minimum light ($T_{obs}$) and its one-sigma uncertainty.  The fourth column gives a running integer, N, which keeps track of the cycle count as used in the ephemeris equations.  The last column gives the O-C value (in units of days) for the observed HJD versus the model time as based on the best-fit post-eruption linear ephemeris.

\section{Orbital Period Change for CI Aql}

The orbital period of a nova will change for many reasons.  Across an eruption, the period will change suddenly ($\Delta P$) due to the loss of the the ejected mass (the effect being sought here) and the drag of the companion star during the common envelope stage.  Between eruptions, the period will have a steady period change ($\dot{P}$) due to the mass transfer and due to angular momentum losses.  These changes can be illustrated and analyzed with the O-C diagram.  Full details on the physics of the period changes and the O-C diagram (with equations and examples) are given in Appendix B.

The time period from 2001 to 2009 (i.e., the post-eruption interval) is by far the best observed for CI Aql.  So I will start by considering an unchanging period and only those eclipse times after the end of the 2000 eruption (Section 6.1).  This will provide the base linear model for use in the O-C diagram.  In Section 6.2, I will consider the steady period change ($\dot{P}$) between eruptions.  I will evaluate $\dot{P}$ by fitting curvature in the O-C diagram for 2001-2009 and by placing constraints from theory.  In Section 6.3, with my post-eruption ephemeris, I consider the eclipse times in the tail of the 2000 eruption.  I will find that the eclipse times in the tail are systematically early, due to the shift of the center of light.  In Section 6.4, I will add in the 1991-1996 eclipse times.  With the well-measured $P$ and $\dot{P}$ for the post-eruption interval, I extrapolate back to the time of the eruption.  This time is the end point for the O-C curve of the pre-eruption interval.  Indeed, this end point will be the best measured point on the pre-eruption O-C curve, and will serve as the anchor for measuring $P_{pre}$.  The pre-eruption O-C curve (from the RoboScope data plus the extrapolated eclipse time at the epoch of the eruption) will then be fit with the allowed range of $\dot{P}$ so as to get the period change across the eruption.  Finally, in Section 6.5, I will make a fit to all 76 non-eruption eclipse times from 1926-2009.  The fits will vary substantially within the allowed range of $\dot{P}$ and whether there are one or two undiscovered eruptions between 1941 and 2000.  This will give us a range of possible period changes across the eruptions.

	\subsection{2001-2009 With No Period Change}

I have 45 post-eruption eclipse times from November 2001 to September 2009 (see Table 7).  The O-C diagram is plotted in Figure 10.  

For the many highly accurate eclipse times, we see a scatter that is significantly larger than any smooth curve we could draw.  This scatter cannot be due to measurement errors, nor to jitter of the position of the stars in their orbit.  The likely cause of the scatter is changes in the brightness of the accretion disk across the duration of the eclipse minimum.  The backside of the accretion disk is still visible during mid-eclipse (Hachisu et al. 2003; Lederle \& Kimeswinger 2003) and that the disk brightness changes on all timescales (see Figs 3 and 4).  So, if the visible part of the disk is brightening across the eclipse minimum then the light curve will be `tilted' such that the apparent minimum occurs early, whereas if the disk is fading across the eclipse minimum then the light curve will have the apparent time of minimum occur late.  This is an easy and expected explanation for intrinsic jitter in the apparent eclipse minimum times.

With this intrinsic scatter, the real uncertainty of each minimum time for indicating some fiducial orbital phase will be larger than the quoted measurement error bars in Table 7.  The intrinsic scatter will be some constant throughout the post-eruption interval, roughly given as the RMS scatter of the best measured points around the best fit smooth curve.  The total uncertainty in the fiducial orbital phase will be the addition in quadrature of the measurement error and the intrinsic scatter.  Formally, I will derive the intrinsic scatter as that value which produces the best fit reduced chi-square equal to unity for the 2001-2009 interval.

For the model that the O-C curve is linear, I have used a chi-square fitting routine to derive the best fit period and epoch.  For this fit, I have chosen the fiducial epoch to be near the peak date for the eruption.  I find an epoch of $E_0=2451669.0575 \pm 0.0002$ and $P_{post}=0.61836051 \pm 0.00000006$.  With an intrinsic scatter of 0.00136 days (1.9 minutes), I get a fit with a chi-square of 43.0 for 43 degrees of freedom.  So the scatter caused by the disk variations is less than two minutes, while the orbital period is known to 0.0043 seconds accuracy.  This period has the fractional accuracy of 81 parts per billion.

The O-C curves can be readily interpreted if the model is linear.  So I should choose some linear model, and I chose to use this best-fit post-eruption linear model as my fiducial linear model.  So all my O-C values and plots are constructed with
\begin{equation}
T_{model}=2451669.0575 + N \times 0.61836051,
\end{equation}
as in Table 7 and Figure 10.  $N$ is an integer that counts the orbital cycles from the original epoch.

	\subsection{Steady Period Change}

The system parameters along with Equation 22 (in Appendix B) can be used to place an upper limit on $\dot{P}$.  From Hachisu, Kato, \& Schaefer (2003),  $M_{comp}=1.7 ^{+0.3}_{-0.7}$ M$_{\odot}$, $M_{WD}=1.2 \pm 0.2$ M$_{\odot}$, and $\dot{M} \sim 1.0 \times 10^{-7}$ M$_{\odot}$ yr$^{-1}$.  From Lederle \& Kimeswinger (2003), $M_{comp}=1.5$ M$_{\odot}$, $M_{WD}=1.2$ M$_{\odot}$, and $\dot{M}$ is $2.5\times10^{-8}$ M$_{\odot}$ yr$^{-1}$ in 1991-1996 and $5.5 \times 10^{-8}$ M$_{\odot}$ yr$^{-1}$ in 2002.  We also have the more restrictive constraint that $q\lesssim 5/6$.  With these inputs, the upper limit on $\dot{P}$ varies from +1 to +9 times $10^{-11}$ days per cycle.  With this, I can only constrain 
\begin{equation}
\dot{P} < +0.9 \times 10^{-10}
\end{equation}
days per cycle.

The 45 post-eruption eclipse times can be fit to Equation 20 in Appendix B, with no ambiguity relating to period changes related to eruptions.  I find the best-fit for $E_0=2451669.0563$, $P_0=0.61836136$ days, and $\dot{P}$= -24$\times$10$^{-11}$ days per cycle.  This gives a chi-square of 42.0.  The one-sigma range (i.e., the range over which the chi-square is within 1.0 of the minimum), is 
\begin{equation}
-52 \times10^{-11} < \dot{P} < 0
\end{equation}
in units of days per cycle.  With this, $\dot{P}$ is likely negative, and this means that the $\dot{J}$ term in Equation 12 must be substantial and will dominate over the effects of the steady mass transfer.

The constraint arising from the 1926 and 1935 eclipse will depend on the derived $\Delta P$ value, which will come from Section 6.4.  In all, the best constraints on $\dot{P}$ are those given in Equation 4, while the best estimate is $\dot{P}$= -24$\times$10$^{-11}$ days per cycle.
 
	\subsection{Time Offsets During Eruption}

The post-eruption ephemeris can be extrapolated to the tail of the eruption with high accuracy, where the acceptable range of $\dot{P}$ makes for no significant change.  I have four eclipse times for the tail of the 2000 eruption.  The three times on the late tail are consistent with the post-eruption ephemeris to within the intrinsic scatter.  But the one eclipse time in the early tail comes 0.022 days (32 minutes or 0.037 in phase) early.  This offset is large, but the uncertainty is not small because no eclipse is visible (see Figure 8).  The chi-square for a sine wave fit to the data from 40 to 200 days after peak with a zero offset is 4.9 worse than the best fit, indicating that the existence of the offset is significant at the 2.2-sigma level.

If this offset is taken at face value, then the eclipse minima occur {\it before} the time expected from the quiescent minima. This shift is too large to be caused by any real change in the orbital period, so it must be caused simply by a shift in the center of light from eruption to quiescence.  During the eruption, the dominant light sources will be centered on the white dwarf and centered on the illuminated hemisphere of the companion, both of which will make for minima at the time of conjunction between the two stars.  During quiescence, the dominant light source will be the hot spot in the accretion disk, with this being offset from the line connecting the centers of the two stars.  The hot spot will be centrally eclipsed at a time roughly 0.03 in phase {\it after} the time of the conjunction (based on the geometry in Lederle \& Kimeswinger, 2003).  So we have a consistent explanation for why the eclipse times are systematically offset from quiescence to the eruption.

	\subsection{Period Change Across the 2000 Eruption}

The eclipse times from 1991 to 1996 can now be used to measure $\Delta P$ across the 2000 eruption.  For now, I will not use the 1926 and 1935 eclipses in the chi-square calculation.  To get a good idea of the situation, let me start by adopting the case that best fits the post-eruption times ($P_{post}=0.61836136$ days, $E_0=2451669.0563$, and $\dot{P}=-24 \times 10^{-11}$ days per cycle).  When $\Delta P=0$, the chi-square is 65.6 for the 74 non-eruption eclipse times from 1991 to 2009.  If $\Delta P$ is allowed to vary freely, the chi-square reaches a minimum of 63.8 at $\Delta P= +5 \times 10^{-7}$ days.  The one-sigma range (where the chi-square is within unity of 63.8) is $\pm3 \times 10^{-7}$ days.  The positive sign for $\Delta P$ means that the $P_{post}>P_{pre}$, which is to say that the period {\it increases} across the eruption.

Next, we can look at the fits for the range of $\dot{P}$ allowed by the post-eruption times.  One extreme of the allowed range has $\dot{P}=0$, for which ($P_{post}=0.61836051$ days and $E_0=2451669.0575$).  For this case, with no period change across the eruption, the chi-square is 77.1 (for 74 non-eruption eclipse times from 1991-2009).  When the $\Delta P$ is allowed to vary, the chi-square reaches a minimum at 64.8 for a period change of $-12 \times 10^{-7}$ days.  The one-sigma error bar is $\pm 0.3 \times 10^{-6}$ days.  For the maximal allowed steady period change ($\dot{P}=-52\times10^{-11}$ days per cycle with $P_{post}=0.61836231$ days and $E_0=2451669.0550$) and zero sudden period change, the chi-square is 112.5.  When $\Delta P$ is allowed to vary, the chi-square reaches a minimum at 64.8 for a period change of $+23\pm 3 \times 10^{-7}$ days.  Characteristically, the best fit value for $\Delta P$ is correlated with $\dot{P}$, such that the more curved the O-C curve (with the parabola concave down) the more positive the required period change.  For the minimal curvature (i.e., flat with $\dot{P}=0$) the period must have {\it decreased} across the 2000 eruption (i.e., $\Delta P <0$), while for the maximal curvature the period must have {\it increased} across the eruption.

As the fits go from minimal to optimal to maximal $\dot{P}$ (as based on the post-eruption times alone), the minimum chi-square goes from 64.8 to 63.8 to 64.8.  With this, we see that we have closely mapped out the one-sigma range as based on the inclusion of the 1991-1996 eclipse times. That is, all the 1991-2009 data is best described by a model with $\dot{P}=-24 \times 10^{-11}$ days per cycle, with the one-sigma range from zero to $-2.6\times10^{-10}$ days per cycle.  The best fit has $\Delta P= +5 \times 10^{-7}$ days, with a one-sigma range from around -12 to +23 times $10^{-7}$ days.

	\subsection{Fit to 1926-2009 Eclipse Times}

Finally, we can now do a fit to all 76 non-eruption eclipse times from 1926-2009.  This is different from the results in the previous section only in that the two eclipses from 1926 and 1935 are included.  These old eclipses have substantial error bars in their timing, but they are around eighty years ago, so their long lever arm in time will provide strong constraints.  An ambiguity in this fit is whether CI Aql had any undiscovered eruptions between the years 1941 and 2000.  Schaefer (2010) puts forth the evidence (based on quiescent magnitudes and discovery probabilities) that one or two eruptions were missed, although zero missed eruptions is certainly possible.

To start this, let us consider the result for the best fit from the 1991-2009 times ($\Delta P= +5 \times 10^{-7}$ days, $\dot{P}=-24 \times 10^{-11}$ days per cycle, $P_{post}=0.61836136$ days, and $E_0=2451669.0563$).  With these values, no adjustments, and no undiscovered eruptions, the chi-square for all 76 non-eruption eclipse times is 79.4.  The O-C curve passes somewhat below the 1935 point.  If I add two eruptions (in 1960 and 1980) with the same $\Delta P$ value for all eruptions, then the fit is improved to a chi-square of 72.4.  If further the $\Delta P$ value is allowed to vary, then a value of $\Delta P= +9.5 \times 10^{-7}$ days, produces a good chi-square of 67.2.  Alternatively, if $\dot{P}$ is allowed to vary, then the minimum chi-square is 68.5 for $\dot{P}=-22 \times 10^{-11}$ days per cycle (still with $\Delta P= +5 \times 10^{-7}$ days).

For the best minimal curvature case ($\dot{P}=0$), with no change from the best fit from 1991-2009, and eruptions in 1960 and 1980, we have a good minimum chi-square of 66.1 with no adjustments.  If I allow the period change across the eruption to vary, we get a minimum chi-square of 65.6 for $\Delta P= -13 \times 10^{-7}$ days.  If I do not allow for any undiscovered eruptions, then the chi-square minimum is poor at 73.8 for $\Delta P= -18 \times 10^{-7}$ days.

For the best maximal curvature case ($\dot{P}=-52\times10^{-11}$ days per cycle), with no changes from the best fit of 1991-2009, and eruptions in 1960 and 1980, the chi-square is bad at 129.3 (the O-C curve passing far below the 1935 eclipse).  By adjusting only the $\Delta P$ value, the best chi-square equals the poor 85.6 for $\Delta P=  +35 \times 10^{-7}$ days.  If I do not include undiscovered eruptions, the best chi-squares are always very bad because the O-C curve falls far below the 1935 point.  With this, the inclusion of the two very old eclipse times places a constraint that the $\dot{P}$ value must be substantially closer to flat than the maximal curvature value.

For the general case, we should try to minimize the chi-square by simultaneously allowing $\Delta P$, $\dot{P}$, and the number of missed eruptions to vary.  For this, the period just after the 2000 eruption ($P_{post}$ and the epoch ($E_0$) will be held constant to the values obtained for the 2001-2009 data alone.  (This is because the statistical weight of the these many and accurate times will dominate over all the other parameters.)  I will further assume that $\dot{P}$ remains constant throughout and that $\Delta P$ is the same for all eruptions.  With this, the global best fit (for all 76 non-eruption eclipse times) has a chi-square of 64.6 with $\Delta P=-3.7 \times 10^{-7}$ days, $\dot{P}=-12\times10^{-11}$ days per cycle, $P_{post}=0.61836092$ days, $E_0=2451669.0570$, all with an adopted intrinsic scatter of 0.00136 days and no undiscovered eruptions.  This is my best result for the period change of CI Aql across eruptions, and hence the main goal of the CI Aql program.

The one-sigma region of parameter space can be determined by seeking those values for which the chi-square is within unity of the global minimum (i.e., $\leq 65.6$).  With this, the allowed range for $\dot{P}$ is -2 to -20 times $10^{-11}$ days per cycle, over which $\Delta P$ varies from -11 to +6 times $10^{-7}$ days.  Over this same range, the $E_0$ varies from 2451669.0574 to 2451669.0565, and $P_{post}$ varies from 0.61836060 to 0.61836122 days.

	\subsection{Summary for CI Aql}

All this work has largely been aimed at getting the one number for CI Aql of $\Delta P = -0.00000037$ days.  The one sigma error range is $-3.7^{+9.2}_{-7.3} \times 10^{-7}$ days.  This measure has an accuracy of close to $0.00000085/P$ days = $1.4\times 10^{-6}$, or 1.4 parts per million.  This period change across the 2000 eruption is very small, and consistent with zero.  The slightly-prefered negative value of $\Delta P$ implies that $P_{post}<P_{pre}$, which is to say that the orbital period of CI Aql {\it decreased} across the 2000 eruption.

The dominant uncertainty comes from not knowing the exact curvature of the O-C curve between eruptions.  That is, we can have the long term shape roughly reproduced with concave-up parabolas plus period decreases through each eruption or by concave-down parabolas plus period increases through each eruption.  At this point all old archival sources have been exhaustively examined, so the only way to improve the measure of $\Delta P$ is to better constrain the $\dot{P}$ measure (by accumulating more eclipse timings).

The global best fit is displayed as the thick curved line in Figures 10-12.  To be explicit, the equation for this best fit ephemeris is
\begin{equation}
T_{eph}=2451669.0570 + 0.61836092 N - 0.5 N^2 12\times10^{-11}, N>0,
\end{equation}
\begin{equation}
T_{eph}=2451669.0570 + 0.61836129 N - 0.5 N^2 12\times10^{-11}, -35045<N<0,
\end{equation}  
\begin{equation}
T_{eph}=2451669.0570 + 0.61836587 (N+35045) - 0.5 (N+35045)^2 12\times10^{-11}, N<-35045,
\end{equation}  
The first part covers the time after the 2000 eruption, the second part from 1941 to 2000, and the third part from 1917 to 1941.  This equation is applicable only when CI Aql is in quiescence.

To illustrate the extreme acceptable solutions, Figures 10-12 also display the model for the $\dot{P}=0$, as well as for the largest acceptable curvature.  To be explicit, the zero curvature model has parameters $\dot{P}=0$, $\Delta P=-13 \times 10^{-7}$ days, $P_{post}=0.61836051$ days, $E_0=2451669.0575$, for eruptions in 1917, 1941, 1960, 1980, and 2000.  The largest acceptable curvature model has $\dot{P}=-20\times10^{-11}$ days per cycle, $\Delta P=+5.5 \times 10^{-7}$ days, $P_{post}=0.61836122$ days, $E_0=2451669.0565$, for eruptions in 1917, 1941, 1960, 1980, and 2000.

In all, I now have a good O-C curve for CI Aql from 1926 to 2009, and from this I get a confident value for the period change across the 2000 eruption ($-3.7^{+9.2}_{-7.3} \times 10^{-7}$ days).  While the best estimate is for a negative $\Delta P$ (the period {\it decreasing} across the eruption), the error bars allow for a zero or positive value.

\section{Orbital Period Change for U Sco}

I have many eclipse times during quiescence from 1989-1997 and 2001-2009, so the period change across the 1999 eruption will be well-measured.  In Section 7.1, I will start by considering only the post-eruption eclipse times with an unchanging period.  The best fit from this will provide the linear ephemeris that will be the basis for all O-C values and plots.  In Section 7.2, I will evaluate the curvature ($\dot{P}$) by two methods.  The first is to look at the post-eruption interval alone and fit for the best curvature.  Second, I will use the best star masses and accretion rates so as to get a theoretical limit on $\dot{P}$ for conservative mass transfer.  In Section 7.3, I will use the best fit post-eruption ephemeris to determine the phase offset during the 1999 and 2010 eruptions.  Again, I will find significant offsets, caused by shifting in the center of light between quiescence and eruption.  In Section 7.4, I will use all non-eruption eclipse times from 1989-1997 and 2001-2009 to fit the O-C curve so as to measure the $\Delta P$ across the 1999 eruption.  This is the whole point of my program begun in 1987.  In Section 7.5, I will consider the 1945 eclipse time.  My conclusion will be that this is evidence for long term variations in the $\dot{P}$, with this being expected.  Section 7.6 will summarize my results for U Sco.

	\subsection{2001-2009 With No Period Change}

I have 29 eclipse times from May 2001 to July 2009 (see Table 8).  The O-C curve is plotted in Figure 13.    This shows the best measured period as well as the curvature caused by any $\dot{P}$ term, all with no complication of period changes across eruptions.

Again, we see that the scatter of the best-measured eclipse times around any smooth curve is substantially larger than the quoted error bars.  I take this to mean that U Sco has some intrinsic scatter in its stability of its eclipse times, likely associated with the ubiquitous variability of the disk which will tilt the light curve outside of the mid-eclipse interval.  To account for this intrinsic variability, I will add the quoted measurement uncertainty (see Table 8) in quadrature with some constant intrinsic uncertainty so as to get the total uncertainty.  This total uncertainty is what will be used in chi-square fits to the O-C curve.  I evaluate the intrinsic uncertainty as that value which yields a reduced chi-square equal to unity in the best line fit for the 2001-2009 interval.  I find that the intrinsic error has its one-sigma equal to 0.00242 days (3.5 minutes).  With this comes the realization that there is little utility in measuring the eclipse times to better than roughly one-minute accuracy, also that the bigger telescopes (with better photometric accuracy and higher time resolution) are not needed.

With the intrinsic error of 0.00242 days added in, the best fit line for the 29 eclipse times (hence 27 degrees of freedom) produces a chi-square of 27.0.  This best fit has the post-eruption period equal to $P_{post}= 1.23054695$ days and the heliocentric Julian date of the epoch equal to $E_0=2451234.5387$.  This epoch was chosen as being at the time of the start of the 1999 eruption, when the system lost most of the ejected mass.  To be explicit, the linear model
\begin{equation}
T_{model}=2451234.5387 + N \times 1.23054695
\end{equation}
will be used for all calculations of O-C for U Sco.   

	\subsection{Steady Period Change}

The O-C curve displayed in Figure 13 does not show any obvious curvature.  Nevertheless, the orbital period of U Sco in quiescence must change at some level, if only from conservative mass transfer.  The O-C curve for 2001-2009 can be used to derive the best fit $\dot{P}$ with no complication from the sudden period change across eruptions. 

The best fit parabola to the 29 eclipse times from 2001-2009 has a chi-square of 26.75, which is only slightly better than a straight line.  This immediately tells us that $\dot{P}$ is near zero and could either be positive or negative.  This best fit parabola is with $\dot{P}=-110\times10^{-11}$ days per cycle, $P_{post}=1.23054915$ days, and $E_0=2451234.5369$.  At the one-sigma level, the smallest acceptable value of $\dot{P}$ is $-330\times10^{-11}$ days per cycle, with $P_{post}=1.23055339$ days, and $E_0=2451234.5335$.  At the one-sigma level, the largest acceptable value of $\dot{P}$ is $+110\times10^{-11}$ days per cycle, with $P_{post}=1.23054488$ days, and $E_0=2451234.5405$.  So we can express $\dot{P}$ as $-110\pm 220\times10^{-11}$ days per cycle.

For conservative mass transfer, the limit on $\dot{P}$ is given by Equation 22, and depends on the masses of the two stars and the accretion rate.  Hachisu et al. (2000) have analyzed the 1999 eruption and conclude that $M_{WD}=1.37 \pm 0.01$ M$_{\odot}$, $M_{comp}=1.5$ M$_{\odot}$ (with 0.8-2.0 M$_{\odot}$ being acceptable), and $\dot{M}\sim 2.5 \times 10^{-7}$ M$_{\odot}$ yr$^{-1}$.  Fortunately, U Sco is a double-lined spectroscopic and eclipsing binary, and Thoroughgood et al. (2001) have measured a radial velocity curve for both stars, concluding that $M_{comp}=0.88 \pm 0.17$ M$_{\odot}$.  With this, the theoretical value for no angular momentum loss is $\dot{P}= +126^{+85}_{-57} \times 10^{-11}$ days per cycle.  Pushing all the values to their extremes and realizing that we only have a limit (because the angular momentum losses are unknown), all I can get is $\dot{P}< +211 \times 10^{-11}$ days per cycle.

Both ranges are comparable in size, similar in magnitude, and one is contained entirely within the other.  The best fit curvature term has $\dot{P} = -110\pm 220\times10^{-11}$ days per cycle.  I will treat the steady period change as a free parameter within the range -330 to +110 times $10^{-11}$ days per cycle.

	\subsection{Eclipse Times in the Tails of Eruptions}

The O-C diagram for the 2001-2009 interval is well determined, and we can extrapolate the ephemeris to the times of the 1999 and 2010 eruptions with negligible error.  With the best fit from the previous section, I have
\begin{equation}
T_{eph}=2451234.5369 + 1.23054915 N - 0.5 N^2 110\times10^{-11}, N>0,
\end{equation}
We see from Figure 13 that the uncertainties in the O-C for the ephemeris is around $\pm$0.0015, depending on the curvature.  With this uncertainty in mind, Equation 10 is a reasonable representation, and thus the O-C values in Table 8 are good for the eclipse times in the tails of the 1999 and 2010 eruptions.  With this, I have plotted the O-C values for the eclipse times during the tails of the eruptions in Figure 14.  We see a strong roughly-linear trend, where the eclipses within 30 days of the peak (i.e., on the long plateau) are {\it early} by up to 0.013 days (19 minutes), while the eclipses late in the tail are {\it late} by up to the same amount.  This trend is highly significant (although with substantial scatter), and is much larger than the uncertainty associated with the exact degree of curvature.

The orbital period of U Sco cannot change fast enough to explain the variations seen in Figure 14.  So the explanation undoubtedly arises from a shift of the center of light in the U Sco system as the eruption progresses.  This is an expected result of changes in the relative brightness of the emission regions centered on the WD and those off the line connecting the stellar centers (like from the hot spot and structure in the accreting material).  I would hope that someone can construct a detailed model of the U Sco eruption that reproduces and explains the variations shown in Figure 14.

The large variations of eclipse times from the quiescent ephemeris means that these times should not be combined with the quiescent eclipse times.  The deviations from zero in Figure 14 are greatly larger than those in Figure 13.  So any naive combination of quiescence and eruption times will be dominated by the offsets during the eruption, and any derived period changes (e.g., Matsumoto et al. 2003) must certainly be wrong.

	\subsection{Period Change Across the 1999 Eruption}

With the base period for the O-C curve and the observed constraints on the $\dot{P}$, I will now use all the 46 non-eruption eclipse times from 1987 to 2009 to determine the sudden period change across the 1999 eruption.  The O-C diagram for these eclipse times is in Figure 15.  For this, I will use a chi-square analysis on the eclipse times when compared against a model with a constant $\dot{P}$ throughout and a sharp $\Delta P$ at the time of the 1999 eruption.  The $\dot{P}$ value will be kept inside the one-sigma constraints from Section 7.2.  The total uncertainty in the eclipse times will be the quoted measurement uncertainty (from Table 8) added in quadrature with 0.00242 days (for the systematic jitter in the eclipse times).

An early report on the $\Delta P$ value from my eclipse times (Martin et al. 2010) quotes a negative value, with this being hard to understand because $\Delta P_{drag}$ must be negligibly small.  For this early report, I had not appreciated the effects of allowing for a non-zero $\dot{P}$.  That is, I had taken the O-C curve in Figure 15, and fitted straight line segments (assuming $\dot{P}=0$) with a break in 1999, and seen that the pre-eruption points fall significantly below the $O-C=0$ axis, and then concluded that $\Delta P$ must be significantly negative.  However, with a substantial concave-down parabola shape between eruptions, we can get a positive $\Delta P$.  With this, the sign of $\Delta P$ is ambiguous.  As such, the mechanism proposed by Martin et al. (2010) is not now required, it is certainly operating at some level, and it might yet be the dominating effect.

Let me report the chi-squares for the 1989-2009 eclipse times for the best and extreme fits as derived from the 2001-2009 data alone (Section 7.2).  The best fit (with $\dot{P}=-110\times10^{-11}$ days per cycle) has a chi-square equal to 60.9 (for 42 degrees of freedom).  For one extreme allowed curvature (with $\dot{P}=+110\times10^{-11}$ days per cycle), the chi-square is 64.9.  For the other extreme allowed curvature (with $\dot{P}=-330\times10^{-11}$ days per cycle), the chi-square is 60.4.

The overall best fit for 1989-2009 is for $\dot{P}=-250\times10^{-11}$ days per cycle, with a chi-square of 60.2.  For this best fit, $\Delta P = +43 \times 10^{-7}$ days.  Again, the best fits have $\Delta P$ and $\dot{P}$ strongly correlated, so that as $\dot{P}$ gets increasingly negative, the $\Delta P$ value gets increasingly positive.  The one-sigma range for $\Delta P$ (i.e., the values over which the chi-square can be less than 61.2) will correspond to a range of $\dot{P}$.  For the most positive $\dot{P}$ ($-80\times10^{-11}$ days per cycle), the chi-square is 61.2 for $\Delta P = -24 \times 10^{-7}$ days.  For the most negative $\dot{P}$ ($-430\times10^{-11}$ days per cycle), the chi-square is 61.2 for $\Delta P = +114 \times 10^{-7}$ days.  Figure 15 displays these three fits as curves, with the best fit curve as the thick line.

These fits from 1989-2009 represent my best answer for $\Delta P$ across the eruption.  I find that $\Delta P = (+43 \pm 69) \times 10^{-7}$ days as $\dot{P}=(-250 \pm 170) \times10^{-11}$ days per cycle.  The best $\Delta P$ value is positive, but the value is small and the uncertainty is such that it might be zero or even negative.  The $\dot{P}$ value is negative, which implies that the $\dot{J}$ term is significant.

	\subsection{The 1945 Eclipse}

The eclipse in the tail of the 1945 eruption (see Figure 9) provides a good time for the O-C diagram (see Figure 16).  The light curve might be poorly sampled by the standards of my modern time series photometry, but the data is more than good enough to show the existence of the eclipse and to determine the time of minimum to $\pm$0.009 days.  The eclipse is in the tail of an eruption, so it should suffer some systematic offset relative to an ephemeris for eclipses during quiescence.  The eclipse minimum is from plates taken 34 days after the 1945 peak, so from Figure 14, we see that the offset should be near zero.  The uncertainty in this zero offset is around $\pm$0.005 days, with this being around half the size of the one-sigma error bar for the eclipse time.  Thus, to all needed accuracy, the quoted eclipse time can be directly compared with the ephemeris from quiescence.

Figure 16 shows the back extrapolation of the best fit ephemeris as based on the 1989-2009 non-eruption eclipse times.  The predicted eclipse time for 1945 is earlier than the observed eclipse time by roughly 0.2 days (around 5 hours).  The early predicted time (with a phase of 0.84) is completely rejected by the 1945 light curve (Figure 9).  The 1945 eclipse time disagrees with the 1989-2009 best fit at close to the three-sigma confidence level.

To within the errors, the 1945 point has a near-zero O-C value, which implies that the long term average orbital period (from 1945 to 2009) is nearly the same as the orbital period from 2001-2009 (that interval used to select the period for constructing the O-C diagram).  That is, the period from 2005 is close to the overall period going back in time.  With the analysis from Section 5.3, this means that the steady period change between eruption must closely balance out the abrupt period change for each eruption.  The period change throughout an inter-eruption interval with $\Delta N$ cycles will be $\Delta N \dot{P}$.  This must be balanced by the period change across any one eruption, so $\Delta P = -\Delta N \dot{P}$.  We expect the $\dot{P}$ to change, so a better notation would be $\langle \dot{P} \rangle$ to indicate the long term average of the period change.  Thus, $\langle \dot{P} \rangle=-\Delta P/\Delta N$.  For the average inter-eruption interval of near ten years, $\Delta N$ will be close to 3000.  For our best fit 
$\Delta P = +43 \times 10^{-7}$ days, we get $\langle \dot{P} \rangle=-140\times10^{-11}$ days per cycle.  This long term period change is nearly identical with the best fit curvature for the 2001-2009 eclipse times alone (Section 7.2).

We now have the case that the long term average period change nearly equals the best fit period change from 2001-2009, and is somewhat different from the best fit period change from 1989-2009.  This merely indicates, with no surprise, that the $\dot{P}$ changes over the years.  Schaefer et al. (2010) calculate a quantity proportional to the average $\dot{M}$ (as based on the B-band flux) and demonstrate variations by a factor of 2.0.  From Equation 21, these changes in $\dot{M}$ will translate into corresponding changes in $\dot{P}$.  With this, we expect that curvatures will vary from 1945-2009, even though the $\Delta P$ values could remain constant.  Figure 17 shows the difference between a steady $\dot{P}$ case and a variable $\dot{P}$ case.  We do not have enough information to measure both $\dot{M}$ and $\dot{J}$ throughout the various inter-eruption intervals, so we cannot construct any approximate O-C curve back to 1945.  Nevertheless, the required $\langle \dot{P} \rangle$ (for the observed $\Delta P$ from 1999) is the same as the observed $\dot{P}$ from 2001-2009, so the 1945 eclipse time is plausible for the expected situation.  In all, the 1945 eclipse time is consistent with the idea that $\dot{P}$ varies somewhat as expected, yet the $\Delta P$ value is constant.

	\subsection{Summary for U Sco}

I have a good measure of the orbital period change across the 1999 eruption.  (The measure of $\Delta P$ for the 2000 eruption of CI Aql is better because there are more points and the intrinsic timing jitter is smaller.)  My best fit value and the one-sigma uncertainty is $\Delta P = (+43 \pm 69) \times 10^{-7}$ days.  Note that the value is within one-sigma of zero.  

The $\Delta P$ value depends primarily on the curvature in the O-C diagram.  The best fit and extreme values of $\Delta P$ correspond to the best fit and extreme values of $\dot{P}=(-250 \pm 170) \times10^{-11}$ days per cycle.  In retrospect, this degeneracy is caused by the lack of eclipses from May 1997 to May 2001, which is the interval where the models disagree the most, as can be seen in Figure 15.  We do have three eclipses at the time of the 1999 eruption, but these have relatively large error bars on the eclipse times and they have offsets from the quiescent ephemeris.  Correcting for the offsets as well-observed in the 2010 eruption and taking a weighted average, I get an O-C value of -0.0067$\pm$0.0041.  This could be plotted in Figure 15.  This composite O-C curve point is close to the curve with the maximal $\Delta P$, however its error bars include the best fit curve, and even the other extreme curve is not confidently rejected.  The post-eruption gap will not be repeated for the 2010 eruption, yet a good measure must await the full analysis that would independently yield a $\Delta P$ across the 2010 eruption.  In all, at this time, I know of no way to break the degeneracy between $\Delta P$ and $\dot{P}$ across the 1999 eruption.

To be explicit, let me quote the full best fit ephemeris for the entire 1989-2009 interval:
\begin{equation}
T_{eph}=2451234.5348 + 1.23055183 N - 0.5 N^2 250\times10^{-11}, 0<N<3200,
\end{equation}
\begin{equation}
T_{eph}=2451234.5348 + 1.2054753 N - 0.5 N^2 250\times10^{-11}, -3600<N<0,
\end{equation}  
This best fit ephemeris is given as the thick curve in Figures 13, 15, and 16.  During the tail of an eruption, there are offsets from this ephemeris by up to 0.010 days due to shifts in the center of light within the binary system.  Extensions before 1987 are substantially uncertain due to the inevitable secular variations in $\dot{P}$.

In all, I have a confident measure of the period change across the 1999 eruption of U Sco, with the value being $(+43 \pm 69) \times 10^{-7}$ days.  This was the observational goal of my program started 24 years ago.

\section{The Ejected Mass}

My observational goal (measure $\Delta P$) was designed to obtain the best possible measure of the mass ejected by the RN eruptions.  Here, I will complete that analysis for both CI Aql and U Sco.

	\subsection{$M_{ejecta}$ for CI Aql}

The basic formula for determining $M_{ejecta}$ is Equation 19.  Here, I will collect the various needed system parameters.  Hachisu, Kato, \& Schaefer (2003) and Lederle \& Kimeswinger (2003) both give the mass of the WD as 1.2 M$_{\odot}$, although the mass cannot be much smaller for the system to have a recurrence time scale known to be as small as 23.6 years.  Hachisu, Kato, \& Schaefer (2003) place only a weak constraint on the mass of the companion star to be greater than 1.0 M$_{\odot}$, while we have a strong constraint that $q\lesssim 5/6$.  $M_{comp}=1.0$ M$_{\odot}$ is consistent with all the constraints.  Hachisu, Kato, \& Schaefer (2003) give $a=4.25$ R$_{\odot}$ and $R_{comp}=1.69$ R$_{\odot}$, so that $\beta=0.04$.  There is no reason to expect that the mass ejected from the system will have a high or low specific angular momentum, so I take $\alpha=1$.  With this, the best value of $A$ from Equation 15 is 1.02, while plausible variations in the input parameters allow $A$ to vary over a range with $\pm$0.06.  The expansion velocity of the shell is 2000-2500 km s$^{-1}$ (Kiss et al. 2001).  The orbital velocity is 230 km s$^{-1}$.  The drag term in the square bracket of Equation 19 is negligibly small at -0.016.

With the drag term being negligible, Equation 19 becomes $M_{ejecta}=M_{WD}(\Delta P/P)/A$.  The measure of $\Delta P=-3.7^{+9.2}_{-7.3} \times 10^{-7}$ days has its one-sigma upper limit at $\Delta P<+5.5 \times 10^{-7}$.  With the best values and the one-sigma upper limit on $\Delta P$, I find that $M_{ejecta}<1.0 \times 10^{-6}$ M$_{\odot}$.  This limit is insensitive to the adopted stellar masses and sizes.  The three-sigma upper limit (with $\Delta P < 24 \times 10^{-7}$ days) is $M_{ejecta}<4.5 \times 10^{-6}$ M$_{\odot}$.  

	\subsection{$M_{ejecta}$ for U Sco}

For U Sco, the mass of the WD must be near the Chandrasekhar mass, with Hachisu et al. (2000) giving $M_{WD}=1.37 \pm 0.01$ M$_{\odot}$.  The companion star was measured by Thoroughgood et al. (2001) to have $M_{comp}=0.88 \pm 0.17$ M$_{\odot}$, and this is consistent with the requirement that $q\lesssim 5/6$.  Hachisu et al. (2000) give $a=6.87$ R$_{\odot}$ and $R_{comp}=2.66$ R$_{\odot}$, so that $\beta=0.04$.  These values give $A=1.11 \pm 0.05$ with the uncertainty dominated by $M_{comp}$.  The expansion velocity of U Sco is quite high, being close to 5000 km s$^{-1}$ (Munari et al. 1999).  The orbital velocity is 195 km s$^{-1}$.  The drag term in the square bracket of Equation 19 is -0.007 and completely negligible.

My measured value of $\Delta P$ is $(+43 \pm 69) \times 10^{-7}$ days.  With this, I find the best value to be $M_{ejecta}=43 \times 10^{-7}$ M$_{\odot}$.  (The equality of $M_{ejecta}$ and $\Delta P$ in these units is coincidental.)  The one-sigma upper limit is $110 \times 10^{-7}$ M$_{\odot}$, while the three-sigma upper limit is $250 \times 10^{-7}$ M$_{\odot}$.

\section{Conclusions}

The Type Ia supernova progenitor problem has long been of high importance, and recurrent novae have long been one of the best candidate progenitors.  The progenitor question revolves around whether the white dwarf is gaining or losing mass, and this depends primarily on whether the ejected mass is larger or smaller than some fairly-well-known critical accreted mass.  Previous methods of estimating $M_{ejecta}$ are uncertain by several orders of magnitude, so a new method is needed.  With a very long term program for measuring the sudden change of the orbital period across the recurrent nova eruption, we can determine $M_{ejecta}$ to the same fractional accuracy at which we can measure $\Delta P$.  That is, the systematic errors are all small for these recurrent novae.

On this basis, in 1987, I started on a program to measure the orbital period change across recurrent nova eruptions.  For CI Aql, I collected 4960 magnitudes over 387 nights (with 280 critical nights from Mennickent \& Honeycutt) recording 91 eclipses from 1926 to 2009.  For U Sco, I collected 2382 magnitudes over 116 nights recording 50 eclipses from 1989 to 2010 in quiescence, plus 15 eclipses during eruptions from 1945 to 2010.  This paper is the first report on my program.

The results are that $\Delta P=-3.7^{+9.2}_{-7.3} \times 10^{-7}$ days for CI Aql and $\Delta P=(+43 \pm 69) \times 10^{-7}$ days for U Sco.  Both of these values are consistent with zero, so it might be better to express the results as upper limits.  The dominant uncertainty in measuring $\Delta P$ was the trade-off with $\dot{P}$.  With the fairly well-known system parameters, I derive one-sigma upper limits on $M_{ejecta}$ of $<10 \times 10^{-7}$ M$_{\odot}$ for CI Aql, and $<110 \times 10^{-7}$ M$_{\odot}$ for U Sco.  

The further question of whether the white dwarfs are increasing in mass (i.e., whether $M_{ejecta} \le \dot{M} \Delta T$) is still a complex one.  For U Sco, with $\dot{M} \approx 0.1 \times 10^{-6}$ M$_{\odot}$ $yr^{-1}$ (Duschl et al. 1990; Hachisu et al. 2000b; Shen \& Bildsten 2007) and $\Delta T = 10 \pm 2$ years, the conclusion remains ambiguous.  For CI Aql, with $\dot{M} \approx 0.1 \times 10^{-6}$ M$_{\odot}$ $yr^{-1}$ (Hachisu, Kato, \& Schaefer 2003) and $\Delta T = 24 \pm 5$ years (Schaefer 2010), the white dwarf appears to be gaining mass.

For the future, I will continue to make frequent eclipse timings each year for U Sco and CI Aql.  For U Sco, we will start to get the $\Delta P$ value across the 2010 eruption, with this taking perhaps a decade (until the $\sim$2020 eruption) to get useful accuracy.  For this eruption, I already have many eclipse timings soon before and after the eruption so as to avoid the limiting mistake of having a large gap around the time of the eruption, so I anticipate that the $\Delta P$ across the 2010 eruption will be substantially more accurate than for the 1999 eruption and should be able to produce an unambiguous conclusion.  For CI Aql, continuing timings will extend the post-eruption interval which will improve the measure of $\dot{P}$ and hence make for a substantially improved $\Delta P$ across the 2000 eruption.  Now, with the recent surprising eruption of T Pyx (with various groups having highly accurate $P$ and $\dot{P}$ measures from the middle 1980's up until 40 days before the eruption), we have a third recurrent nova that will soon be able to have an accurate measure of $\Delta P$.  Over the next decade, I anticipate the discovery of 1-3 new recurrent novae from amongst the current list of so-called classical novae, and some of these might already have a good pre-eruption period.

~

~

This work is a summary of observations made at many telescopes from 1987 to this year, and I have had much support from queue observers, time allocation committees, and observatory staffs, despite the goal being only far in the future.  So I thank Suzanne Tourtellotte, Charles Bailyn, Michelle Buxton, Rebecca Winnick, Mario Hamuy, Arturo Gomez, Alison Doane, Eli Rykoff, Lonique Coots, Fergal Mullally, and Hye-Sook Park.  Kent Honeycutt provided the data for CI Aql from 1991 to 1996, which is critical for the analysis of the period change.  Richard Wade took spectra and magnitudes of U Sco at Cerro Tololo in June 1990, and I am thankful for his recovery of this information from his old notebooks.  I am a Visiting Astronomer at the Cerro Tololo Inter-American Observatory and the Kitt Peak National Observatory, National Optical Astronomy Observatories, which are operated by the Association of Universities for Research in Astronomy, under contract with the National Science Foundation.  The Royal Observatory in Edinburgh provided the scans of the deep Schmidt survey plates.  This work is supported under a grant from the National Science Foundation (AST 0708079).

~

~
\begin{center}   {\bf Appendix A}  \end{center} 

\begin{center}   {\bf Prior Estimates of $M_{ejecta}$}   \end{center} 

In the normal course of research on these two stars, prior workers have published a variety of estimates of $M_{ejecta}$ from both CI Aql and U Sco.  All of these are based either on emission line fluxes or on theory.  In this appendix, I will summarize and evaluate these prior estimates.
	
The traditional method for estimating the mass ejected is to measure one of the hydrogen emission lines, where the flux should be proportional to the ejected mass and other parameters.  The basic idea is to count the number of, say, H$\alpha$ photons being emitted each second, and with a known density this can be translated into the number of hydrogen atoms in the nova shell and hence a mass of the ejected material.  The observed flux in the H$\alpha$ line is
\begin{equation}
F_{H\alpha} = (h \nu_{H\alpha} \alpha_{H\alpha} / 4 \pi D^2) \int n_e n_H dV.
\end{equation}
Here, $h$ is the Planck constant, $\nu_{H\alpha}$ is the frequency of the light, $\alpha_{H\alpha}$ is the recombination coefficient for the line, $D$ is the distance to the star, $n_e$ is the electron number density, $n_H$ is the hydrogen number density, and the integral is performed over the volume element $dV$.  This equation can also be applied for the H$\beta$ line.  For nova shells, the ionization fraction is high, so $n_e=n_H$ to a good approximation.  The number densities are assumed to be constant within some volume, taken to be $\epsilon (4\pi /3) R_{shell}^3$, where $\epsilon$ is some filling factor and the shell radius is $R_{shell}= v \Delta t$, where $v$ is the expansion velocity of the shell (determined from the line profiles) and $\Delta t$ is the time since the eruption.  The above equation can be solved for $n_H$, with the mass of the shell being $m_H \int n_H dV$.  The measured quantities are $F_{H\alpha}$, $D$, $v$, and $\Delta t$, while the guessed input is either $n_e$ or $\epsilon$ (depending on the way these formulae are applied).

For CI Aql, no prior observational estimate of $M_{ejecta}$ has been published.  For U Sco, four such estimates have been made.  Barlow et al. (1981) found $M_{ejecta}=72 \times 10^{-7}$ M$_{\odot}$ yr$^{-1}$ from the H$\beta$ line fluxes on days 6 and 12 after the peak of the 1979 eruption.  Williams et al. (1981) found $0.1-1 \times 10^{-7}$ M$_{\odot}$ yr$^{-1}$ from line fluxes on days 4 to 16 after the peak of the 1979 outburst.  Anupama \& Degawan (2000) found $\sim 1 \times 10^{-7}$ M$_{\odot}$ yr$^{-1}$ from the H$\alpha$ and H$\beta$ line fluxes on days 11-12 after the peak of the 1999 eruption.  Banerjee et al. (2010) report that the mass range is $7.2-23 \times 10^{-7}$ M$_{\odot}$ yr$^{-1}$ from the infrared hydrogen lines for 1.5 days after the peak of the 2010 eruption.  (All of these eruption light curves are identical to within measurement errors, Schaefer 2010, and nova trigger theory also strongly points to the eruptions being identical, Schaefer 2005, so the ejected masses should be the same for all these eruptions.)  In units of $10^{-6}$ M$_{\odot}$ yr$^{-1}$, the four values are 7.2, 0.01-0.1, 0.1, and 0.72-2.3.  We have the discouraging reality that the reported $M_{ejecta}$ values have a range of almost three orders of magnitude.

I can also estimate the total uncertainty by considering the errors arising from each of the assumptions and input values.  Here, I will neglect the errors arising from the implicit assumption of Case B recombination (which is a good approximation), the implicit assumption that the shell is completely ionized (another good approximation), the number of electrons per nucleon (1 for pure hydrogen and 2 for pure helium), and the extinction correction (this notorious problem is actually negligibly small compared to many other uncertainties).  Here are six huge sources of uncertainty:  

(1) The $\alpha_{H\alpha}$ depends critically on the temperature of the shell, which is only known approximately.  Even for the detailed physical analysis of the 1979 eruption of U Sco, Williams et al. (1981) could only consider temperatures over the range of 10,000 to 25,000, and the resulting shell masses varied by a factor of 1,500.  So already we know that any $M_{ejecta}$ measures by this method have real uncertainties of orders-of-magnitude.  Later papers have ignored this problem, instead simply adopting some one temperature and hence its corresponding $\alpha_{H\alpha}$.  

(2) The filling factor, $\epsilon$, is based entirely on guesses, with no factual basis.  No calculations are made based on the theory of turbulence in the shell, and there is no way to measure the filling factor based on line ratios.  No account is ever taken of the hollow inside the shell caused by the turn-off of ejection, the inevitable bipolar outflows caused by the accretion disk, or the equatorial plane of enhanced emission (as causes the triple peaked lines observed for U Sco).  While the value of $\epsilon$ cannot be gotten from theory or observation, workers can only resort to guesses.  For U Sco, published values for $\epsilon$ range from $<$0.001 to 0.1, but I see no reason to think that larger or smaller values are not unreasonable.  Even with this, we have an uncertainty of two orders of magnitude in $\epsilon$ and hence also in $M_{ejecta}$.  

(3) The nova distance enters in as a square, because we have to convert the observed line flux into a line luminosity.  For expansion parallaxes, Wade et al. (2000) have shown that uncertainties related to the unknown shape of the ejecta will lead to errors by up to a factor of 2.5, while they report further major problems with knowing what isophot to use, and knowing what velocity in the line profile (e.g., HWZI or HWHM) matches to the selected isophot.  For the particularly good case of RS Oph, in light of these systematic problems with the expansion parallax, the real uncertainty in distance is over a factor of two (Schaefer 2009).  For distances based on the `maximum magnitude versus rate of decline' (MMRD) relation (Downes \& Duerbeck 2000), the one-sigma scatter is 0.6 mag with greatest deviations of 1.6 mag, while even the different versions of the relation differ by up to 1.0 mag.  It might be possible to improve the calibration of the MMRD by the use of extragalactic novae, for which the one-sigma scatter is roughly 0.3 mag with greatest deviations of 1.1 mag (Della Valle \& Livio 1995).  However, Kasliwal et al. (2011) have gathered a new set of nova light curves (with very high quality, deep limits, daily cadence, and spectral confirmations) to demonstrate that the MMRD actually does {\it not} hold for extragalactic novae, as well as that the MMRD does not work either for recurrent novae or for detailed nova explosion models.  In general, the distances to novae are only known to a factor of two or so, with the resulting distance uncertainty making for a $\sim$4$\times$ error in $M_{ejecta}$.  In particular, CI Aql has its best distance to be $5000^{+5000}_{-2500}$ pc with a factor of two error, while U Sco has the unique blackbody distance for its companion star allowed for by its total eclipse to be $12\pm2$ kpc (Schaefer 2010).  

(4) The traditional method with line fluxes implicitly assumes that the shell is optically thin, so that we can see {\it all} the line photons being emitted.  Early during the eruption, the shell is certainly optically thick, and it becomes thin only after substantial expansion.  We know in the case of U Sco, that the inner binary is hidden behind an optically thick photosphere until about 13 days after the peak, when the eclipses suddenly start and when the supersoft source becomes first visible (Schlegel et al. 2010; Schaefer et al. 2010).  Presumably, the outer volume of the shell can be optically thin, while the inner volume (and the volume it shadows) is not included in the mass estimate.  All U Sco estimates are made based on spectra from dates when the shell is still optically thick.  A correction factor is needed, and this will be large when the nova is near peak and will then decrease to near unity as the eruption ends.  I have seen no consideration or calculation of this correction factor in the literature.  

(5) The line flux scales as the square of the electron density for complete ionization, so the derived ejecta mass has a strong dependency on $n_e$.  In the literature, the line ratios cannot give any accurate density measure, so the best we have for U Sco is constraints like  $2 \times 10^8 \lesssim n_e \lesssim 10^{10}$ cm$^{-3}$ (Barlow et al. 1981) and $10^7 \lesssim n_e \lesssim 10^{11}$ cm$^{-3}$ (Williams et al. 1981).  The need to know the value of $n_e$ can be eliminated (e.g., Banerjee et al. 2010), but then the uncertainty is simply transfered to the filling factor, for which the uncertainty is still very large.  In all, $n_e$ is uncertain by three orders of magnitude, with a corresponding error in $M_{ejecta}$.  

(6) The adopted volume of the shell scales as the cube of the expansion velocity, with $v$ taken from some emission line profile.  However, there is no understanding of {\it which} velocity to take from the line profile so as to correlate with the outer edge of the shell.  Should we take the HWHM or the HWZI of the line profile, or some other velocity?  The line widths change substantially with time as the photosphere recedes, so should we take the line width at the nova peak, at the time of the observation, or at some other time?  For U Sco, Zwitter \& Munari (2000) gives the HWZI to vary as $5015-130 \Delta t$ km s$^{-1}$, while Anupama \& Dewangan (2000) report HWZI values of 5065 and 3262 km s$^{-1}$ for $\Delta t$ values of 0.5 and 12 days.  We also expect differences in velocity by perhaps a factor of 2.5 with the angle from the orbital pole due to bipolar ejection (e.g., Walder et al. 2008).  All these problems certainly lead to errors in $v$ of a factor of $\gtrsim$2.  With this, the resulting error in $M_{ejecta}$ will be a factor of $\gtrsim$8.

We see that the $M_{ejecta}$ values from the traditional method has many large errors that cannot be recovered.  When many of the problems {\it are} considered, the allowed range becomes very large, for example the very detailed physical model for V838 Her gives a final range of $1.6 \times 10^{-7} < M_{ejecta} < 9.6 \times 10^{-3}$ M$_{\odot}$ (Vanlandingham et al. 1996), a factor of 60000.  The sizes of the range in $M_{ejecta}$ are factors of 1500 for $\alpha_{H\alpha}$, 100 for $\epsilon$, $\sim$4 for $D$, 1000 for $n_e$, and $\gtrsim$8 for $v$.  Depending on the formulation, only one of the uncertainties for $\epsilon$ or $n_e$ should be applied.  Added in quadrature (logarithmically), the real total uncertainty, given all the problems with the input parameters, is four orders of magnitude.  In summary, the published $M_{ejecta}$ values from the traditional method have a real uncertainty of greater than three orders of magnitude.
	
Various estimates of $M_{ejecta}$ have been published with the basis being theoretical models.  We have to be careful with such estimates, because models and ideas change, and we do not want to evaluate later models on the basis of earlier models.  Importantly, the theoretical models all have made many untested assumptions, any of which could easily be wrong.  Nevertheless, in the void of any other useful information, theoretical estimates of $M_{ejecta}$ can be of interest for some purposes.

For CI Aql, the only theoretical estimate of $M_{ejecta}$ is from a very detailed model of the nova event by Hachisu, Kato, \& Schaefer (2003) and Hachisu \& Kato (2000).  They find ejected masses of $47-66 \times 10^{-7}$ M$_{\odot}$.

For U Sco, a variety of theoretical estimates have been published.  Detailed physical models of the U Sco system have resulted in estimates of $2.1 \times 10^{-7}$ M$_{\odot}$ (Kato 1990), $\sim 18 \times 10^{-7}$ M$_{\odot}$ (Hachisu et al. 2000a), and  $4.3 \times 10^{-7}$ M$_{\odot}$ (Starrfield et al. 1988).   Yaron et al. (2005) have constructed a generic series of nova models, for which I have take $\dot{M}=10^{-7}$ M$_{\odot}$ yr$^{-1}$ and interpolated to $M_{WD}=1.37$ M$_{\odot}$, for a model value of $M_{ejecta}=44 \times 10^{-7}$ M$_{\odot}$.  These values stretch over a range of a factor of 22.

A tempting, but ultimately circular, argument comes from observational estimates of the mass accreted onto the WD along with some theoretical factor for how much of that mass is actually ejected.  So for example, U Sco has $\dot{M} \sim 10^{-7}$ M$_{\odot}$ yr$^{-1}$ and inter-eruption intervals of around 10 years, so $M_{ejecta} \sim10^{-6}$ M$_{\odot}$ if it is all blown off.  Critically, we have no real measure of the ratio of the accreted mass to the ejected mass.  A key issue for the application in this paper (measuring $M_{ejecta}$) is that any assumption of this ratio is circular in that is assumes the answer.
	
The only prior method of measuring $M_{ejecta}$ is to use the line fluxes.  However, the evidence is overwhelming that the real uncertainties in this method lead to errors spanning a range of larger than three orders of magnitude.  From theory, the span is better, being only a factor of 22, with the price being that we are captives of the unknown validity of the many model assumptions.  For factors of $>$1000 or 22, the uncertainties are much too large to be of any use for the question of whether $M_{ejecta}>\dot{M} \Delta T$.

~

~
\begin{center}   {\bf Appendix B}  \end{center} 
\begin{center}   {\bf Period Changes and the O-C Curve}   \end{center} 

The observational goal of this paper is to measure the period changes across the RN eruptions.  For this, I must present the background and calculations for the physics of the period changes in the system, as well as my primary tool of the O-C diagram.
	
The orbital period of any nova system will suffer an abrupt period change during a nova eruption simply because mass is ejected from the system.  This is easy to see from Kepler's Law, where the orbital period is a function of the total system mass, so any ejection must change the period.  With this in combination with conservation of angular momentum, Schaefer \& Patterson (1983) derive the period change from the mass loss to be 
\begin{equation}
\Delta P_{ml} = A P M_{ejecta} / M_{WD},
\end{equation}
\begin{equation}
A = (2q+3q^2-3q^2\alpha + 3q^2\alpha \beta -3\beta -2\beta q)/(q+q^2),
\end{equation}
\begin{equation}
q=M_{comp}/M_{WD},
\end{equation}
\begin{equation}
\beta = - \dot{M}_{comp} / \dot{M}_{WD}.
\end{equation}
The mass ratio ($q$) is the mass of the companion star ($M_{comp}$) divided by the mass of the white dwarf ($M_{WD}$).  The $\alpha$ parameter is the average specific angular momentum of the ejected material in units of the specific orbital angular momentum of the WD.  Under ordinary circumstances, we have a strong expectation that $\alpha$ will be unity, and this is consistent with approximate constraints derived for the nova BT Mon (Schaefer \& Patterson 1983).  The $\beta$ parameter is a measure of the fraction of the ejected mass that ends up on the companion star.  The shell's expansion velocity is much greater than the escape velocity, so $\beta$ will be determined by the star's geometrical cross section.  Closely, $\beta = (\pi [R_{comp}/a]^2)/(4 \pi)$, where $R_{comp}$ is the radius of the companion star and $a$ is the semimajor axis of the orbit.

For $\alpha=1$ and the approximation that $\beta \approx 0$, then $A=2/(1+q)$.  With this, $\Delta P_{ml}/P = 2 M_{ejecta}/(M_{comp}+M_{WD})$.  From the ejection, $\Delta P_{ml}$ must be positive, which is to say that $P_{post}>P_{pre}$.  So the effect of the mass ejection is to lengthen the period.  This will also slightly increase the separation between the two stars.

The loss of mass by the system will increase the orbital period, whereas other effects might serve to abruptly decrease the orbital period across an eruption.  Livio (1991) provides calculations of the period decrease due to angular momentum loss caused by the drag of the companion star as it moves through the nova envelope (somewhat like in a common envelope stage).  This was applied to the case of the RN T Pyx, with the result that the period decrease arising from the frictional angular momentum losses were somewhat larger than the period increase arising from the mass ejection.  So if we are to isolate the effects of the mass ejection, then we must be able to calculate the drag effects.  Schaefer et al. (2009, Eq. 4) give the period change from the common envelope drag during the eruption to be 
\begin{equation}
\Delta P_{drag} = -0.75 P (M_{ejecta}/M_{comp})(V_{orb}/V_{exp})(R_{comp}/a)^2.
\end{equation}
Here, $V_{orb}$ is the orbital velocity of the companion star and $V_{exp}$ is the expansion velocity of the ejected shell.  For both CI Aql and U Sco, the $V_{orb}/V_{exp}$ factor is small.

Other mechanisms might also be able to produce angular momentum losses during the eruption.  Martin et al. (2010) proposes one reasonable mechanism.  For this paper, I will only consider the effects of the mass loss and frictional drag.  The observed sudden period change across an eruption will be the sum of all the effects, so $\Delta P = \Delta P_{ml} + \Delta P_{drag}$.  With Equations 14 and 18, 
\begin{equation}
\Delta P/P = (M_{ejecta}/M_{WD})[A-0.75 (V_{orb}/V_{exp})(R_{comp}/a)^2/q].
\end{equation}
With the various system parameters known with good accuracy, if we measure $\Delta P$, then we can derive $M_{ejecta}$.  This is the key point for converting the observational goal of my program (measuring $\Delta P$) into a science result (deriving $M_{ejecta}$) of importance for the Type Ia supernova progenitor question.
	
In the quiescence between eruption, the orbital period must change, if only due to the steady transfer of mass between the stars.  That is, for conservative mass transfer, the simple shifting of mass will move the system's center of mass and cause a steady period change.  In addition, there might be mechanisms (for example, a wind from the companion star) causing a steady loss of angular momentum from the system, with this leading to a steady period change.  For both cases, the change of the orbital period across each orbital cycle ($\dot{P}$ measured in units of days per cycle) can be idealized as being constant throughout quiescence.

In quiescence, the period is given as 
\begin{equation}
P=P_0+N\dot{P},
\end{equation}
where $N$ is the cycle count of orbits from some fiducial time (i.e., a measure of time) and $P_0$ is the period at the epoch when $N=0$.  The observed eclipse times will be
\begin{equation}
T_{obs}=E_0 + N P_0 + 0.5N^2 \dot{P}.
\end{equation}
I will constrain $\dot{P}$ by three different methods.  First, I will use the eclipse times from 2001-2009 to measure the quadratic term.  Second, I will use independent measures of the mass transfer rate and the stellar masses so as to calculate the theoretical $\dot{P}$ for conservative mass transfer.  This will actually only place an upper limit on $\dot{P}$, because we cannot make any real estimate the rate of angular momentum loss from the system.   Third, I can use the eclipse times from the first half of the 1900's so as to constrain the long term average $\dot{P}$.

With Kepler's Law, the basic equation for conservative mass transfer (where no mass is lost from the binary system) is
\begin{equation}
\dot{P}/P = (3\dot{J}/J) + (3 \dot{M}/M_{comp})(1-q),
\end{equation}
where the mass transfer rate ($\dot{M}$) is a positive quantity, $J$ is the total angular momentum in the system, and $\dot{J}$ is the time derivative of $J$ (Frank et al. 2002).  For the usual units of days, solar mass, solar mass, and solar mass per year, respectively, the unit of $\dot{P}$ will be days per year.  To get to days per cycle (as in Eq. 19), we have to multiply by $P/365.25$.  In the limiting case of $\dot{J}=0$, we have $\dot{P}>0$ when $q<1$.  That is, when the WD is more massive than the companion star, the orbital period should be steadily increasing.  The angular momentum of the system cannot be increasing, so $\dot{J} \leq 0$, so this can set an upper limit on $\dot{P}$,  
\begin{equation}
\dot{P} \leq (3P \dot{M}/M_{comp})(1-q).
\end{equation}
If the angular momentum loss is small then $\dot{P}$ will be positive, otherwise $\dot{P}$ can be negative.  The size of $\dot{J}$ cannot now be known with any reliability, so all we really have is a {\it limit} on $\dot{P}$.

Both U Sco and CI Aql have relatively steady accretion on the time scale of a century or so.  Certainly, the accretion is not secularly increasing on a dynamical or thermal time scale.  This can be used to place a limit on the mass of the companion stars.  Frank et al. (2002, Eq. 4.15) derives a limit that a binary with conservative mass transfer and with $q \gtrsim 5/6$ will undergo runaway accretion.  Immediately, we see from Equation 22 that the upper limit on $\dot{P}$ will be positive, so the $\dot{P}$ value might be either positive or negative.  For our two RNe, the limit on $q$ says $M_{comp} < (\sim5/6)M_{WD}$.  Both of the stars are RNe, so $M_{WD} \lesssim 1.4$ M$_{\odot}$ and hence $M_{comp}<1.2$ M$_{\odot}$.
	
The long term change in the orbital period will be the result of the steady change between eruptions punctuated by the abrupt change across each eruption.  As used below, the base orbital period ($P_0$) is the average period over the last decade.  In the time interval from the last eruption until now, the period will be changing linearly with time, with $P_0$ corresponding to the period for a time near 2005. Below, I will be keeping track of time by means of a cycle count from the time of the eruption a decade ago (in 2000 for CI Aql and 1999 for U Sco).  The cycle count will be an integer, $N$, that counts the cycles from that epoch.  This cycle count up to the time of the average period (around 2005) will be labeled $N_{2005}$.  With this, the period across the last decade or so will be $P=P_0+(N-N_{2005})\dot{P}$.  Just after the time of the previous eruption (with $N\approx 0$), the period is $P_0-N_{2005}\dot{P}$. 

Now we can consider the eruption in the preceding inter-eruption interval.  (For U Sco, this interval is from 1987 to 1999.)  Just before the eruption, the orbital period is smaller by $\Delta P$, so $P=P_0-N_{2005}\dot{P}-\Delta P$.  For the time before this eruption (as measured by the cycle count N, with $N<0$), the period is $P=P_0-N_{2005}\dot{P}-\Delta P + N\dot{P}$.  This inter-eruption interval is started at the time of the preceding eruption, with a cycle count of $N_{-1}$.  The period just after this prior eruption is $P_0-N_{2005}\dot{P}-\Delta P +N_{-1}\dot{P}$.

In the preceding inter-eruption interval (for U Sco, from 1979 to 1987), the cycle count will go from $N_{-2}$ to $N_{-1}$.  Over this interval, $P=P_0-N_{2005}\dot{P}-2\Delta P +N\dot{P}$.

We can generalize this to all earlier intervals as $P=P_0+(N-N_{2005})\dot{P}-n\Delta P$.  Here, $n$ is the number of eruptions between the time and the year 2005 (with $n$ being a positive number).  As time increases, the $-n\Delta P$ term will be getting less negative.  If $\dot{P}>0$, then the $(N-N_{2005})\dot{P}$ term will also be getting less negative as time increases.  With this case, both steady period change and the mass ejection will work in the same direction, with the period secularly increasing.  Alternatively, if $\dot{P}<0$, then the period will alternatively increase across eruptions and decrease between eruptions.  The long term trend will have the period being roughly constant for the case where $\dot{P} = -\Delta P/ \Delta N$, where $\Delta N$ is the average number of cycles between eruptions.

A substantial problem with the idealization in the preceding paragraph is that the quiescent brightness of nova change on all time scales with surprisingly large amplitude (e.g., Collazzi et al. 2009; Schaefer 2010; Honeycutt et al. 1998; Kafka \& Honeycutt 2004).  The brightness is largely a measure of the accretion rate and hence of the $\dot{P}$.  So the $\dot{P}$ is changing on all time scales with substantial amplitude.  For recurrent novae in particular, Schaefer (2005; 2010) and Schaefer et al. (2009) have demonstrated that factor of two variations are common, and T Pyx even has varied by a factor of $\sim30$ over the last century.  This allows and requires that the steady period changes vary across and within inter-eruption intervals.  Without having highly accurate measures with great coverage, about all we can do is deal with the averages over whatever interval is in question.  Such averages should not be applied to other time intervals.

For CI Aql, Schaefer (2010) has already pointed out that the quiescent brightness level has only small changes.  This impies that CI Aql should have only relatively small secular changes of $\dot{P}$.  For U Sco, Schaefer (2005; 2010) and Schaefer et al. (2010) have already pointed out that the accretion rate varies by up to factors of two.  This implies that U Sco should have substantial secular changes of $\dot{P}$.

The O-C curve is a convenient and traditional means to represent deviations between observations (`$O$') and the model calculations (`$C$').  For a linear model with no period changes, the ephemeris predicts that the time of minimum is $T_{model}=E_0 + N P$, where $E_0$ is some epoch (in HJD), $N$ is an integer that counts cycles from the epoch, and $P$ is the orbital period.  The O-C simply equals $T_{obs}-T_{model}$.  The O-C diagram is a plot of the observed O-C values as a function of the time.  Nonlinear models can be used in principle (for example to remove some known effect), but for this paper I will use a linear model for constructing the O-C curve.

If the observations perfectly followed the model, then O-C would be zero for all times.  With the usual observational uncertainties and a perfect model, the values of O-C will scatter uniformly around the horizontal axis.  If the epoch of the model is wrong, then the O-C values will follow a horizontal line that is offset from the horizontal axis by an amount that equals the error in $E_0$.  If the period of the model is wrong, then the plotted values will  have a non-zero slope, a positive slope if the true period is longer than the model period, and a negative slope if the true period is shorter than the model period.  If the orbital period changes uniformly, then the O-C curve will show a parabola.  For the case where the orbital period decreases with a constant period derivative, the parabola will be concave down.  For the case where the period undergoes a sudden change, then the O-C curve will show a broken line, with the kink at the time of the change and the before and after slopes appropriate for the periods.  The larger the angle at the kink, the larger the period change.  

For the case of a nova eruption that increases the orbital period and for a convention that the (better measured) post-eruption period is that used in the linear ephemeris, then the O-C curve should show the pre-eruption segment as a line falling with negative slope, touching the horizontal axis at the time of the eruption, then breaking to a horizontal line that follows the axis off to the right.  With the reality that the binaries will suffer some small steady period change due to mass transfer, the simple broken line segment picture must be made more complex by having each of the line segments actually being a shallow parabola.  The O-C curve will always be continuous, as any discontinuity could only be caused by the stars jumping forward in their orbit.  The derivative of the O-C curve can change discontinuously with a sudden period change (like when the nova blows of a shell), and indeed this is the very effect I am looking for.

To illustrate the O-C curves, I have made some schematic plots.  Figure 17 shows the idealized case of a nova that erupts in 1999 for various values of $\Delta P$ for the case with $\dot{P}=0$.  The linear ephemeris is from the best fit linear period after the eruption.  The period before the eruption is shorter, so the O-C curve is a descending line segment which meets the post-eruption segment at the time of the eruption.  The amount of the kink at the time of the eruption is determined by $\Delta P$.  The uppermost line is for the case of a large $\Delta P$ caused by a large $M_{ejecta}$, such that the WD is losing more mass each eruption than it gains between eruptions so that the WD is losing mass in the long term.  (For an RN system, the implication is that the system then is {\it not} a Type Ia supernova progenitor.)  For RN systems, each will have some critical $\Delta P$ for which the WD is (on average) neither gaining nor losing mass, and this is illustrated by the middle line in Figure 17.  For lines below this, those with a smaller-than-critical $\Delta P$, the corresponding mass ejection is so small that the WD is {\it gaining} mass over each eruption cycle, and any RN system like that will inevitably have the WD rise to the Chandrasekhar mass and become a Type Ia supernova.  The various broken lines segments in Figure 17 are the theoretical predictions for the various models as described.  These predictions are to be compared against the actual eclipse times, which can be plotted as points with error bars on the same O-C diagram.  This comparison can be made (between theory and observation) by means of the usual chi-square minimization, in this case so as to derive the best fit value for $\Delta P$.  To illustrate, simulated data points have been included in the left side of Figure 17, and we see that a small $\Delta P$ is indicated such that the system has its WD mass increasing.  The right side of Figure 17 illustrates the case where $\dot{P}$ is substantially negative.

The long term O-C diagram will depend on the interplay of $\Delta P$ and $\dot{P}$.  Figure 18 shows an illustrative O-C diagram for a case like U Sco.  Each dashed vertical line indicates a known eruption, while the undoubted eruption around the year 1957 is also included for its $\Delta P$ even though no line is drawn for this event.  The $\Delta P$ is held constant for all eruptions.  The upper and the lower curves are for cases where $\dot{P}$ is held constant, with the upper curve approximately satisfying the relation $\dot{P} = -\Delta P/ \Delta N$.  The idealized case of constant $\dot{P}$ is certainly wrong for novae and RNe, with Schaefer (2005; 2010) and Schaefer et al. (2009) demonstrating typical variations of a factor of two and extreme variations of a factor of 30 over a century.  To illustrate how this might look, the middle curve in Figure 18 shows a typical case where $\dot{P}$ changes randomly between each eruption.

The measured error bars of the eclipse times varies greatly.  For CI Aql, it varies from $\pm$0.0001 day for a wonderfully observed full eclipse on the large McDonald 2.1-m telescope all the way to $\pm$0.0263 day for a single magnitude at the start of an ingress.  This creates a visual perception problem in the display of the O-C points on a plot.  The usual representation is to give all observed points the same size symbol and draw the one-sigma error bars.  But the human eye tends to make judgments by the amount of black ink on a plot, so the worst measured points (with large error bars) are what define the judged curve and uncertainty regions, while the best observed points are overlooked due to their vanishing error bars.  To avoid these inevitable perception problems, I have presented each point with the size of the symbol and the prominence of the error bar scaling with the uncertainty.  Thus, the many highly accurate measures are plotted as large symbols (with error bars that vanish under the symbol) so that people can readily follow where the best fit lines should go and can readily see the scatter in the real data.  The poor-accuracy measures are plotted as small symbols and thin error bars without the ending line segments, with this giving the correct visual impression that the these points have very low statistical weight.  Rather than make the symbol sizes adjusted for every measure, I have made four groups which share symbol size and error bar style.

~
~

{}

\begin{deluxetable}{lllllllll}
\tabletypesize{\scriptsize}
\tablewidth{0pc}
\tablecaption{Journal of Observations for CI Aql in Quiescence}
\tablehead{\colhead{UT Start Date} & \colhead{End Date} & \colhead{JD Start} & \colhead{JD End} & \colhead{Telescope} & \colhead{Filters} & \colhead{Eclipses} & \colhead{Magnitudes} & \colhead{Nights}}
\startdata
1903 Aug 15	&	1935 Aug 1	&	2416342	&	2428016	&	HCO plates	&	B	&	2	&	13	&	11	\\
1950 Jul 18	&	1997 Jul 1	&	2433480	&	2450630	&	Schmidt surveys	&	BRI	&	0	&	12	&	11	\\
1988 Aug 31	&	1988 Aug 31	&	2447404	&	2447404	&	KPNO 0.9m\tablenotemark{a}	&	BVR	&	0	&	3	&	1	\\
1988 Sep 6	&	1988 Sep 6	&	2447410	&	2447410	&	KPNO 1.3m\tablenotemark{a}	&	JK	&	0	&	2	&	1	\\
1991 Jun 4	&	1996 Sep 29	&	2448412	&	2450356	&	Roboscope\tablenotemark{b}	&	V	&	37	&	285	&	280	\\
2001 Apr 19	&	2001 Jun 3	&	2452018	&	2452032	&	McDonald 2.1m	&	V	&	3	&	118	&	6	\\
2001 Aug 4	&	2001 Aug 10	&	2452125	&	2452131	&	McDonald 2.1m	&	V	&	2	&	400	&	7	\\
2001 Nov 11	&	2001 Nov 25	&	2452224	&	2452238	&	SuperLOTIS	&	V	&	1	&	499	&	10	\\
2002 May 25	&	2002 May 25	&	2452419	&	2452419	&	McDonald 0.8m	&	R	&	1	&	34	&	1	\\
2002 May 31	&	2002 Jun 3	&	2452425	&	2452428	&	McDonald 2.1m	&	BV	&	2	&	243	&	4	\\
2002 Sep 4	&	2002 Sep 4	&	2452521	&	2452521	&	McDonald 0.9m	&	CV	&	1	&	52	&	1	\\
2003 May 6	&	2003 May 23	&	2452765	&	2452785	&	McDonald 0.8m	&	UBVRI	&	2	&	168	&	5	\\
2003 May 26	&	2003 May 26	&	2452785	&	2452785	&	McDonald 2.1m	&	V	&	1	&	164	&	1	\\
2003 Jul 2	&	2003 Jul 25	&	2452822	&	2452845	&	ROTSE 3b	&	CR	&	2	&	209	&	2	\\
2004 Feb 19	&	2004 Mar 18	&	2453082	&	2453083	&	McDonald 0.8m	&	BVRI	&	0	&	11	&	2	\\
2004 Apr 17	&	2004 Apr 17	&	2453112	&	2453112	&	McDonald 2.1m	&	V	&	1	&	473	&	1	\\
2004 Jul 4	&	2004 Jul 9	&	2453190	&	2453542	&	CTIO 1.0m	&	V	&	3	&	351	&	3	\\
2004 Jul 23	&	2004 Sep 13	&	2453210	&	2453262	&	CTIO 1.0m	&	V	&	5	&	253	&	5	\\
2005 Jun 19	&	2005 Jun 22	&	2453540	&	2453543	&	CTIO 0.9m	&	V	&	1	&	65	&	2	\\
2005 Aug 24	&	2005 Sep 10	&	2453607	&	2453624	&	CTIO 0.9m	&	BVRI	&	2	&	76	&	3	\\
2006 Mar 11	&	2006 Aug 27	&	2453806	&	2453975	&	CTIO 0.9m	&	BVRI	&	5	&	275	&	6	\\
2007 Jun 1	&	2007 Aug 15	&	2454253	&	2454328	&	CTIO 0.9m	&	V	&	4	&	246	&	4	\\
2008 Apr 25	&	2008 Jul 9	&	2454582	&	2454657	&	CTIO 1.3m	&	V	&	6	&	498	&	10	\\
2009 May 1	&	2009 Aug 31	&	2454953	&	2455075	&	CTIO 1.3m	&	V	&	10	&	510	&	10	\\
\enddata
\tablenotetext{a}{From Szkody (1994).}  \tablenotetext{b}{From Mennickent \& Honeycutt (1995).}
\label{Table1}
\end{deluxetable}

\begin{deluxetable}{lllllllll}
\tabletypesize{\scriptsize}
\tablewidth{0pc}
\tablecaption{Journal of Observations for U Sco in Quiescence}
\tablehead{\colhead{UT Start Date} & \colhead{End Date} & \colhead{JD Start} & \colhead{JD End} & \colhead{Telescope} & \colhead{Filters} & \colhead{Eclipses} & \colhead{Magnitudes} & \colhead{Nights}}
\startdata
1954 Jun 29	&	1954 Jun 29	&	2434923	&	2434923	&	Palomar 48"	&	BR	&	0	&	2	&	1	\\
1979 Aug 15	&	1996 May 10	&	2444099	&	2450212	&	UK Schmidt	&	BRI	&	0	&	7	&	7	\\
1982 Jun 18	&	1982 Jul 9	&	2445139	&	2445159	&	AAT 3.9m\tablenotemark{a}	&	VJHK	&	0	&	4	&	2	\\
1988 Jun 25	&	1988 Jul 5	&	2447338	&	2447348	&	CTIO 0.9m	&	B	&	0	&	57	&	4	\\
1989 Jan 29	&	1989 Jan 30	&	2447556	&	2447557	&	KPNO 0.9m	&	B	&	0	&	14	&	2	\\
1989 Apr 14	&	1989 Apr 14	&	2447630	&	2447630	&	KPNO 0.9m\tablenotemark{b}	&	BVR	&	0	&	3	&	1	\\
1989 Jul 8	&	1989 Jul 20	&	2447716	&	2447727	&	CTIO 0.9m	&	B (V)	&	5	&	157	&	13	\\
1990 Jun 1	&	1990 Jun 2	&	2448043	&	2448044	&	CTIO 4.0m	&	0.47-0.48 $\mu$	&	1	&	22	&	2	\\
1993 Jun 15	&	1993 Jun 17	&	2449154	&	2449156	&	KPNO 0.9m	&	BVR	&	1	&	14	&	3	\\
1994 Jul 16	&	1994 Jul 30	&	2449550	&	2449564	&	CTIO 0.9m	&	B	&	3	&	56	&	6	\\
1995 Jun 25	&	1995 Jul 4	&	2449894	&	2449901	&	CTIO 0.9m	&	BVI	&	3	&	74	&	3	\\
1996 Jul 24	&	1996 Jul 24	&	2450289	&	2450290	&	CTIO 0.9m	&	I	&	1	&	29	&	1	\\
1997 May 10	&	1997 May 15	&	2450579	&	2450584	&	CTIO 0.9m	&	B	&	3	&	31	&	3	\\
2001 Apr 17	&	2001 May 2	&	2452017	&	2452032	&	McDonald 2.1m	&	VR	&	1	&	42	&	5	\\
2001 Aug 4	&	2001 Aug 12	&	2452126	&	2452132	&	McDonald 2.1m	&	BVI	&	2	&	90	&	5	\\
2002 May 31	&	2002 Jun 6	&	2452426	&	2452432	&	McDonald 2.1m	&	BVI	&	3	&	87	&	4	\\
2003 May 6	&	2003 May 6	&	2452766	&	2452766	&	McDonald 0.8m	&	BI	&	0	&	2	&	1	\\
2003 May 8	&	2003 May 8	&	2452768	&	2452768	&	McDonald 2.7m	&	B	&	1	&	104	&	1	\\
2003 May 10	&	2003 May 29	&	2452770	&	2452789	&	McDonald 0.8m	&	BVI	&	1	&	45	&	3	\\
2003 Jul 5	&	2003 Jul 5	&	2452826	&	2452826	&	McDonald 2.7m	&	BG40	&	1	&	573	&	1	\\
2004 Mar 18	&	2004 Mar 18	&	2453082	&	2453082	&	McDonald 0.8m	&	UBVRI	&	0	&	5	&	1	\\
2004 Jul 1	&	2004 Jul 13	&	2453187	&	2453199	&	CTIO 1m	&	BVRI	&	2	&	247	&	12	\\
2004 Aug 7	&	2004 Aug 7	&	2453225	&	2453225	&	CTIO 1.0m	&	VI	&	0	&	29	&	1	\\
2005 Jun 9	&	2005 Jun 19	&	2453530	&	2453540	&	CTIO 0.9m	&	I	&	2	&	44	&	2	\\
2005 Sep 20	&	2005 Sep 20	&	2453634	&	2453634	&	CTIO 1.3m	&	BVRI	&	0	&	4	&	1	\\
2006 Mar 4	&	2006 Mar 4	&	2453799	&	2453799	&	CTIO 1.3m	&	BVRI	&	0	&	4	&	1	\\
2006 May 5	&	2006 Sep 9	&	2453861	&	2453988	&	CTIO 0.9m	&	I	&	4	&	209	&	4	\\
2007 May 19	&	2007 Sep 7	&	2454240	&	2454351	&	CTIO 0.9m	&	VI	&	2	&	80	&	2	\\
2008 Mar 1	&	2008 May 21	&	2454527	&	2454608	&	CTIO 1.3m	&	B	&	0	&	14	&	14	\\
2008 May 5	&	2008 Jun 28	&	2454592	&	2454645	&	CTIO 0.9m	&	I	&	4	&	135	&	4	\\
2009 Mar 28	&	2009 Jul 28	&	2454918	&	2455041	&	CTIO 0.9m	&	I	&	4	&	164	&	5	\\
2009 Jul 12	&	2009 Jul 12	&	2455024	&	2455024	&	McDonald 2.1m	&	VR	&	1	&	34	&	1	\\
\enddata
\tablenotetext{a}{From Hanes (1985).} \tablenotetext{b}{From Szkody (1994).}
\label{Table2}
\end{deluxetable}

\begin{deluxetable}{lllll}
\tabletypesize{\scriptsize}
\tablecaption{All Quiescent Magnitudes for CI Aql
\label{tbl3}}
\tablewidth{0pt}
\tablehead{
\colhead{HJD}   &
\colhead{Band}   &
\colhead{Magnitude}   &
\colhead{Sigma}  &
\colhead{Phase}
}
\startdata

2416342.6092	&	B	&	17.22	&	0.15	&	0.826	\\
2424681.8210	&	B	&	17.10	&	0.15	&	0.821	\\
2424711.7603	&	B	&	17.00	&	0.15	&	0.238	\\
2424712.7343	&	B	&	$>$17.5	&	0.15	&	0.813	\\
2424730.6628	&	B	&	17.05	&	0.15	&	0.807	\\
2424732.6617	&	B	&	17.05	&	0.15	&	0.040	\\
2424766.5305	&	B	&	16.98	&	0.15	&	0.812	\\
2424766.5625	&	B	&	17.20	&	0.15	&	0.863	\\
2424766.5935	&	B	&	17.20	&	0.15	&	0.913	\\
2427686.2807	&	B	&	17.20	&	0.15	&	0.569	\\
2427688.2995	&	B	&	17.32	&	0.15	&	0.834	\\
2428012.3059	&	B	&	17.75	&	0.15	&	0.810	\\
\enddata
\end{deluxetable}

\begin{deluxetable}{lllll}
\tabletypesize{\scriptsize}
\tablecaption{All Quiescent Magnitudes for U Sco
\label{tbl4}}
\tablewidth{0pt}
\tablehead{
\colhead{HJD}   &
\colhead{Band}   &
\colhead{Magnitude}   &
\colhead{Sigma}  &
\colhead{Phase}
}
\startdata

2434923.2171	&	R	&	18.00	&	0.18	&	0.657	\\
2434923.2171	&	B	&	18.80	&	0.15	&	0.657	\\
2444099.0000	&	B	&	18.01	&	0.08	&	0.328	\\
2445139.8000	&	V	&	17.85	&	0.10	&	0.130	\\
2445159.0000	&	J	&	16.88	&	0.10	&	0.733	\\
2445159.0000	&	H	&	16.46	&	0.10	&	0.733	\\
2445159.0000	&	K	&	16.45	&	0.10	&	0.733	\\
2445515.4708	&	B	&	18.21	&	0.10	&	0.418	\\
2446169.3183	&	R	&	18.72	&	0.05	&	0.765	\\
2447337.5887	&	B	&	18.59	&	0.04	&	0.156	\\
2447337.5982	&	B	&	18.69	&	0.04	&	0.164	\\
2447337.6056	&	B	&	18.73	&	0.04	&	0.170	\\
\enddata
\end{deluxetable}

\begin{deluxetable}{lllll}
\tabletypesize{\scriptsize}
\tablecaption{Quiescent Light Curve Templates
\label{tbl5}}
\tablewidth{0pt}
\tablehead{
\colhead{Phase}   &
\colhead{V (mag)}   &
\colhead{V (mag)}   &
\colhead{B (mag)}  &
\colhead{I (mag)}
}
\startdata

	&	CI AQL	&	CI AQL	&	U SCO	&	U SCO	\\
	&	1991-1996	&	2002-2009	&	1988-2008	&	1995-2009	\\
0.00	&	16.74	&	16.74	&	19.98	&	18.18	\\
0.01	&	16.73	&	16.72	&	19.98	&	18.18	\\
0.02	&	16.67	&	16.65	&	19.80	&	18.05	\\
0.03	&	16.58	&	16.57	&	19.61	&	17.92	\\
0.04	&	16.50	&	16.47	&	19.40	&	17.80	\\
0.05	&	16.43	&	16.39	&	19.22	&	17.69	\\
0.06	&	16.38	&	16.32	&	19.03	&	17.59	\\
0.07	&	16.34	&	16.27	&	18.82	&	17.48	\\
0.08	&	16.31	&	16.24	&	18.70	&	17.38	\\
0.09	&	16.28	&	16.22	&	18.65	&	17.29	\\
0.10	&	16.27	&	16.20	&	18.60	&	17.27	\\
0.11	&	16.26	&	16.19	&	18.57	&	17.25	\\
0.12	&	16.25	&	16.17	&	18.55	&	17.24	\\
0.13	&	16.24	&	16.16	&	18.55	&	17.24	\\
0.14	&	16.23	&	16.15	&	18.55	&	17.24	\\
0.15	&	16.22	&	16.13	&	18.55	&	17.24	\\
0.16	&	16.21	&	16.12	&	18.55	&	17.24	\\
0.17	&	16.20	&	16.11	&	18.55	&	17.24	\\
0.18	&	16.19	&	16.10	&	18.55	&	17.24	\\
0.20	&	16.18	&	16.09	&	18.55	&	17.24	\\
0.24	&	16.14	&	16.08	&	18.55	&	17.24	\\
0.28	&	16.11	&	16.06	&	18.55	&	17.24	\\
0.32	&	16.09	&	16.05	&	18.55	&	17.24	\\
0.36	&	16.10	&	16.06	&	18.55	&	17.24	\\
0.40	&	16.12	&	16.08	&	18.55	&	17.27	\\
0.44	&	16.16	&	16.13	&	18.55	&	17.34	\\
0.48	&	16.23	&	16.21	&	18.55	&	17.44	\\
0.50	&	16.24	&	16.26	&	18.55	&	17.50	\\
0.52	&	16.23	&	16.25	&	18.55	&	17.44	\\
0.56	&	16.16	&	16.20	&	18.55	&	17.34	\\
0.60	&	16.12	&	16.15	&	18.55	&	17.27	\\
0.64	&	16.10	&	16.13	&	18.55	&	17.24	\\
0.68	&	16.09	&	16.13	&	18.55	&	17.24	\\
0.72	&	16.11	&	16.15	&	18.55	&	17.24	\\
0.76	&	16.14	&	16.17	&	18.55	&	17.24	\\
0.80	&	16.18	&	16.20	&	18.55	&	17.24	\\
0.82	&	16.19	&	16.21	&	18.55	&	17.24	\\
0.83	&	16.20	&	16.22	&	18.55	&	17.24	\\
0.84	&	16.21	&	16.21	&	18.55	&	17.24	\\
0.85	&	16.22	&	16.22	&	18.55	&	17.24	\\
0.86	&	16.23	&	16.22	&	18.55	&	17.24	\\
0.87	&	16.24	&	16.22	&	18.55	&	17.24	\\
0.88	&	16.25	&	16.23	&	18.55	&	17.24	\\
0.89	&	16.26	&	16.24	&	18.57	&	17.25	\\
0.90	&	16.27	&	16.25	&	18.60	&	17.27	\\
0.91	&	16.28	&	16.26	&	18.65	&	17.29	\\
0.92	&	16.31	&	16.28	&	18.70	&	17.38	\\
0.93	&	16.34	&	16.30	&	18.82	&	17.48	\\
0.94	&	16.38	&	16.32	&	19.03	&	17.59	\\
0.95	&	16.43	&	16.40	&	19.22	&	17.69	\\
0.96	&	16.50	&	16.49	&	19.40	&	17.80	\\
0.97	&	16.58	&	16.58	&	19.61	&	17.92	\\
0.98	&	16.67	&	16.65	&	19.80	&	18.05	\\
0.99	&	16.73	&	16.72	&	19.98	&	18.18	\\
1.00	&	16.74	&	16.74	&	19.98	&	18.18	\\
\enddata
\end{deluxetable}

\begin{deluxetable}{llllllll}
\tabletypesize{\scriptsize}
\tablewidth{0pc}
\tablecaption{Offsets for Ingress and Egress Observations}
\tablehead{\colhead{RN} & \colhead{Filter} & \colhead{Branch} & \colhead{$m_{fid}$} & \colhead{$\Delta T$ (days)} & \colhead{$A$ (days)} & \colhead{$B$ (mag/day)} & \colhead{$\sigma$ (days)}}
\startdata
CI Aql	&	V	&	Ingress	&	$m_{min}-0.4$	&	\ldots	&	0.0503	&	-0.00106	&	0.0020	\\
CI Aql	&	V	&	Ingress	&	$m_{min}-0.3$	&	\ldots	&	0.0376	&	-0.00065	&	0.0038	\\
CI Aql	&	V	&	Ingress	&	$m_{min}-0.2$	&	\ldots	&	0.0242	&	-0.00017	&	0.0020	\\
CI Aql	&	V	&	Ingress	&	$m_{min}-0.1$	&	0.0138	&	\ldots	&	\ldots	&	0.0036	\\
CI Aql	&	V	&	Egress	&	$m_{min}-0.1$	&	-0.0139	&	\ldots	&	\ldots	&	0.0020	\\
CI Aql	&	V	&	Egress	&	$m_{min}-0.2$	&	\ldots	&	-0.0279	&	-0.00050	&	0.0019	\\
CI Aql	&	V	&	Egress	&	$m_{min}-0.3$	&	\ldots	&	-0.0334	&	-0.00053	&	0.0030	\\
CI Aql	&	V	&	Egress	&	$m_{min}-0.4$	&	\ldots	&	-0.0510	&	-0.00135	&	0.0024	\\
U Sco	&	B	&	Ingress	&	$m_{min}-0.4$	&	0.0329	&	\ldots	&	\ldots	&	0.0032	\\
U Sco	&	B	&	Ingress	&	$m_{min}-0.3$	&	0.0284	&	\ldots	&	\ldots	&	0.0032	\\
U Sco	&	B	&	Ingress	&	$m_{min}-0.2$	&	0.0243	&	\ldots	&	\ldots	&	0.0031	\\
U Sco	&	B	&	Ingress	&	$m_{min}-0.1$	&	0.0181	&	\ldots	&	\ldots	&	0.0033	\\
U Sco	&	B	&	Egress	&	$m_{min}-0.1$	&	-0.0118	&	\ldots	&	\ldots	&	0.0038	\\
U Sco	&	B	&	Egress	&	$m_{min}-0.2$	&	-0.0217	&	\ldots	&	\ldots	&	0.0051	\\
U Sco	&	B	&	Egress	&	$m_{min}-0.3$	&	-0.0297	&	\ldots	&	\ldots	&	0.0036	\\
U Sco	&	B	&	Egress	&	$m_{min}-0.4$	&	-0.0369	&	\ldots	&	\ldots	&	0.0047	\\
U Sco	&	V	&	Ingress	&	$m_{min}-0.3$	&	0.0282	&	\ldots	&	\ldots	&	\ldots	\\
U Sco	&	V	&	Ingress	&	$m_{min}-0.2$	&	0.0228	&	\ldots	&	\ldots	&	\ldots	\\
U Sco	&	V	&	Ingress	&	$m_{min}-0.1$	&	0.0169	&	\ldots	&	\ldots	&	\ldots	\\
U Sco	&	V	&	Egress	&	$m_{min}-0.1$	&	-0.0142	&	\ldots	&	\ldots	&	\ldots	\\
U Sco	&	V	&	Egress	&	$m_{min}-0.2$	&	-0.0220	&	\ldots	&	\ldots	&	\ldots	\\
U Sco	&	V	&	Egress	&	$m_{min}-0.3$	&	-0.0302	&	\ldots	&	\ldots	&	\ldots	\\
U Sco	&	V	&	Egress	&	$m_{min}-0.4$	&	-0.0365	&	\ldots	&	\ldots	&	\ldots	\\
U Sco	&	I	&	Ingress	&	$m_{min}-0.4$	&	\ldots	&	0.0664	&	-0.00131	&	0.0063	\\
U Sco	&	I	&	Ingress	&	$m_{min}-0.3$	&	\ldots	&	0.0639	&	-0.00200	&	0.0066	\\
U Sco	&	I	&	Ingress	&	$m_{min}-0.2$	&	\ldots	&	0.0490	&	-0.00153	&	0.0022	\\
U Sco	&	I	&	Ingress	&	$m_{min}-0.1$	&	0.0220	&	\ldots	&	\ldots	&	0.0032	\\
U Sco	&	I	&	Egress	&	$m_{min}-0.1$	&	-0.0209	&	\ldots	&	\ldots	&	0.0037	\\
U Sco	&	I	&	Egress	&	$m_{min}-0.2$	&	\ldots	&	-0.0394	&	-0.00074	&	0.0036	\\
U Sco	&	I	&	Egress	&	$m_{min}-0.3$	&	\ldots	&	-0.0501	&	-0.00072	&	0.0042	\\
U Sco	&	I	&	Egress	&	$m_{min}-0.4$	&	\ldots	&	-0.0668	&	-0.00132	&	0.0046	\\
\enddata
\end{deluxetable}

\begin{deluxetable}{lllll}
\tabletypesize{\scriptsize}
\tablecaption{CI Aql Eclipse Minimum Times
\label{tbl7}}
\tablewidth{0pt}
\tablehead{
\colhead{UT Date}   &
\colhead{Telescope}   &
\colhead{$T_{obs}$ (HJD)}   &
\colhead{N}  &
\colhead{O-C (days)\tablenotemark{a}}
}
\startdata

1926 Jul 16	&	MF10536	&	2424712.7343	$\pm$	0.0300	&	-43593	&	-0.1335	\\
1935 Jul 28	&	A17852	&	2428012.3059	$\pm$	0.0300	&	-38257	&	-0.1335	\\
1991 Jun 13	&	RoboScope	&	2448420.8002	$\pm$	0.0080	&	-5253	&	-0.0095	\\
1991 Aug 1	&	RoboScope	&	2448469.6472	$\pm$	0.0065	&	-5174	&	-0.0130	\\
1991 Aug 27	&	RoboScope	&	2448495.6270	$\pm$	0.0080	&	-5132	&	-0.0044	\\
1992 Jul 29	&	RoboScope	&	2448832.6369	$\pm$	0.0046	&	-4587	&	-0.0010	\\
1992 Sep 24	&	RoboScope	&	2448889.5182	$\pm$	0.0090	&	-4495	&	-0.0088	\\
1992 Oct 15	&	RoboScope	&	2448910.5564	$\pm$	0.0105	&	-4461	&	0.0052	\\
1993 Jun 26	&	RoboScope	&	2449164.6980	$\pm$	0.0049	&	-4050	&	0.0006	\\
1993 Jul 27	&	RoboScope	&	2449195.6173	$\pm$	0.0071	&	-4000	&	0.0018	\\
1993 Aug 7	&	RoboScope	&	2449206.7433	$\pm$	0.0083	&	-3982	&	-0.0026	\\
1993 Aug 9	&	RoboScope	&	2449208.5959	$\pm$	0.0062	&	-3979	&	-0.0051	\\
1993 Oct 8	&	RoboScope	&	2449268.5828	$\pm$	0.0043	&	-3882	&	0.0008	\\
1994 May 14	&	RoboScope	&	2449486.8647	$\pm$	0.0080	&	-3529	&	0.0014	\\
1994 May 19	&	RoboScope	&	2449491.7943	$\pm$	0.0040	&	-3521	&	-0.0159	\\
1994 May 22	&	RoboScope	&	2449494.9276	$\pm$	0.0263	&	-3516	&	0.0256	\\
1994 Jul 2	&	RoboScope	&	2449535.7081	$\pm$	0.0043	&	-3450	&	-0.0057	\\
1994 Aug 10	&	RoboScope	&	2449574.6676	$\pm$	0.0062	&	-3387	&	-0.0028	\\
1994 Aug 28	&	RoboScope	&	2449592.5903	$\pm$	0.0099	&	-3358	&	-0.0126	\\
1994 Aug 28	&	RoboScope	&	2449592.5932	$\pm$	0.0049	&	-3358	&	-0.0097	\\
1994 Oct 6	&	RoboScope	&	2449631.5632	$\pm$	0.0111	&	-3295	&	0.0036	\\
1995 Mar 3	&	RoboScope	&	2449779.9606	$\pm$	0.0065	&	-3055	&	-0.0055	\\
1995 Mar 11	&	RoboScope	&	2449788.0055	$\pm$	0.0117	&	-3042	&	0.0006	\\
1995 Apr 3	&	RoboScope	&	2449810.8865	$\pm$	0.0077	&	-3005	&	0.0024	\\
1995 Jun 7	&	RoboScope	&	2449875.8131	$\pm$	0.0053	&	-2900	&	0.0011	\\
1995 Jul 26	&	RoboScope	&	2449924.6552	$\pm$	0.0065	&	-2821	&	-0.0073	\\
1995 Aug 3	&	RoboScope	&	2449932.7030	$\pm$	0.0105	&	-2808	&	0.0018	\\
1995 Sep 3	&	RoboScope	&	2449963.6099	$\pm$	0.0065	&	-2758	&	-0.0093	\\
1995 Sep 29	&	RoboScope	&	2449989.5856	$\pm$	0.0071	&	-2716	&	-0.0047	\\
1996 Apr 21	&	RoboScope	&	2450194.8848	$\pm$	0.0053	&	-2384	&	-0.0013	\\
1996 Aug 8	&	RoboScope	&	2450303.7166	$\pm$	0.0068	&	-2208	&	-0.0009	\\
2000 Sep 5	&	AAVSO	&	2451792.7070	$\pm$	0.0105	&	200	&	-0.0226	\\
2001 Jun 21	&	Lederle \& Kimeswenger	&	2452081.5022	$\pm$	0.0046	&	667	&	-0.0018	\\
2001 Aug 6	&	McDonald 2.1-m	&	2452127.8787	$\pm$	0.0003	&	742	&	-0.0023	\\
2001 Aug 8	&	McDonald 2.1-m	&	2452129.7349	$\pm$	0.0003	&	745	&	-0.0012	\\
2001 Nov 17	&	SuperLOTIS	&	2452230.5241	$\pm$	0.0027	&	908	&	-0.0047	\\
2001 Nov 20	&	SuperLOTIS	&	2452233.6255	$\pm$	0.0027	&	913	&	0.0049	\\
2001 Nov 25	&	SuperLOTIS	&	2452238.5680	$\pm$	0.0027	&	921	&	0.0005	\\
2002 May 25	&	McDonald 0.8-m	&	2452419.7457	$\pm$	0.0009	&	1214	&	-0.0015	\\
2002 May 31	&	McDonald 2.1-m	&	2452425.9300	$\pm$	0.0004	&	1224	&	-0.0008	\\
2002 Jun 2	&	McDonald 2.1-m	&	2452427.7867	$\pm$	0.0001	&	1227	&	0.0009	\\
2002 Sep 4	&	McDonald 0.9-m	&	2452521.7752	$\pm$	0.0006	&	1379	&	-0.0014	\\
2003 May 11	&	McDonald 0.8-m	&	2452770.9798	$\pm$	0.0027	&	1782	&	0.0039	\\
2003 May 23	&	McDonald 0.8-m	&	2452782.7229	$\pm$	0.0027	&	1801	&	-0.0019	\\
2003 May 26	&	McDonald 2.1-m	&	2452785.8151	$\pm$	0.0027	&	1806	&	-0.0015	\\
2004 Apr 17	&	McDonald 2.1-m	&	2453112.9299	$\pm$	0.0001	&	2335	&	0.0006	\\
2004 Jul 4	&	CTIO 1.0-m	&	2453190.8433	$\pm$	0.0007	&	2461	&	0.0006	\\
2004 Jul 6	&	CTIO 1.0-m	&	2453192.6976	$\pm$	0.0022	&	2464	&	-0.0002	\\
2004 Jul 9	&	CTIO 1.0-m	&	2453195.7871	$\pm$	0.0004	&	2469	&	-0.0025	\\
2004 Aug 14	&	CTIO 1.0-m	&	2453231.6566	$\pm$	0.0005	&	2527	&	0.0021	\\
2004 Sep 1	&	CTIO 1.0-m	&	2453249.5878	$\pm$	0.0004	&	2556	&	0.0008	\\
2004 Sep 14	&	CTIO 1.0-m	&	2453262.5735	$\pm$	0.0004	&	2577	&	0.0010	\\
2005 Jun 19	&	CTIO 0.9-m	&	2453540.8383	$\pm$	0.0022	&	3027	&	0.0035	\\
2005 Aug 25	&	CTIO 0.9-m	&	2453607.6188	$\pm$	0.0003	&	3135	&	0.0011	\\
2005 Sep 7	&	CTIO 1.3-m	&	2453620.6007	$\pm$	0.0027	&	3156	&	-0.0026	\\
2006 Jun 30	&	CTIO 0.9-m	&	2453916.7972	$\pm$	0.0003	&	3635	&	-0.0008	\\
2006 Jul 28	&	CTIO 0.9-m	&	2453944.6247	$\pm$	0.0002	&	3680	&	0.0005	\\
2006 Jul 31	&	CTIO 0.9-m	&	2453947.7174	$\pm$	0.0004	&	3685	&	0.0014	\\
2006 Aug 14	&	CTIO 0.9-m	&	2453962.5563	$\pm$	0.0007	&	3709	&	-0.0003	\\
2006 Aug 28	&	CTIO 0.9-m	&	2453975.5401	$\pm$	0.0002	&	3730	&	-0.0021	\\
2007 Jun 2	&	CTIO 0.9-m	&	2454253.8017	$\pm$	0.0005	&	4180	&	-0.0027	\\
2007 Jun 7	&	CTIO 0.9-m	&	2454258.7536	$\pm$	0.0002	&	4188	&	0.0023	\\
2007 Aug 13	&	CTIO 0.9-m	&	2454325.5345	$\pm$	0.0004	&	4296	&	0.0002	\\
2007 Aug 15	&	CTIO 0.9-m	&	2454328.6274	$\pm$	0.0007	&	4301	&	0.0013	\\
2008 Apr 26	&	CTIO 1.3-m	&	2454582.7739	$\pm$	0.0004	&	4712	&	0.0017	\\
2008 May 17	&	CTIO 1.3-m	&	2454603.7965	$\pm$	0.0003	&	4746	&	0.0000	\\
2008 May 22	&	CTIO 1.3-m	&	2454608.7449	$\pm$	0.0006	&	4754	&	0.0015	\\
2008 June 8	&	CTIO 1.3-m	&	2454626.6749	$\pm$	0.0027	&	4783	&	-0.0009	\\
2008 Jul 5	&	CTIO 1.3-m	&	2454652.6427	$\pm$	0.0022	&	4825	&	-0.0043	\\
2008 Jul 9	&	CTIO 1.3-m	&	2454657.5937	$\pm$	0.0003	&	4833	&	-0.0001	\\
2009 May 2	&	CTIO 1.3-m	&	2454953.7887	$\pm$	0.0003	&	5312	&	0.0002	\\
2009 May 10	&	CTIO 1.3-m	&	2454961.8280	$\pm$	0.0005	&	5325	&	0.0008	\\
2009 May 15	&	CTIO 1.3-m	&	2454966.7755	$\pm$	0.0004	&	5333	&	0.0014	\\
2009 May 18	&	CTIO 1.3-m	&	2454969.8651	$\pm$	0.0006	&	5338	&	-0.0008	\\
2009 May 23	&	CTIO 1.3-m	&	2454974.8120	$\pm$	0.0003	&	5346	&	-0.0008	\\
2009 Aug 6	&	CTIO 1.3-m	&	2455049.6344	$\pm$	0.0005	&	5467	&	0.0000	\\
2009 Aug 11	&	CTIO 1.3-m	&	2455054.5808	$\pm$	0.0005	&	5475	&	-0.0005	\\
2009 Aug 13	&	CTIO 1.3-m	&	2455057.6715	$\pm$	0.0008	&	5480	&	-0.0016	\\
2009 Aug 24	&	CTIO 1.3-m	&	2455067.5689	$\pm$	0.0005	&	5496	&	0.0020	\\
2009, Sep 1	&	CTIO 1.3-m	&	2455075.6033	$\pm$	0.0005	&	5509	&	-0.0022	\\

\enddata
\tablenotetext{a}{O-C is calculated as $T_{obs}-T_{model}$, where here I use the linear model $T_{model}=2451669.0575 + N \times 0.61836051$.}
\end{deluxetable}

\begin{deluxetable}{lllll}
\tabletypesize{\scriptsize}
\tablecaption{U Sco Eclipse Minimum Times
\label{tbl8}}
\tablewidth{0pt}
\tablehead{
\colhead{UT Date}   &
\colhead{Telescope}   &
\colhead{$T_{obs}$ (HJD)}   &
\colhead{N}  &
\colhead{O-C (days)\tablenotemark{a}}
}
\startdata

1945 Jul 2	&	Harvard plates	&	2431639.3000	$\pm$	0.0090	&	-15924	&	-0.0091	\\
1989 Jul 10	&	CTIO 0.9-m	&	2447717.6064	$\pm$	0.0062	&	-2858	&	-0.0291	\\
1989 Jul 11	&	CTIO 0.9-m	&	2447718.8481	$\pm$	0.0084	&	-2857	&	-0.0180	\\
1989 Jul 15	&	CTIO 0.9-m	&	2447722.5406	$\pm$	0.0018	&	-2854	&	-0.0171	\\
1989 Jul 16	&	CTIO 0.9-m	&	2447723.7675	$\pm$	0.0030	&	-2853	&	-0.0208	\\
1989 Jul 19	&	CTIO 0.9-m	&	2447727.4707	$\pm$	0.0052	&	-2850	&	-0.0092	\\
1990 Jun 1	&	CTIO 4.0-m	&	2448043.7262	$\pm$	0.0042	&	-2593	&	-0.0043	\\
1993 Jun 16	&	KPNO 0.9-m	&	2449154.9116	$\pm$	0.0026	&	-1690	&	-0.0028	\\
1994 Jul 25	&	CTIO 0.9-m	&	2449558.5258	$\pm$	0.0020	&	-1362	&	-0.0080	\\
1994 Jul 26	&	CTIO 0.9-m	&	2449559.7605	$\pm$	0.0046	&	-1361	&	-0.0038	\\
1994 Jul 29	&	CTIO 0.9-m	&	2449563.4515	$\pm$	0.0069	&	-1358	&	-0.0045	\\
1995 Jun 25	&	CTIO 0.9-m	&	2449894.4733	$\pm$	0.0023	&	-1089	&	0.0002	\\
1995 Jun 27	&	CTIO 0.9-m	&	2449895.6939	$\pm$	0.0020	&	-1088	&	-0.0097	\\
1995 Jul 2	&	CTIO 0.9-m	&	2449900.6196	$\pm$	0.0011	&	-1084	&	-0.0062	\\
1996 Jul 24	&	CTIO 0.9-m	&	2450289.4682	$\pm$	0.0022	&	-768	&	-0.0104	\\
1997 May 10	&	CTIO 0.9-m	&	2450578.6517	$\pm$	0.0012	&	-533	&	-0.0055	\\
1997 May 11	&	CTIO 0.9-m	&	2450579.8951	$\pm$	0.0034	&	-532	&	0.0074	\\
1997 May 15	&	CTIO 0.9-m	&	2450583.5677	$\pm$	0.0035	&	-529	&	-0.0117	\\
1999 Mar 16	&	Kyoto 0.25-m\tablenotemark{b}	&	2451254.2110	$\pm$	0.0100	&	16	&	-0.0165	\\
1999 Mar 27	&	VSNET (Ouda)	&	2451265.3060	$\pm$	0.0100	&	25	&	0.0036	\\
1999 Apr 17	&	AAT 3.9-m\tablenotemark{c}	&	2451286.2143	$\pm$	0.0050	&	42	&	-0.0074	\\
2001 May 2	&	McDonald 2.1-m	&	2452031.9339	$\pm$	0.0014	&	648	&	0.0008	\\
2001 Aug 5	&	McDonald 2.1-m	&	2452126.6874	$\pm$	0.0010	&	725	&	0.0022	\\
2001 Aug 10	&	McDonald 2.1-m	&	2452131.5975	$\pm$	0.0090	&	729	&	-0.0099	\\
2002 May 31	&	McDonald 2.1-m	&	2452425.7093	$\pm$	0.0029	&	968	&	0.0012	\\
2002 Jun 5	&	McDonald 2.1-m	&	2452430.6296	$\pm$	0.0030	&	972	&	-0.0008	\\
2002 Jun 6	&	McDonald 2.1-m	&	2452431.8552	$\pm$	0.0041	&	973	&	-0.0057	\\
2003 May 8	&	McDonald 2.7-m	&	2452767.7993	$\pm$	0.0007	&	1246	&	-0.0009	\\
2003 May 29	&	McDonald 0.8-m	&	2452788.7162	$\pm$	0.0011	&	1263	&	-0.0033	\\
2003 Jul 5	&	McDonald 2.7-m	&	2452825.6420	$\pm$	0.0029	&	1293	&	0.0061	\\
2004 Jul 2	&	CTIO 1.0-m	&	2453188.6481	$\pm$	0.0019	&	1588	&	0.0008	\\
2004 Jul 7	&	CTIO 1.0-m	&	2453193.5675	$\pm$	0.0011	&	1592	&	-0.0019	\\
2005 Jun 9	&	CTIO 0.9-m	&	2453530.7350	$\pm$	0.0030	&	1866	&	-0.0043	\\
2005 Jun 19	&	CTIO 0.9-m	&	2453540.5830	$\pm$	0.0038	&	1874	&	-0.0007	\\
2006 May 6	&	CTIO 0.9-m	&	2453861.7560	$\pm$	0.0009	&	2135	&	-0.0004	\\
2006 Jun 7	&	CTIO 0.9-m	&	2453893.7513	$\pm$	0.0019	&	2161	&	0.0006	\\
2006 Jul 3	&	CTIO 0.9-m	&	2453919.5932	$\pm$	0.0018	&	2182	&	0.0011	\\
2006 Sep 10	&	CTIO 0.9-m	&	2453988.5038	$\pm$	0.0009	&	2238	&	0.0010	\\
2007 May 20	&	CTIO 0.9-m	&	2454240.7660	$\pm$	0.0008	&	2443	&	0.0011	\\
2007 Sep 8	&	CTIO 0.9-m	&	2454351.5159	$\pm$	0.0005	&	2533	&	0.0018	\\
2008 May 6	&	CTIO 0.9-m	&	2454592.7013	$\pm$	0.0008	&	2729	&	0.0000	\\
2008 May 17	&	CTIO 0.9-m	&	2454603.7787	$\pm$	0.0015	&	2738	&	0.0025	\\
2008 Jun 7	&	CTIO 0.9-m	&	2454624.6950	$\pm$	0.0011	&	2755	&	-0.0005	\\
2008 Jun 28	&	CTIO 0.9-m	&	2454645.6201	$\pm$	0.0005	&	2772	&	0.0053	\\
2009 Mar 28	&	CTIO 0.9-m	&	2454918.7959	$\pm$	0.0005	&	2994	&	-0.0004	\\
2009 Apr 18	&	CTIO 0.9-m	&	2454939.7113	$\pm$	0.0009	&	3011	&	-0.0043	\\
2009 May 4	&	CTIO 0.9-m	&	2454955.7163	$\pm$	0.0010	&	3024	&	0.0036	\\
2009 Jun 26	&	CTIO 0.9-m	&	2455008.6305	$\pm$	0.0048	&	3067	&	0.0043	\\
2009 Jul 12	&	McDonald 2.1-m	&	2455024.6217	$\pm$	0.0027	&	3080	&	-0.0016	\\
2009 Jul 28	&	CTIO 0.9-m	&	2455040.6145	$\pm$	0.0007	&	3093	&	-0.0059	\\
2010 Feb 12	&	AAVSO (Stein)\tablenotemark{e}	&	2455239.9600	$\pm$	0.0200	&	3255	&	-0.0090	\\
2010 Feb 17	&	AAVSO (Oksanen)\tablenotemark{d}\tablenotemark{e}	&	2455244.8778	$\pm$	0.0005	&	3259	&	-0.0134	\\
2010 Feb 19	&	AAVSO (Tan)\tablenotemark{e}	&	2455247.3505	$\pm$	0.0018	&	3261	&	-0.0018	\\
2010 Feb 22	&	AAVSO (Oksanen)\tablenotemark{e}	&	2455249.8047	$\pm$	0.0008	&	3263	&	-0.0087	\\
2010 Feb 24	&	AAVSO (Tan, Stockdale)\tablenotemark{e}	&	2455252.2681	$\pm$	0.0013	&	3265	&	-0.0064	\\
2010 Mar 5	&	AAVSO (Oksanen)\tablenotemark{e}	&	2455260.8838	$\pm$	0.0010	&	3272	&	-0.0045	\\
2010 Mar 10	&	AAVSO (Oksanen)\tablenotemark{e}	&	2455265.8097	$\pm$	0.0015	&	3276	&	-0.0008	\\
2010 Mar 12	&	AAVSO (Stockdale)\tablenotemark{e}	&	2455268.2625	$\pm$	0.0020	&	3278	&	-0.0091	\\
2010 Mar 15	&	AAVSO (Oksanen)\tablenotemark{e}	&	2455270.7446	$\pm$	0.0009	&	3280	&	0.0119	\\
2010 Mar 16	&	AAVSO (Krajci)\tablenotemark{e}	&	2455271.9637	$\pm$	0.0031	&	3281	&	0.0005	\\
2010 Mar 26	&	AAVSO (Oksanen)\tablenotemark{e}	&	2455281.8158	$\pm$	0.0012	&	3289	&	0.0082	\\
2010 Mar 31	&	AAVSO (Oksanen)\tablenotemark{e}	&	2455286.7411	$\pm$	0.0025	&	3293	&	0.0113	\\
2010 May 18	&	CTIO 0.9-m\tablenotemark{e}	&	2455334.7211	$\pm$	0.0009	&	3332	&	0.0000	\\
2010 Jun 29	&	CTIO 0.9-m\tablenotemark{e}	&	2455376.5650	$\pm$	0.0035	&	3366	&	0.0053	\\
2010 Jul 5	&	MDM 2.4-m (Lepine)\tablenotemark{e}	&	2455382.7126	$\pm$	0.0008	&	3371	&	0.0001	\\
2010 Jul 10	&	CTIO 0.9-m\tablenotemark{e}	&	2455387.6395	$\pm$	0.0010	&	3375	&	0.0048	\\
2010 Aug 16	&	AAVSO (Oksanen)\tablenotemark{e}	&	2455424.5565	$\pm$	0.0010	&	3405	&	0.0054	\\

\enddata
\tablenotetext{a}{O-C is calculated as $T_{obs}-T_{model}$, where here I use the linear model $T_{model}=2451234.5387 + N \times 1.23054695$.}
\tablenotetext{b}{Matsumoto et al. 2003}
\tablenotetext{c}{Thoroughgood et al. 2001}
\tablenotetext{d}{Observers covering the egress are Stein, Harris, Krajci, and Henden}
\tablenotetext{e}{Schaefer et al. 2011}
\end{deluxetable}

\clearpage
\begin{figure}
\epsscale{1.0}
\plotone{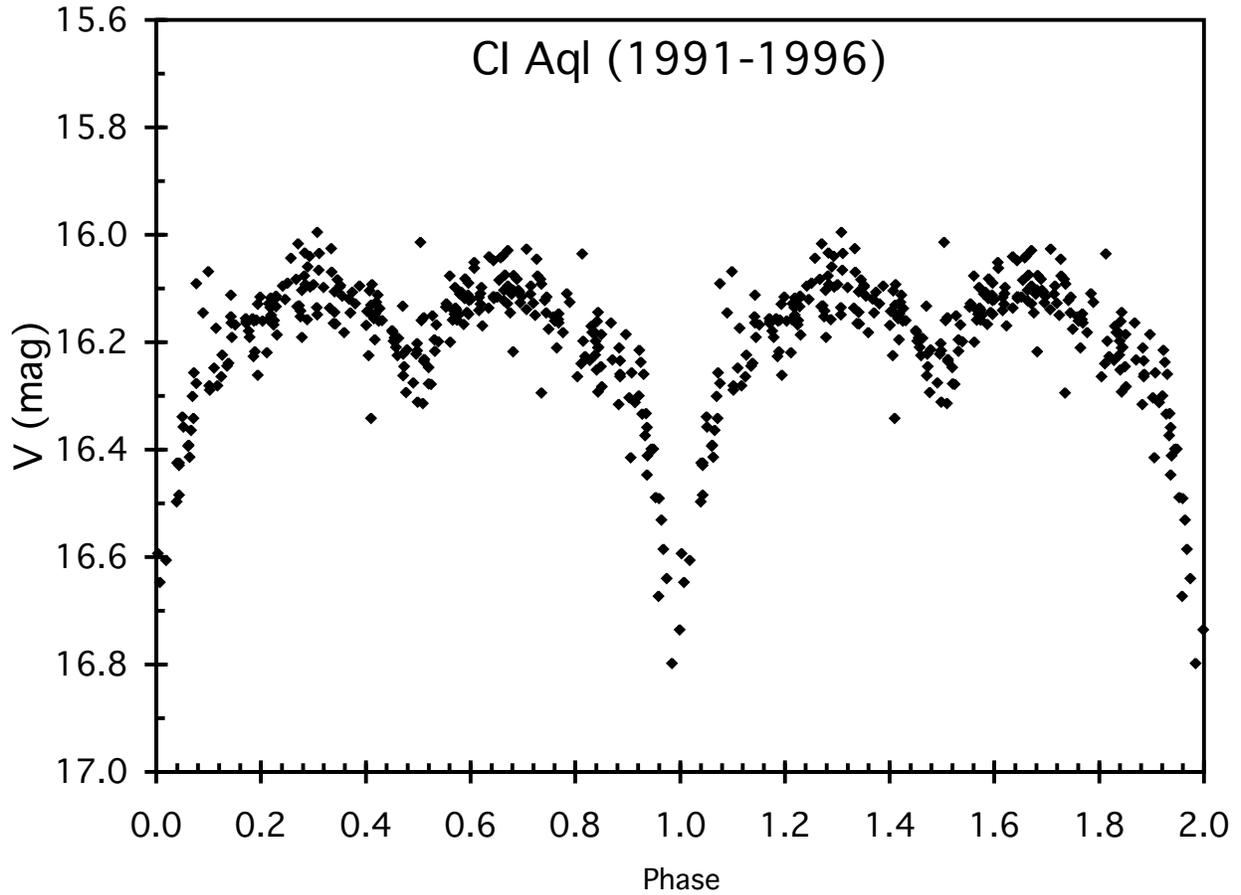}
\caption{
CI Aql 1991-1996.  This light curve folded on the orbital period is essentially taken from Mennickent \& Honeycutt (1995), except that magnitudes in 1996 have been added and a (somewhat uncertain) constant offset has been added in.  This light curve is substantially different from the post-eruption light curve (see Fig. 3).  First, the ellipsoidal modulation and secondary eclipse are very pronounced, with an amplitude of about 0.17 mag.  Second, there is no asymmetry between the elongations at phases 0.25 and 0.75.  Third, large amplitude flickering is not visible.}
\end{figure}

\clearpage
\begin{figure}
\epsscale{1.0}
\plotone{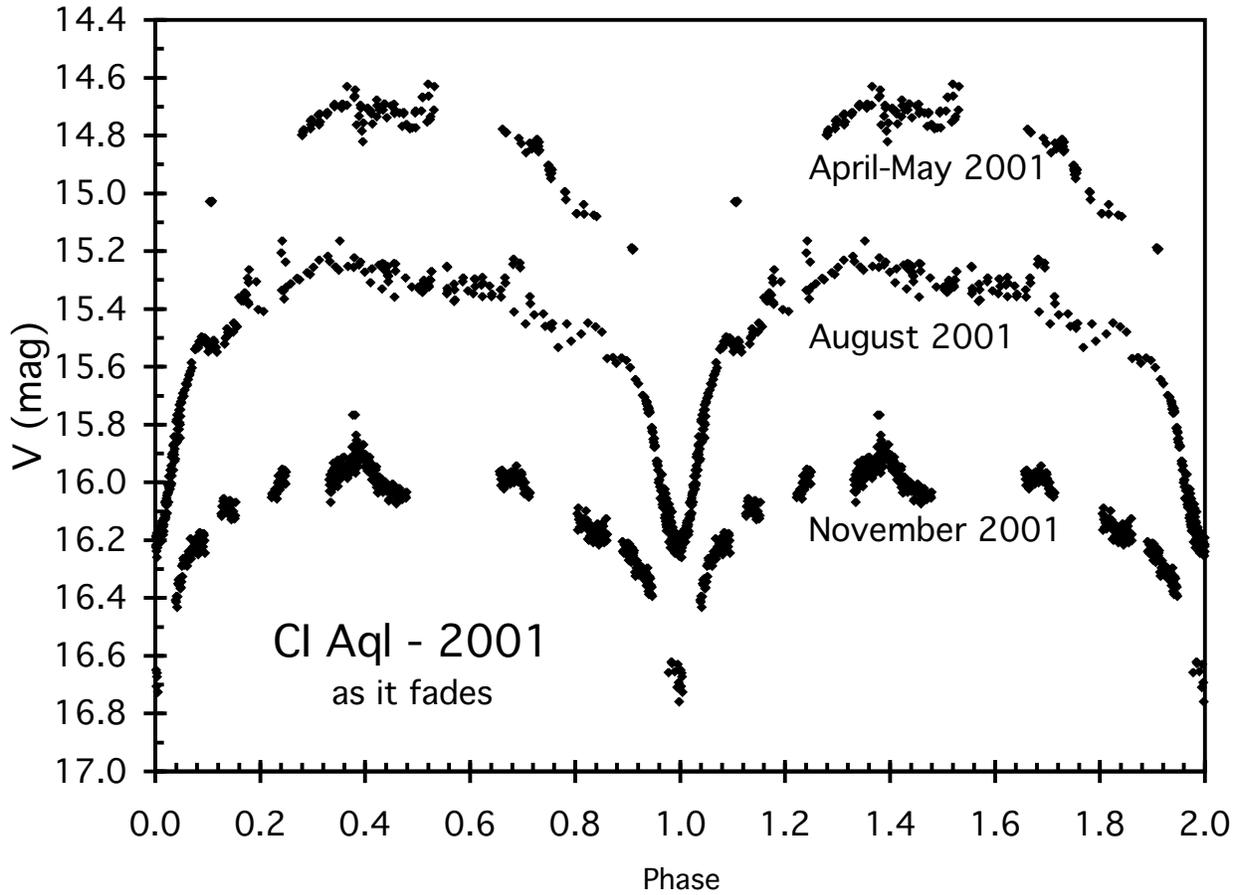}
\caption{
CI Aql fading in 2001.  These V-band light curves for three intervals in 2001 show CI Aql fading in the late tail of its eruption, which peaked in May 2000.  Each magnitude is plotted twice, with phases 0-1 being duplicated in phases 1-2, so as to allow the eclipse at phase 1.0 to be readily visible.  During the decline, no secondary eclipse is apparent, the system displays flickering, it is brighter at the phase 0.25 elongation than the phase 0.75 elongation, and the eclipse duration is the same as in quiescence.}
\end{figure}

\clearpage
\begin{figure}
\epsscale{1.0}
\plotone{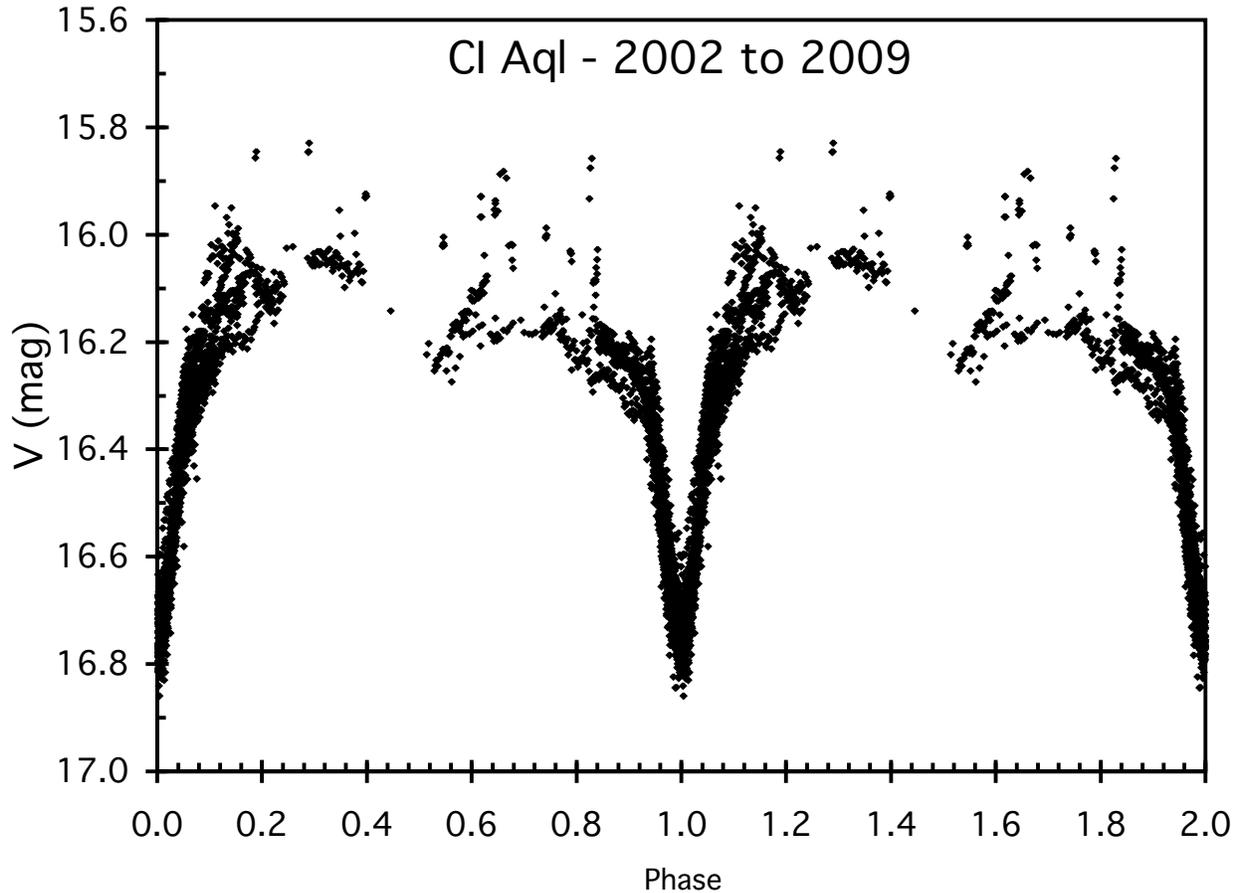}
\caption{
CI Aql 2002-2009.  This light curve is folded around the period 0.61836045 days, with each magnitude plotted twice, with the doubling to allow the eclipse at phase 1.0 to be not shown only in two halves.  The primary observational task was to get many time series through the eclipses, so the phase range 0.9-1.1 has 76\% of the 3221 V-band magnitudes in this plot.  The flickering is apparent only from phases 0.13-0.84, with no flickering during eclipses.  We see a shallow secondary minimum where the accretion disk covers part of the companion star.  The phase 0.25 elongation is 0.08 mag brighter than the phase 0.75 elongation, and this asymmetry might be due to the positioning and radiation pattern of the hot spot (where the accretion stream hits the accretion disk).  We see no secular changes from 2002 to 2009.}
\end{figure}

\clearpage
\begin{figure}
\epsscale{1.0}
\plotone{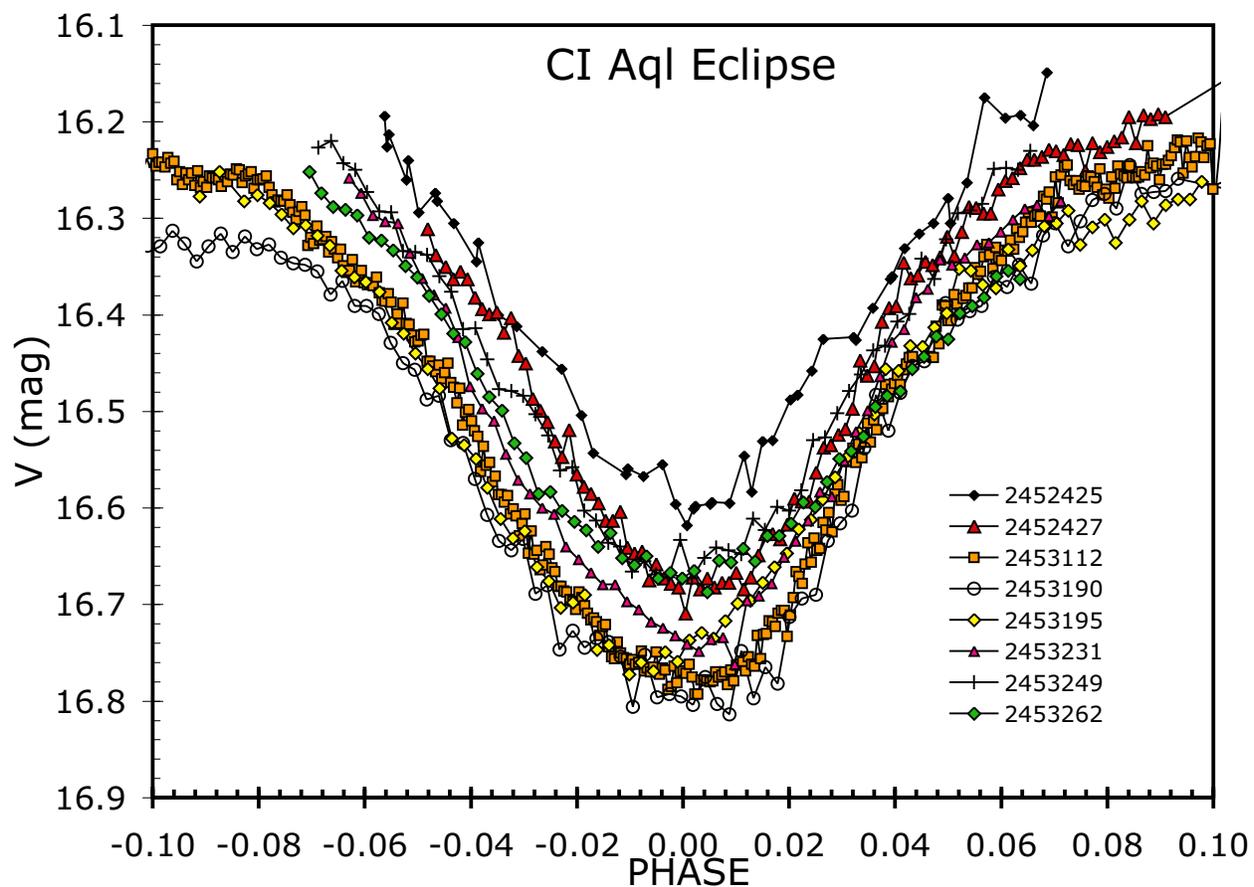}
\caption{
CI Aql eclipses.  These V-band light curves for eight well-observed eclipses show that the basic shape is constant with no apparent flickering.  However, the basic shape is variously shifted up or down, with minima from 16.6 to 16.8 mag.  The interpretation of these shifts is that the usual changes in the accretion rate get translated into changes in the brightness of the outer edge of the accretion disk, with the back part of the accretion disk barely peeping over the top of the companion star at mid-eclipse.}
\end{figure}

\clearpage
\begin{figure}
\epsscale{1.0}
\plotone{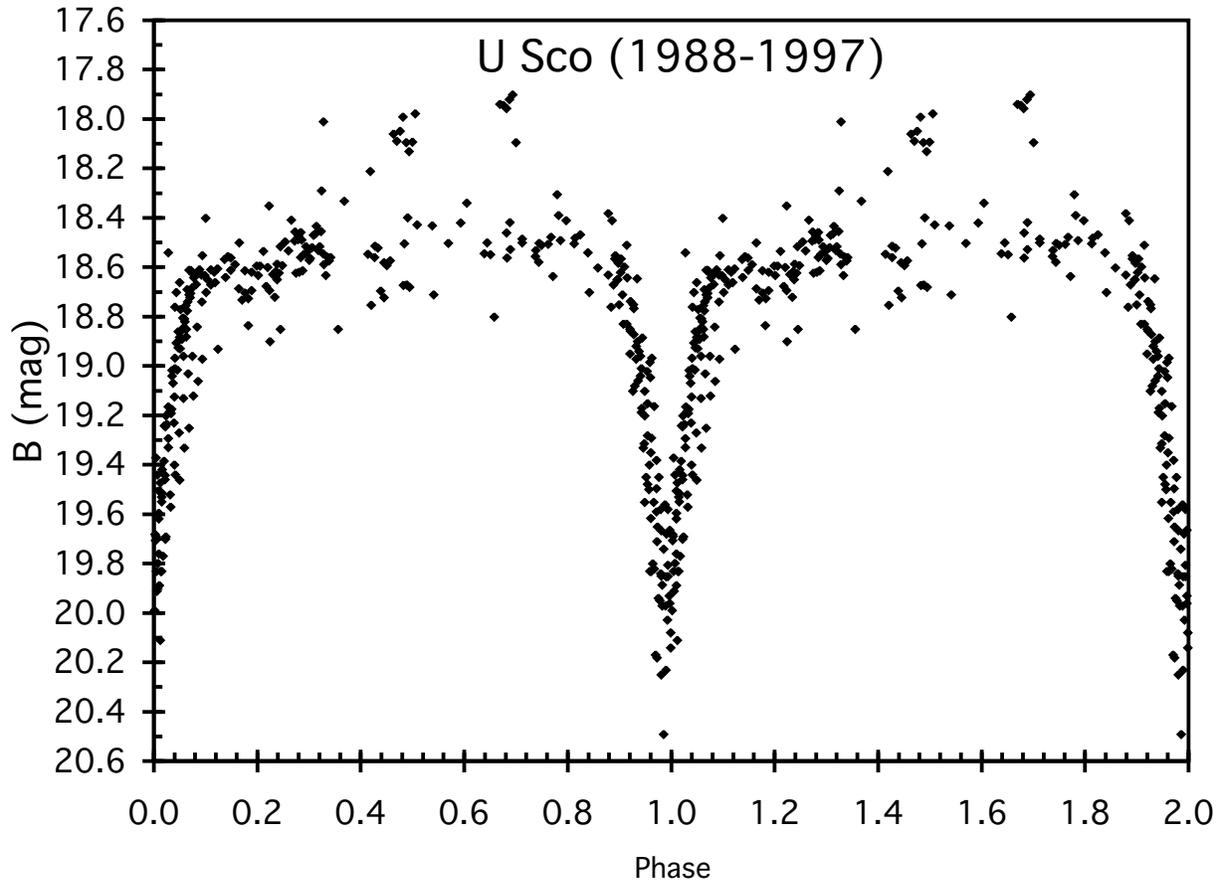}
\caption{
U Sco light curve 1988-1997.  This B-band light curve shows a deep primary eclipse (with no flickering superposed) and no secondary eclipse (with substantial flickering outside eclipses).}
\end{figure}

\clearpage
\begin{figure}
\epsscale{1.0}
\plotone{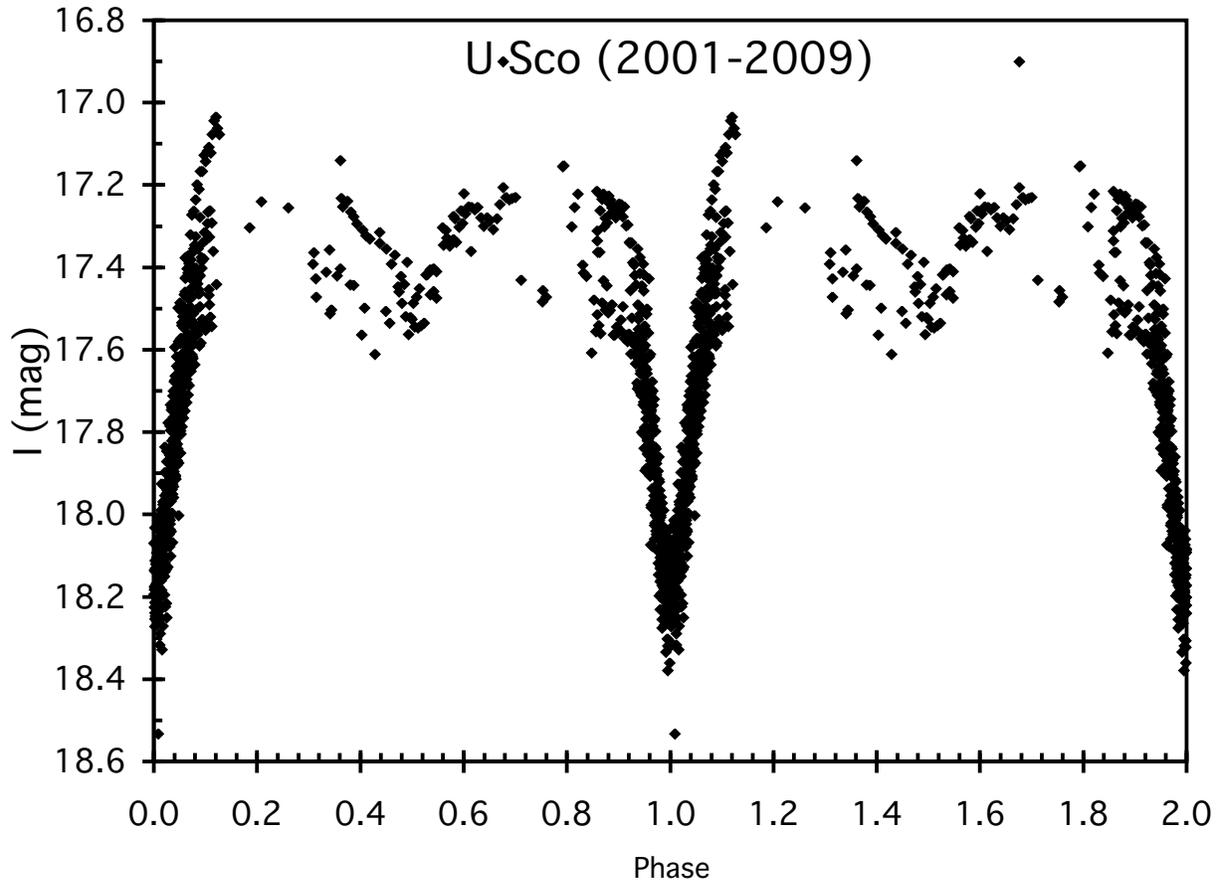}
\caption{
U Sco light curve 2001-2009.  This I-band light curve shows a deep primary eclipse (with no flickering superposed) and a secondary eclipse with depth around a quarter of a magnitude (visible under flickering with a comparable amplitude).}
\end{figure}

\clearpage
\begin{figure}
\epsscale{1.0}
\plotone{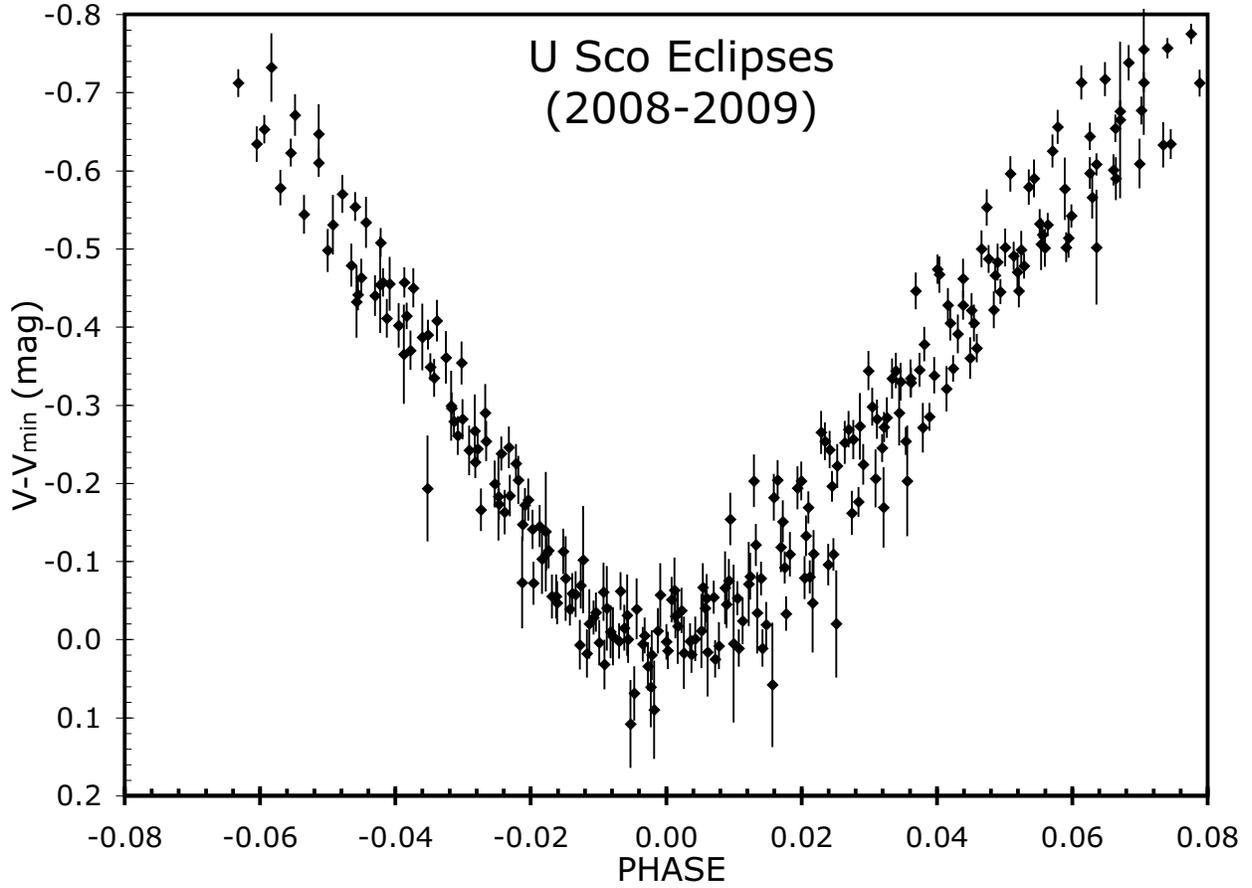}
\caption{
U Sco eclipses 2008-2009.  This I-band folded light curve is zoomed in on the eclipse for the eight eclipses in 2008 and 2009 (the most homogenous set of eclipses).  We see a flat-bottomed eclipse from phase -0.010 to +0.010, with this showing that the eclipse is total.}
\end{figure}

\clearpage
\begin{figure}
\epsscale{1.0}
\plotone{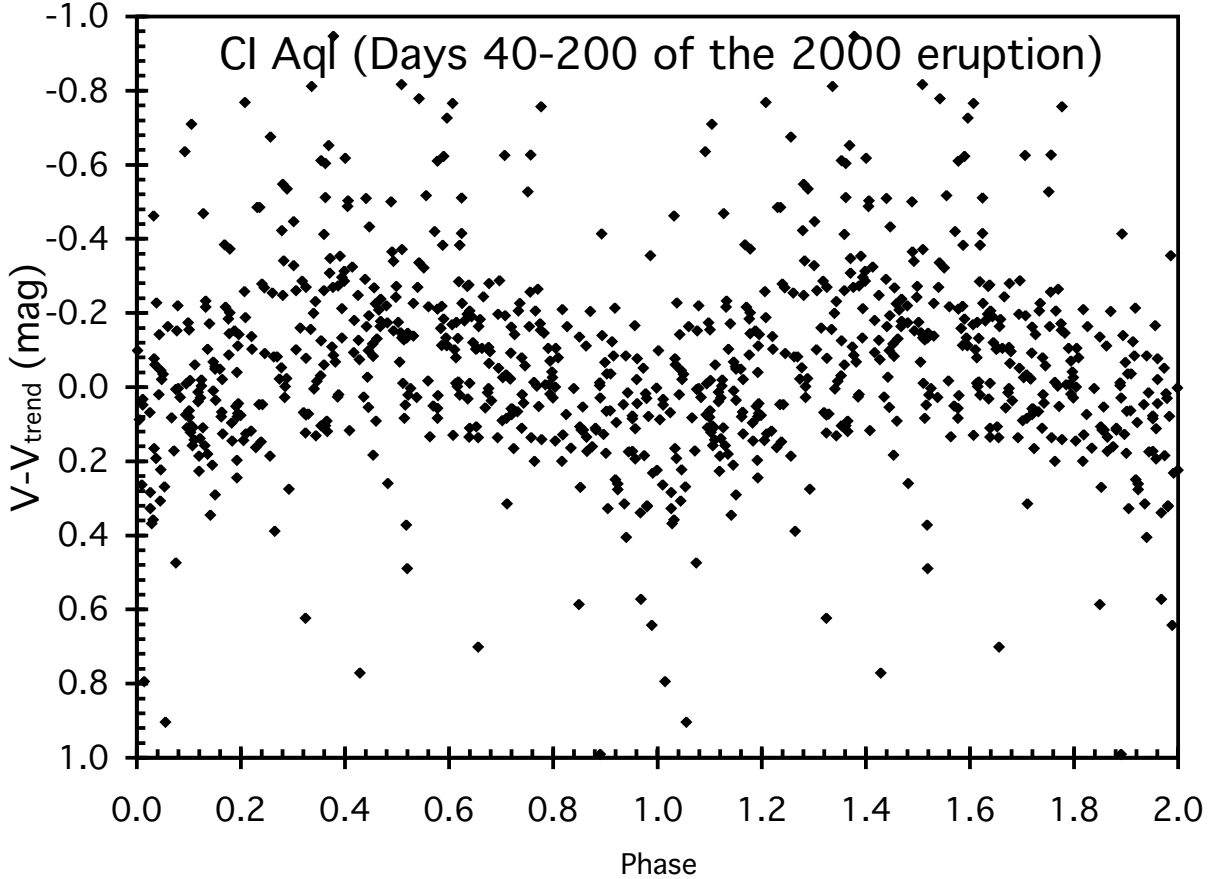}
\caption{
CI Aql during the early tail of the 2000 eruption.  These 460 V-band magnitudes are from days 40-200 after the peak of the 2000 eruption, corresponding to times just after the end of the fast initial drop in the light curve.  The magnitudes primarily come from the AAVSO data base, as made by visual observers with a typical one-sigma uncertainty of 0.15 mag.  The light curve has been detrended to subtract out the overall smoothed decline light curve ($V_{trend}$).  The large scatter is due to imperfections in the detrending, including fast intrinsic variations of CI Aql that are not removed by the slowly-changing trend line.  We see a distinct sinusoidal modulation with a fitted amplitude of 0.16$\pm$0.02 mag (half) amplitude.  I interpret this modulation on the orbital period as due to the varying visibility of the hot inner hemisphere of the companion star as it rotates around the white dwarf, while most of the light comes from a luminous, extended, and transparent envelope caused by the nova wind from the on-going nuclear burning on the surface of the white dwarf.  For the purposes of this paper, the important point is that the sine wave has a minimum that is offset from the zero phase (as based on the post-eruption ephemeris) by -0.037$\pm$0.017.  As usual for eclipse times during the tails of eclipses, this time has a significant offset to early times when compared to the predictions based on the quiescent times. This shift in phase is caused by the shift in the center of light between eruption and quiescence.  As such, eclipse times in quiescence cannot be combined with eclipse times in eruption so as to derive changes in the orbital period.}
\end{figure}

\clearpage
\begin{figure}
\epsscale{1.0}
\plotone{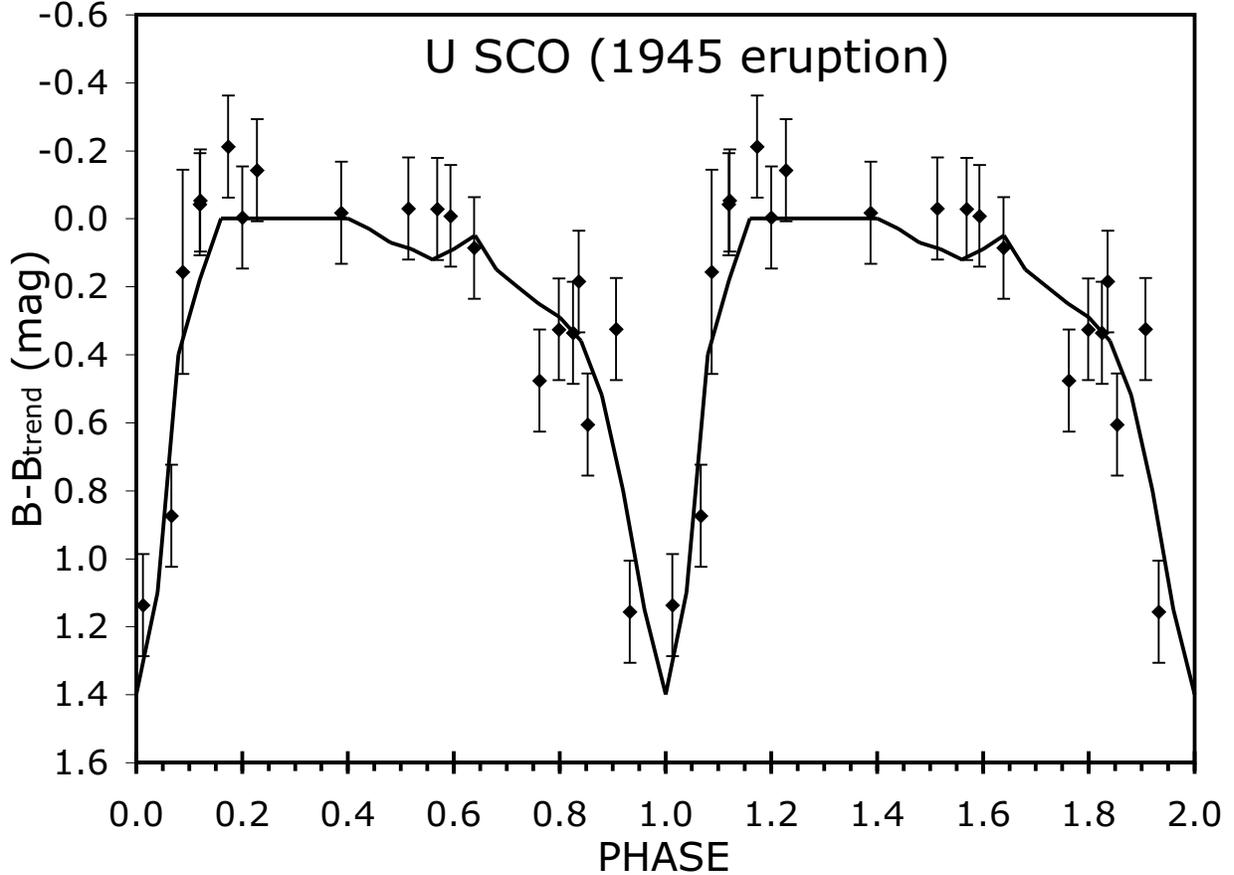}
\caption{
U Sco eclipses during 1945 eruption.  This B-band folded light curve is for days 33 to 47 after the peak of the 1945 eruption.  The declining light curve of the eruption has been subtracted out, and all points are plotted twice (once from phase 0-1 and a second time with 1.0 added to the phase).  We see an obvious eclipse around zero phase, with the expected shallow ingress. The minimum and egress all come from one night (2 July 1945).  The curve is the best fit light curve template as derived independently from the well-observed light curves of the same number of days after the peak of the 2010 eruption.  The uncertainty in the minimum time is $\pm$0.0090 days.  From Section 7.4, an extrapolation of the best fit ephemeris back to 1945 predicts the eclipse should happen around phase 0.84.  So an important point from this figure is that we have a confident eclipse from 1945 and that the eclipse time is around zero phase, with this phase certainly being far different from the extrapolated phase.}
\end{figure}

\clearpage
\begin{figure}
\epsscale{1.0}
\plotone{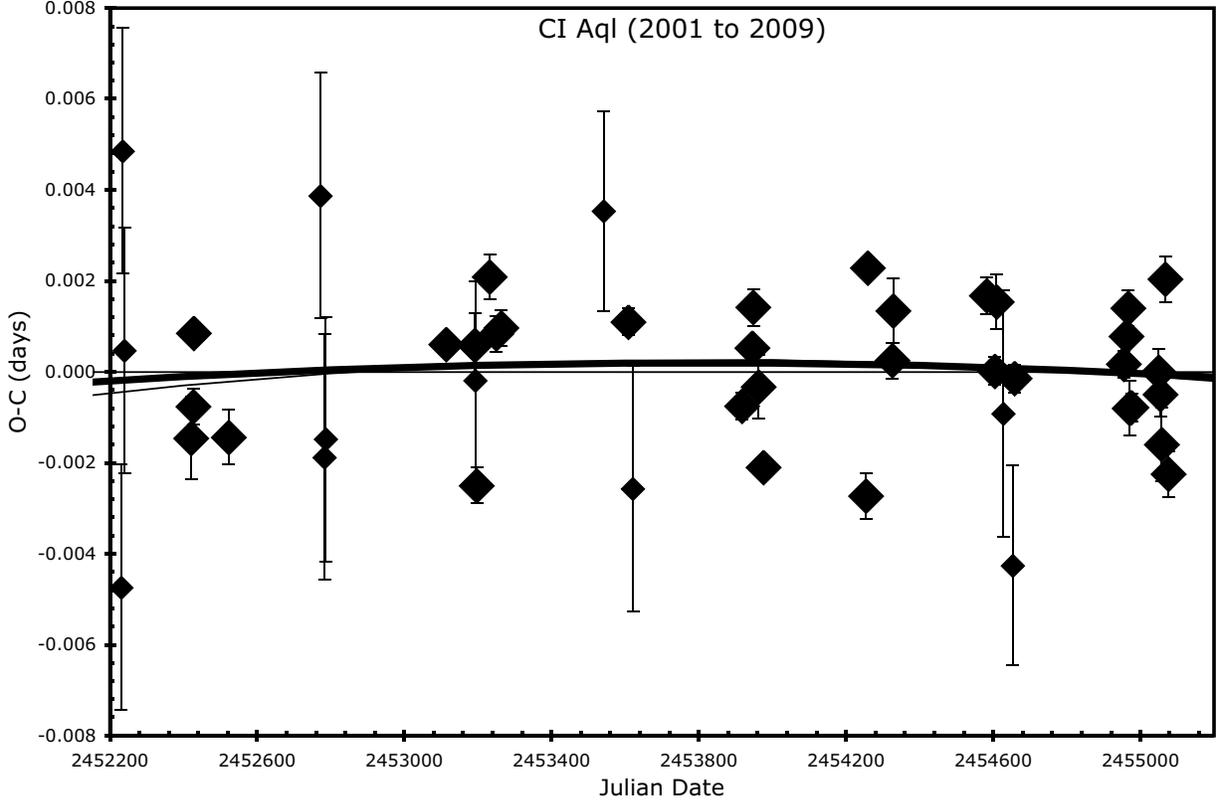}
\caption{
CI Aql O-C curve for 2001-2009.  The O-C curve shows deviations in the eclipse times ($T_{obs}-T_{model}$) from the linear ephemeris of $T_{model}=2451669.0575 + N \times 0.61836051$.  The large diamonds are for the eclipses followed completely through the minimum, for which the measurement uncertainty is $\leq0.0009$ days.  The smaller diamonds are for eclipses with coverage over either the ingress or egress for which the measurement uncertainty is either 0.0022 day for where the minimum is covered or 0.0027 day where it is not.  The scatter of the well-observed eclipse times are substantially larger than the quoted measurement uncertainties, so CI Aql has an intrinsic scatter (likely due to secular variations in the accretion disk brightness through the duration of the eclipse) with an RMS of 0.00136 days (1.9 minutes).  The global best fit model (with $\dot{P}=-12\times10^{-11}$ days per cycle) is plotted as the slightly curved (concave down) thick line.  Also displayed is the best fit model with the minimal acceptable steady period change (indeed, $\dot{P}=0$), with this appearing as a thin flat line along the O-C equal to zero line.  The model with the maximum acceptable steady period change ($\dot{P}=-20\times10^{-11}$ days per cycle) is displayed as the thin curved line.  The point to take from this figure is that I have a very well measured post-eruption period.}
\end{figure}

\clearpage
\begin{figure}
\epsscale{1.0}
\plotone{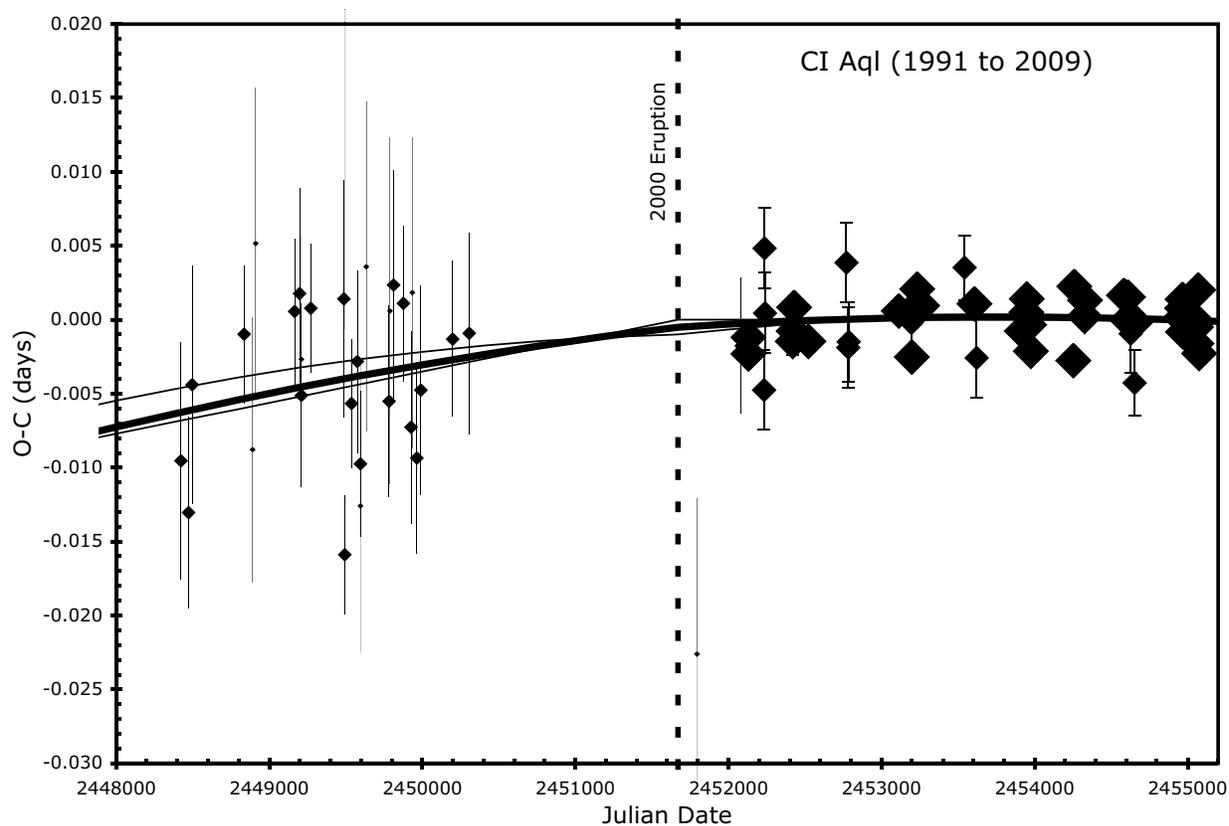}
\caption{
CI Aql O-C curve for 1991-2009.  The O-C curve is as in Figure 12.  The two sizes of small diamonds indicate eclipse times with one-sigma uncertainties of 0.4-0.8 and $\geq$0.8 day.  The added data is for the 1991-1996 Roboscope measures of Mennickent \& Honeycutt (1995) plus some eclipse times during the tail of the eruption.  The only measure during the early tail of the eruption (days 40-200 after peak) show a large offset in time, being roughly 32 minutes early, with this being due to the expected shift in the center of light between the eruption and quiescence.  The vertical dashed line represents the date of the eruption in the year 2000.  The global best fit model (with a period change of -0.00000037) goes through the middle of all the data.  As in Fig. 6, the maximal and minimal acceptable curvatures are also plotted as thin curves.  Characteristically, for the minimum curvature case the period change across the eruption is negative (for a {\it decreasing} period), while for the maximal curvature case the period change is positive (for an {\it increasing} period).  Even though the best fit $\Delta P$ is negative, values that are zero or positive are also possible, with all this going to say that the real $\Delta P$ value for CI Aql is very small and near zero.}
\end{figure}

\clearpage
\begin{figure}
\epsscale{1.0}
\plotone{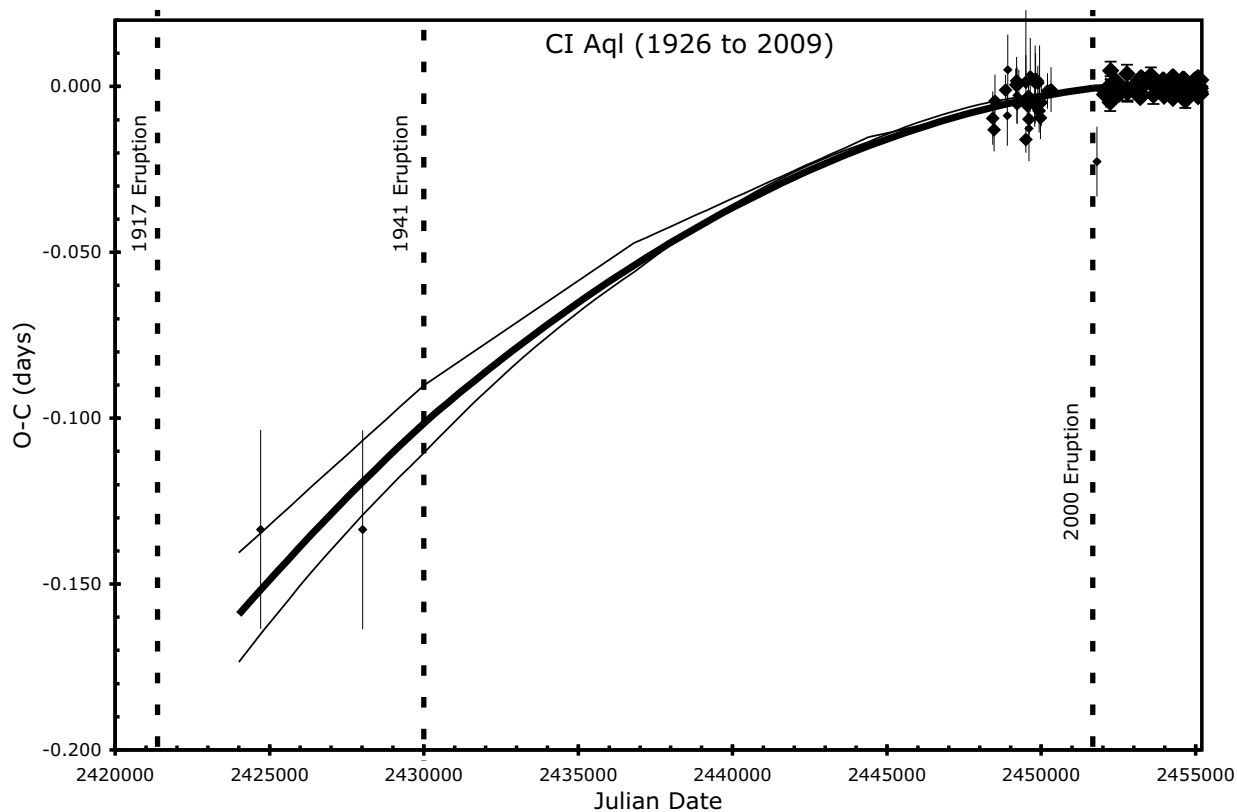}
\caption{
CI Aql O-C curve for 1926-2009.  The O-C curve is as in Figures 12 and 13.  We see that the best fit model (the thick curve) passes right through the 1926 and 1937 eclipse times.  The models with maximal and minimal $\dot{P}$ both require undiscovered eruptions from around 1960 and 1980.  The constraints on $\dot{P}$ (and hence the constraints on $\Delta P$) imposed by the 1926 and 1935 eclipses are close to identical to those obtained from just considering the 1991-2009 data alone.  Thus, the final period change does not depend on the old eclipses, but these do provide valuable and close confirmation.}
\end{figure}

\clearpage
\begin{figure}
\epsscale{1.0}
\plotone{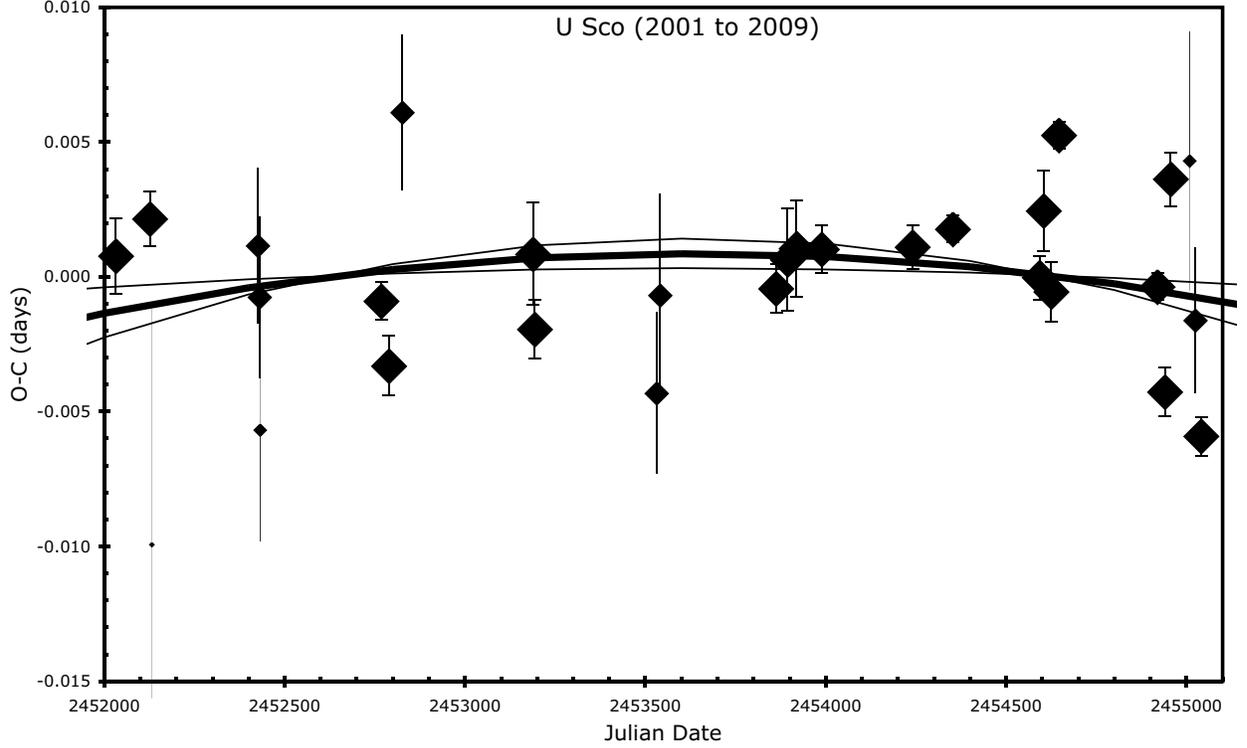}
\caption{
U Sco O-C curve for 2001-2009.  This O-C curve is for the best observed inter-eruption interval, and thus provides the baseline period ($T_{model}=2451234.5387 + N \times 1.23054695$) as well as the best constraints on the curvature.  The large diamonds are for the eclipses with measurement uncertainty $\leq0.0020$ days.  The smaller diamonds are for eclipses with measurement uncertainty 0.0022-0.0040 days.  The scatter of the well-observed eclipse times are substantially larger than the quoted measurement uncertainties, so U Sco has an intrinsic timing jitter with an RMS of 0.00242 days (3.5 minutes).  The global best fit model (with $\dot{P}=-250\times10^{-11}$ days per cycle) is plotted as the slightly curved (concave down) thick line.  The two other curves (with thin lines) are for the extreme acceptable curvatures (as based on the 2001-2009 times alone), with $\dot{P}=+110\times10^{-11}$ days per cycle for the curve that is slightly concave up and $\dot{P}=-330\times10^{-11}$ days per cycle for the curve that is concave down.  We see that the eclipse times are consistent with either a positive, zero, or negative $\dot{P}$.}
\end{figure}

\clearpage
\begin{figure}
\epsscale{1.0}
\plotone{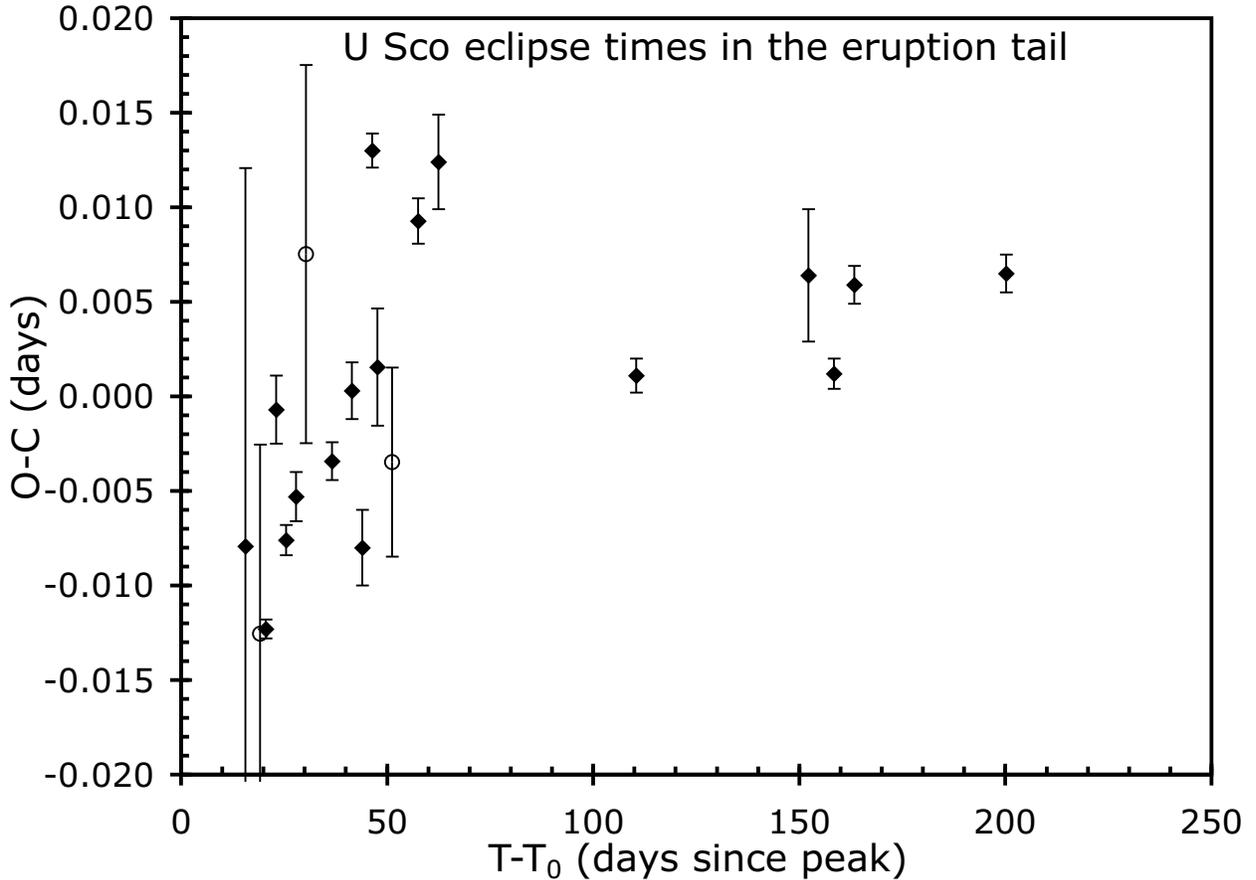}
\caption{
Times for U Sco eruption eclipses.  Deep eclipses are seen in the tails of the 1999 and 2010 eruptions (empty squares and filled circles respectively).  These eclipse times deviate greatly from the ephemeris for eclipses in quiescence.  What is going on is that the center of light shifts from the case for the quiescent system to the case for the system in eruption.  In this figure, the deviations shift systematically throughout the eclipse, with the eclipse minima being {\it early} during the plateau phase (roughly 15-33 days after the peak) and being {\it late} during the post-plateau phase.}
\end{figure}

\clearpage
\begin{figure}
\epsscale{1.0}
\plotone{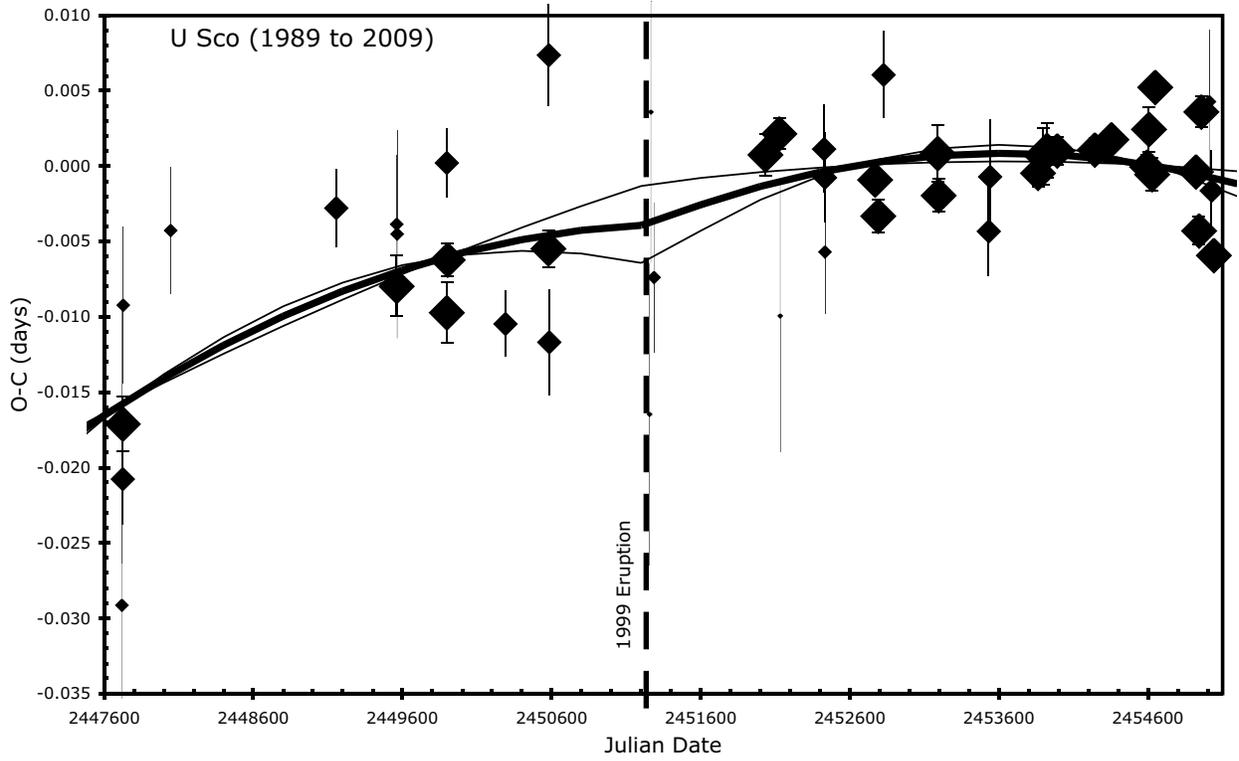}
\caption{
O-C curve for U Sco from 1989-2009.  The eclipse times from before the 1999 eruption and the post-eruption eclipse times together gives a best fit value for the abrupt period change across the eruption.  With a chi-square analysis, I find $\Delta P = (+43 \pm 66) \times 10^{-7}$ days in correlated changes with $\dot{P}=(-250 \pm 170) \times10^{-11}$ days per cycle.  This is one of the two primary results from this paper.   The $\Delta P$ value is positive, but is sufficiently small compared to the uncertainty that it might be zero or negative.}
\end{figure}

\clearpage
\begin{figure}
\epsscale{1.0}
\plotone{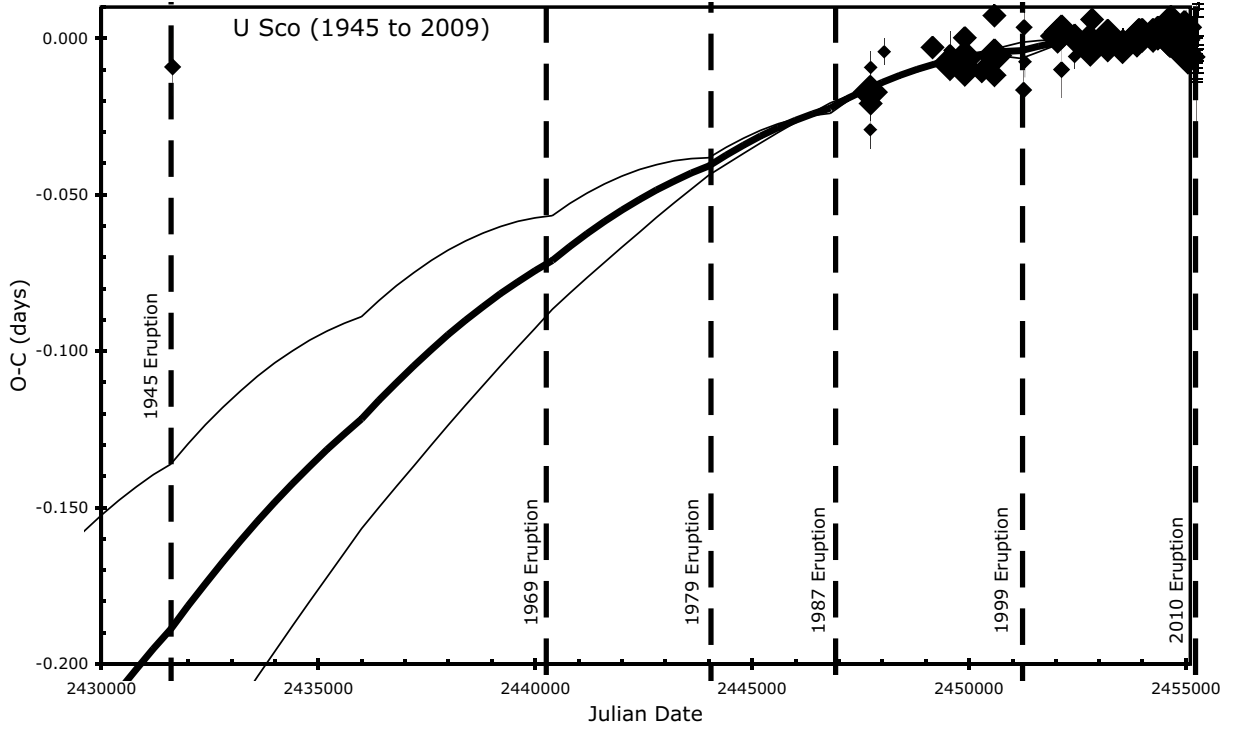}
\caption{
O-C curve for U Sco from 1945-2009.  The Harvard archival plates show an eclipse of U Sco in the tail of the 1945 eruption, and the time of this eclipse produces a point in the O-C plot in the upper left part of this plot, where the point is in a gap of the dashed line indicating the date of the 1945 eruption.  This plot shows vertical dashed lines for the discovered eruptions (in 1945, 1969, 1979, 1987, 1999, and 2010).  The thick curve is the best fit for the 1989-2009 non-eruption eclipse times, while the two thin curves are for the one-sigma extreme solutions.  Especially in the upper curve, we can see the kink caused by the $\Delta P$ at each eruption.  I have assumed that there was a missed eruption in the year 1957, so we see a kink in that year that does not have an associated vertical dashed line.  The extrapolation of the various ephemerides to dates earlier than 1989 all fall low, with O-C values of $\sim-0.2$ in 1945.  The extrapolated ephemerides all fall far below the 1945 eclipse time.  We see from Figure 9 that there is no possibility of the eclipse time being at a phase of 0.84 (corresponding to O-C=-0.2 days).  I interpret this difference as evidence that mass accretion rate changes on all time scales, so the $\dot{P}$ must suffer secular variations, and so the progression of the O-C curve backwards in time must have breaks up and down.  The middle curve in Figure 11 illustrates the sort of O-C curve that is expected in such a case.}
\end{figure}

\clearpage
\begin{figure}
\epsscale{1.1}
\plottwo{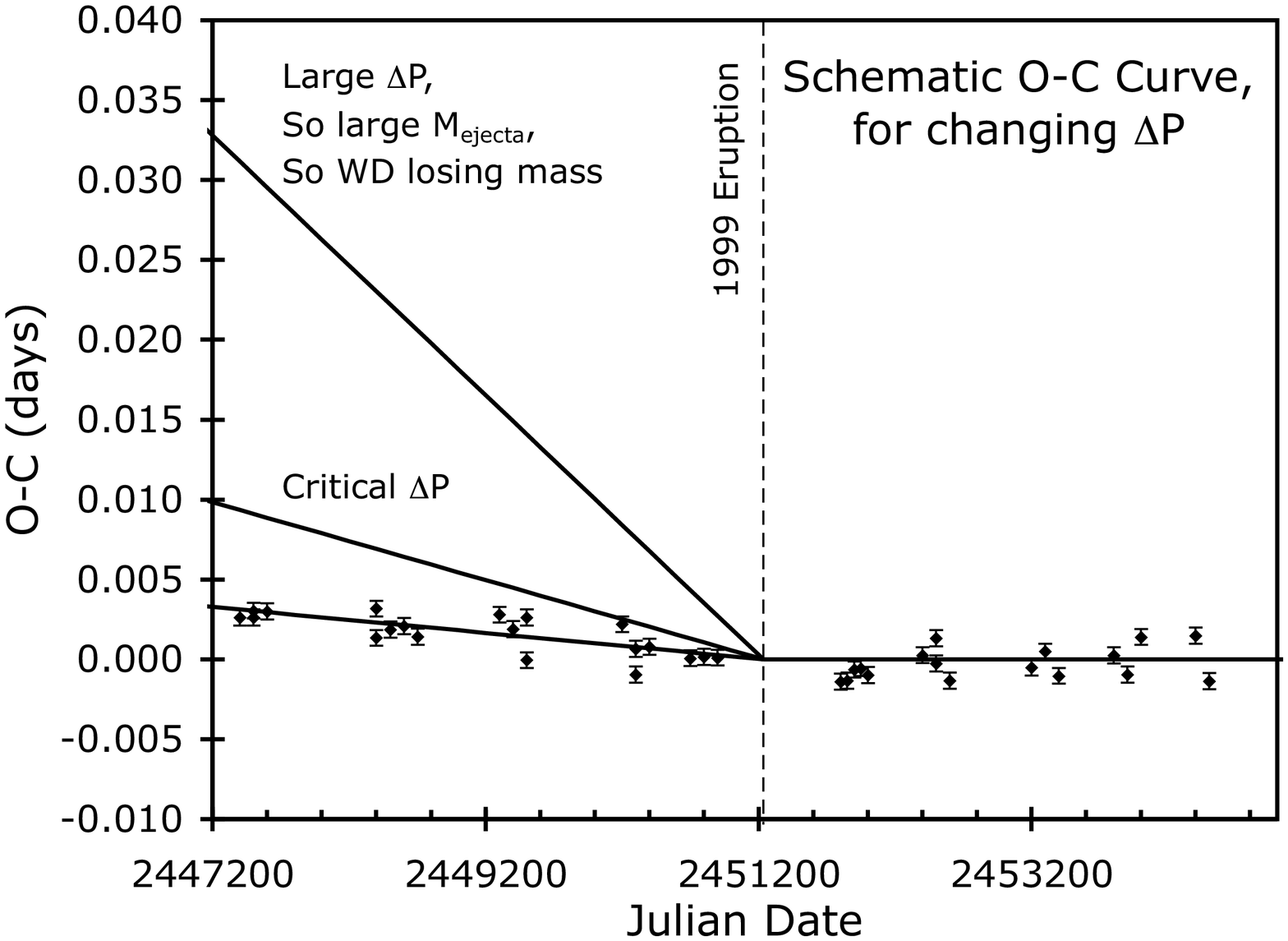}{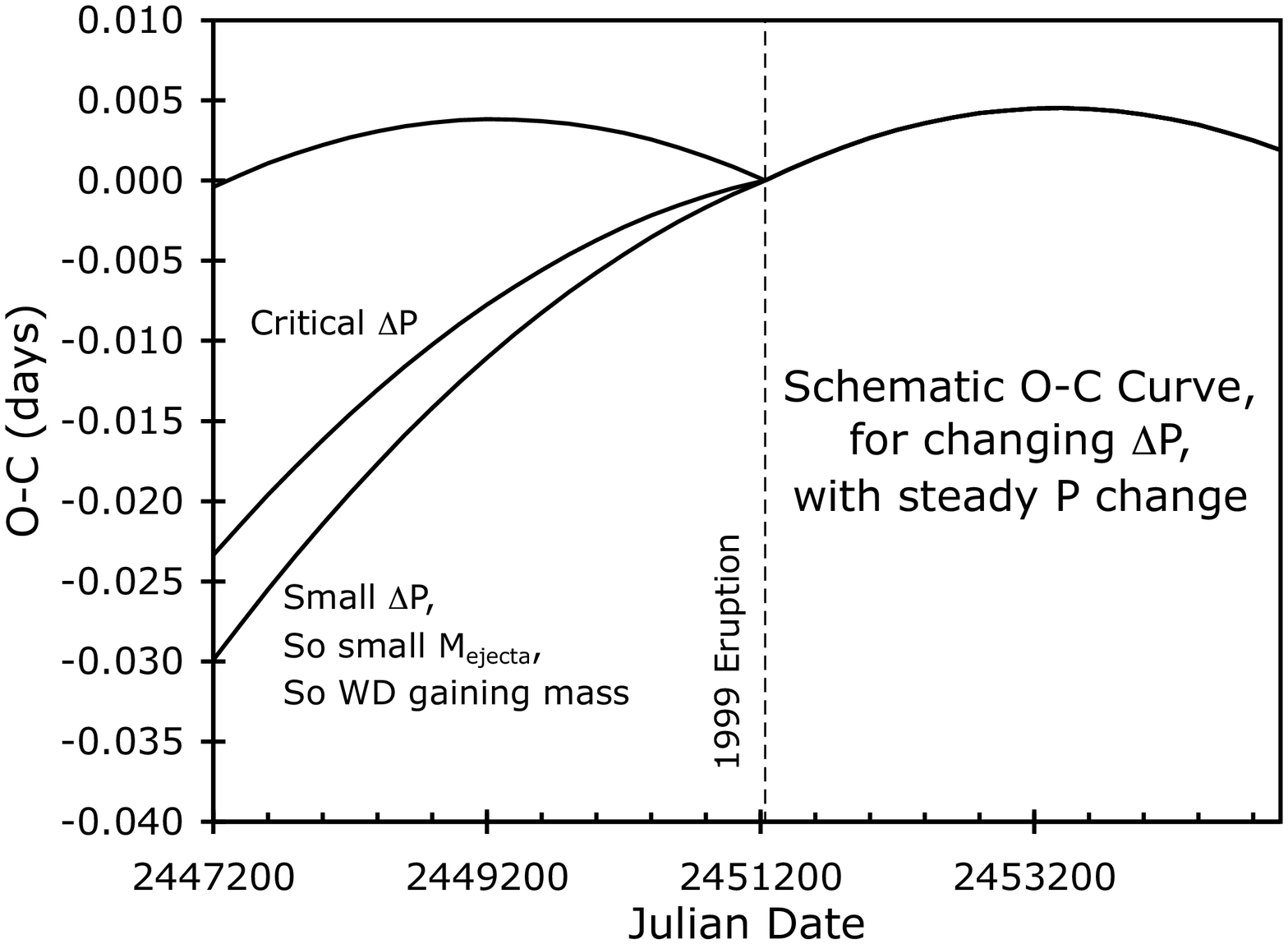}
\caption{
Schematic O-C curves.  The two panels illustrate the effects of $\Delta P$ and $\dot{P}$ on the O-C curve for a nova across an eruption.  The data of the eruption is indicated by a vertical dashed line.  The three theoretical tracks are for three models each with different values of $\Delta P$, with the middle being for the critical value where the ejected mass just equals the mass accreted over the last inter-eruption cycle.  The linear ephemeris used for constructing the O-C curve is taken as the best fitting linear model for the entire post-eruption interval.  The left panel shows the case with $\dot{P}=0$, so the post-eruption and pre-eruption segments are both straight line segments.  The left panel also has idealized observed eclipse times plotted, in this case indicating that the period change was small, so the mass ejected was small, so that the WD is gaining mass through each eruption cycle.  The right panel shows the case for a negative $\dot{P}$, so each segment has a parabolic shape with the concave side down.  The top curve has a large $\Delta P$, as shown by the large change in slope at the time of the eruption, and so the overall period remains similar across the eruption.  The bottom curve has a small change in slope across the eruption, yet we can see that it lies below the middle curve with the critical $\Delta P$, so we would know that the ejected mass is too small for the WD to be gaining mass over the eruption cycle.}
\end{figure}

\clearpage
\begin{figure}
\epsscale{1.0}
\plotone{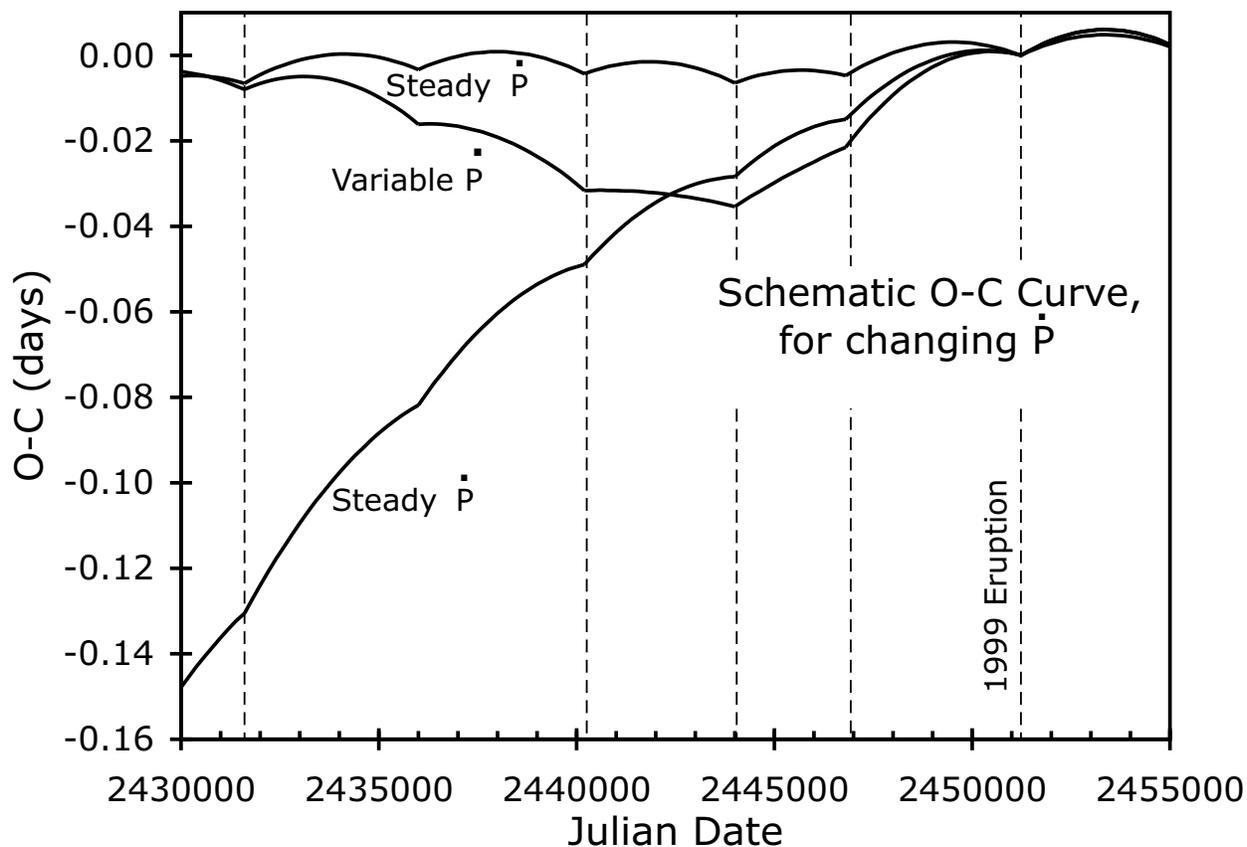}
\caption{
Schematic long term O-C curve.  This O-C curve illustrates the interplay of the steady period change during quiescence (measured by $\dot{P}$) and the abrupt period change across each eruption (measured by $\Delta P$).  As time increases, a negative $\dot{P}$ makes for parabolic shaped curves with the concave side down, while the mass ejection gives a positive $\Delta P$ and an upward kink.  The combination of the parabolas and kinks will determine the long term period change.  The upper curve shows the case with relatively small $\dot{P}$, such that the long term period remains roughly constant.  The lower curve shows the case where the $\dot{P}$ is relatively large, and so this dominates over $\Delta P$ in the long term changes.  The $\Delta P$ value is held constant for all the curves.  Novae and RNe have their brightness varying with surprisingly large amplitude on all time scales, and this implies that the $\dot{P}$ will change on all time scales too.  The middle curve illustrates the case where the $\dot{P}$ changes from eruption-to-eruption.}
\end{figure}

\end{document}